\UseRawInputEncoding
\documentclass[%
 reprint,
superscriptaddress,
 amsmath,amssymb,
 aps,
floatfix,
]{revtex4-2}

\usepackage{graphicx}
\usepackage{dcolumn}
\usepackage[usenames,dvipsnames,svgnames,table]{xcolor}	
\usepackage[normalem]{ulem}
\usepackage{color}
\usepackage{bm}
\RequirePackage{graphicx}
\RequirePackage{color}
\usepackage{mathtools}
\usepackage{extarrows}
\usepackage{bm}
\RequirePackage{latexsym}
\usepackage{caption}
\usepackage{subcaption}
\RequirePackage{xspace}

\usepackage{amsmath,amsfonts,bm}
\usepackage{wasysym}
\usepackage{color}
\usepackage{lineno}
\usepackage[usenames,dvipsnames,svgnames,table]{xcolor}	
\usepackage{appendix}

\newcommand{\M}{\mathcal{M}}
\newcommand{\m}{$\mathcal{M}$\xspace}

\newcommand{\mone}{$\mathcal{M}_{1}$\xspace}

\newcommand{\vbb}{$\nu\beta\beta$\xspace}
\newcommand{\nbb}{$\nu\beta\beta$\xspace}
\newcommand{\bb}{$\beta\beta$\xspace}
\newcommand{\es}{e.s.\xspace}
\newcommand{\gammazero}{(8.9\pm 1.0 \ \text{(stat.+syst.)} ) \times 10^{-22} \, \mathrm{yr^{-1}}}

\newcommand{\gammazeroSSD}{(9.2\pm 1.0 \ \text{(stat.+syst.)}  )\times 10^{-22} \, \mathrm{yr^{-1}}}

\newcommand{\gammazerostat}{(9.1\pm 1.0\ \text{(stat.)}  ) \times 10^{-22} \, \mathrm{yr^{-1}}}
\newcommand{\gammazerostatandsyst}{(9.2\pm 1.0\ \text{(stat.)} \  ^{+0.3}_{-0.5} \ \text{(syst.)}  )\times 10^{-22} \, \mathrm{yr^{-1}}}

\newcommand{\Thalftwonu}{(7.5\pm 0.8 \ \text{(stat.)} \ ^{+ 0.4}_{-0.3} \ \text{(syst.)}  )\times 10^{20} \ \mathrm{yr}}
\newcommand{\Limtwoplus}{4.4\times 10^{21} \ \mathrm{yr} \ \text{(90\% c.i.)}}
\newcommand{\Senstwonu}{3.8\times 10^{21} \ \mathrm{yr}}
\newcommand{\Senszero}{9.8\times 10^{22} \ \mathrm{yr}}
\newcommand{\Senstwo}{3.1\times 10^{23} \ \mathrm{yr}}

\newcommand{\Tzeronutwo}{2.1\times10^{23} \ \mathrm{yr}\ \text{(90\% c.i.)}} 
\newcommand{\Tzeronuzero}{1.2\times10^{23} \ \mathrm{yr}\ \text{(90\% c.i.)}} 
\newcommand{\twonuprob}{0.49} 

\newcommand{\SpectralThalf}{(7.3\pm 1.0 \ \text{(stat.)} )\times 10^{20} \, \mathrm{yr}}
\newcommand{\SpectralLimit}{3.6 \times 10^{21}\,\mathrm{yr}}

\newcommand{\Mtwonu}{0.143\pm 0.008\ \text{(stat.+syst.)}}

\newcommand{\pSSD}{0.51}\newcommand{\pHSD}{0.49}

\begin{document}

\preprint{APS/123-QED}

\title{New measurement of double beta decays of $^{100}$Mo to excited states of $^{100}$Ru with the CUPID-Mo experiment}

\collaboration{CUPID-Mo collaboration}

\date{\today}


\author{C.~Augier }
\affiliation{Univ Lyon, Universit\'{e} Lyon 1, CNRS/IN2P3, IP2I-Lyon, F-69622, Villeurbanne, France }

\author{A.~S.~Barabash }
\affiliation{National Research Centre Kurchatov Institute, Institute of Theoretical and Experimental Physics, 117218 Moscow, Russia }

\author{F.~Bellini }
\affiliation{Dipartimento di Fisica, Sapienza Universit\`a di Roma, P.le Aldo Moro 2, I-00185, Rome, Italy }
\affiliation{INFN, Sezione di Roma, P.le Aldo Moro 2, I-00185, Rome, Italy}

\author{G.~Benato }
\affiliation{INFN, Laboratori Nazionali del Gran Sasso, I-67100 Assergi (AQ), Italy }

\author{M.~Beretta }
\affiliation{ University of California, Berkeley, California 94720, USA }

\author{L.~Berg\'e }
\affiliation{Universit\'e Paris-Saclay, CNRS/IN2P3, IJCLab, 91405 Orsay, France }

\author{J.~Billard }
\affiliation{Univ Lyon, Universit\'{e} Lyon 1, CNRS/IN2P3, IP2I-Lyon, F-69622, Villeurbanne, France }

\author{Yu.~A.~Borovlev }
\affiliation{Nikolaev Institute of Inorganic Chemistry, 630090 Novosibirsk, Russia }

\author{L.~Cardani }
\affiliation{INFN, Sezione di Roma, P.le Aldo Moro 2, I-00185, Rome, Italy}

\author{N.~Casali }
\affiliation{INFN, Sezione di Roma, P.le Aldo Moro 2, I-00185, Rome, Italy}

\author{A.~Cazes }
\affiliation{Univ Lyon, Universit\'{e} Lyon 1, CNRS/IN2P3, IP2I-Lyon, F-69622, Villeurbanne, France }

\author{M.~Chapellier }
\affiliation{Universit\'e Paris-Saclay, CNRS/IN2P3, IJCLab, 91405 Orsay, France }

\author{D.~Chiesa}
\affiliation{Dipartimento di Fisica, Universit\`{a} di Milano-Bicocca, I-20126 Milano, Italy }
\affiliation{INFN, Sezione di Milano-Bicocca, I-20126 Milano, Italy}

\author{I.~Dafinei }
\affiliation{INFN, Sezione di Roma, P.le Aldo Moro 2, I-00185, Rome, Italy}

\author{F.~A.~Danevich }
\affiliation{Institute for Nuclear Research of NASU, 03028 Kyiv, Ukraine }

\author{M.~De~Jesus }
\affiliation{Univ Lyon, Universit\'{e} Lyon 1, CNRS/IN2P3, IP2I-Lyon, F-69622, Villeurbanne, France }

\author{T.~Dixon}\email{toby.dixon@universite-paris-saclay.fr}
\affiliation{Universit\'e Paris-Saclay, CNRS/IN2P3, IJCLab, 91405 Orsay, France }
\affiliation{IRFU, CEA, Universit\'{e} Paris-Saclay, F-91191 Gif-sur-Yvette, France }
\author{L.~Dumoulin }
\affiliation{Universit\'e Paris-Saclay, CNRS/IN2P3, IJCLab, 91405 Orsay, France }

\author{K.~Eitel }
\affiliation{Karlsruhe Institute of Technology, Institute for Astroparticle Physics, 76021 Karlsruhe, Germany }

\author{F.~Ferri }
\affiliation{IRFU, CEA, Universit\'{e} Paris-Saclay, F-91191 Gif-sur-Yvette, France }

\author{B.~K.~Fujikawa }
\affiliation{ Lawrence Berkeley National Laboratory, Berkeley, California 94720, USA }

\author{J.~Gascon }
\affiliation{Univ Lyon, Universit\'{e} Lyon 1, CNRS/IN2P3, IP2I-Lyon, F-69622, Villeurbanne, France }

\author{L.~Gironi }
\affiliation{Dipartimento di Fisica, Universit\`{a} di Milano-Bicocca, I-20126 Milano, Italy }
\affiliation{INFN, Sezione di Milano-Bicocca, I-20126 Milano, Italy}

\author{A.~Giuliani} 
\affiliation{Universit\'e Paris-Saclay, CNRS/IN2P3, IJCLab, 91405 Orsay, France }

\author{V.~D.~Grigorieva }
\affiliation{Nikolaev Institute of Inorganic Chemistry, 630090 Novosibirsk, Russia }

\author{M.~Gros }
\affiliation{IRFU, CEA, Universit\'{e} Paris-Saclay, F-91191 Gif-sur-Yvette, France }

\author{D.~L.~Helis }
\affiliation{IRFU, CEA, Universit\'{e} Paris-Saclay, F-91191 Gif-sur-Yvette, France }
\affiliation{INFN, Laboratori Nazionali del Gran Sasso, I-67100 Assergi (AQ), Italy }

\author{H.~Z.~Huang }
\affiliation{Key Laboratory of Nuclear Physics and Ion-beam Application (MOE), Fudan University, Shanghai 200433, PR China }

\author{R.~Huang }
\affiliation{ University of California, Berkeley, California 94720, USA }

\author{L.~Imbert}
\affiliation{Universit\'e Paris-Saclay, CNRS/IN2P3, IJCLab, 91405 Orsay, France }

\author{J.~Johnston }
\affiliation{Massachusetts Institute of Technology, Cambridge, MA 02139, USA }

\author{A.~Juillard }
\affiliation{Univ Lyon, Universit\'{e} Lyon 1, CNRS/IN2P3, IP2I-Lyon, F-69622, Villeurbanne, France }

\author{H.~Khalife }
\affiliation{Universit\'e Paris-Saclay, CNRS/IN2P3, IJCLab, 91405 Orsay, France }

\author{M.~Kleifges }
\affiliation{Karlsruhe Institute of Technology, Institute for Data Processing and Electronics, 76021 Karlsruhe, Germany }

\author{V.~V.~Kobychev }
\affiliation{Institute for Nuclear Research of NASU, 03028 Kyiv, Ukraine }

\author{Yu.~G.~Kolomensky }
\affiliation{ University of California, Berkeley, California 94720, USA }
\affiliation{ Lawrence Berkeley National Laboratory, Berkeley, California 94720, USA }%

\author{S.I.~Konovalov }
\affiliation{National Research Centre Kurchatov Institute, Institute of Theoretical and Experimental Physics, 117218 Moscow, Russia }

\author{J.~Kotilla}
\affiliation{Department of Physics, University of Jyv\"{a}skyl\"{a}, PO Box 35, FI-40014, Jyv\"{a}skyl\"{a}, Finland }
\affiliation{Finnish Institute for Educational Research, University of Jyv\"{a}skyl\"{a}, P.O. Box 35, FI-40014 Jyva\"{a}skyl\"{a}, Finland }
\affiliation{Center for Theoretical Physics, Sloane Physics Laboratory, Yale University, New Haven, Connecticut 06520-8120, USA}
\author{P.~Loaiza }
\affiliation{Universit\'e Paris-Saclay, CNRS/IN2P3, IJCLab, 91405 Orsay, France }

\author{L.~Ma }
\affiliation{Key Laboratory of Nuclear Physics and Ion-beam Application (MOE), Fudan University, Shanghai 200433, PR China }

\author{E.~P.~Makarov }
\affiliation{Nikolaev Institute of Inorganic Chemistry, 630090 Novosibirsk, Russia }

\author{P.~de~Marcillac }
\affiliation{Universit\'e Paris-Saclay, CNRS/IN2P3, IJCLab, 91405 Orsay, France }

\author{R.~Mariam}
\affiliation{Universit\'e Paris-Saclay, CNRS/IN2P3, IJCLab, 91405 Orsay, France }

\author{L.~Marini }
\affiliation{ University of California, Berkeley, California 94720, USA }
\affiliation{ Lawrence Berkeley National Laboratory, Berkeley, California 94720, USA }
\affiliation{INFN, Laboratori Nazionali del Gran Sasso, I-67100 Assergi (AQ), Italy }

\author{S.~Marnieros }
\affiliation{Universit\'e Paris-Saclay, CNRS/IN2P3, IJCLab, 91405 Orsay, France }

\author{X.-F.~Navick }
\affiliation{IRFU, CEA, Universit\'{e} Paris-Saclay, F-91191 Gif-sur-Yvette, France }

\author{C.~Nones }
\affiliation{IRFU, CEA, Universit\'{e} Paris-Saclay, F-91191 Gif-sur-Yvette, France }

\author{E.~B.~Norman}
\affiliation{ University of California, Berkeley, California 94720, USA }

\author{E.~Olivieri }
\affiliation{Universit\'e Paris-Saclay, CNRS/IN2P3, IJCLab, 91405 Orsay, France }

\author{J.~L.~Ouellet }
\affiliation{Massachusetts Institute of Technology, Cambridge, MA 02139, USA }

\author{L.~Pagnanini }
\affiliation{INFN, Gran Sasso Science Institute, I-67100 L'Aquila, Italy}
\affiliation{INFN, Laboratori Nazionali del Gran Sasso, I-67100 Assergi (AQ), Italy }

\author{L.~Pattavina }
\affiliation{INFN, Laboratori Nazionali del Gran Sasso, I-67100 Assergi (AQ), Italy }
\affiliation{Physik Department, Technische Universit\"at M\"unchen, Garching D-85748, Germany }

\author{B.~Paul }
\affiliation{IRFU, CEA, Universit\'{e} Paris-Saclay, F-91191 Gif-sur-Yvette, France }

\author{M.~Pavan }
\affiliation{Dipartimento di Fisica, Universit\`{a} di Milano-Bicocca, I-20126 Milano, Italy }
\affiliation{INFN, Sezione di Milano-Bicocca, I-20126 Milano, Italy}

\author{H.~Peng }
\affiliation{Department of Modern Physics, University of Science and Technology of China, Hefei 230027, PR China }

\author{G.~Pessina }
\affiliation{INFN, Sezione di Milano-Bicocca, I-20126 Milano, Italy}

\author{S.~Pirro }
\affiliation{INFN, Laboratori Nazionali del Gran Sasso, I-67100 Assergi (AQ), Italy }

\author{D.~V.~Poda }
\affiliation{Universit\'e Paris-Saclay, CNRS/IN2P3, IJCLab, 91405 Orsay, France }

\author{O.~G.~Polischuk }
\affiliation{Institute for Nuclear Research of NASU, 03028 Kyiv, Ukraine }

\author{S.~Pozzi }
\affiliation{INFN, Sezione di Milano-Bicocca, I-20126 Milano, Italy}

\author{E.~Previtali }
\affiliation{Dipartimento di Fisica, Universit\`{a} di Milano-Bicocca, I-20126 Milano, Italy }
\affiliation{INFN, Sezione di Milano-Bicocca, I-20126 Milano, Italy}

\author{Th.~Redon }
\affiliation{Universit\'e Paris-Saclay, CNRS/IN2P3, IJCLab, 91405 Orsay, France }

\author{A.~Rojas }
\affiliation{LSM, Laboratoire Souterrain de Modane, 73500 Modane, France }

\author{S.~Rozov }
\affiliation{Laboratory of Nuclear Problems, JINR, 141980 Dubna, Moscow region, Russia }

\author{V.~Sanglard }
\affiliation{Univ Lyon, Universit\'{e} Lyon 1, CNRS/IN2P3, IP2I-Lyon, F-69622, Villeurbanne, France }

\author{J.~A.~Scarpaci}
\affiliation{Universit\'e Paris-Saclay, CNRS/IN2P3, IJCLab, 91405 Orsay, France }

\author{B.~Schmidt } \thanks{Now at: Northwestern University, Evanston, IL 60208, USA }
\affiliation{ Lawrence Berkeley National Laboratory, Berkeley, California 94720, USA }

\author{Y.~Shen }
\affiliation{Key Laboratory of Nuclear Physics and Ion-beam Application (MOE), Fudan University, Shanghai 200433, PR China }

\author{V.~N.~Shlegel }
\affiliation{Nikolaev Institute of Inorganic Chemistry, 630090 Novosibirsk, Russia }

\author{V.~Singh }
\affiliation{ University of California, Berkeley, California 94720, USA }

\author{C.~Tomei }
\affiliation{INFN, Sezione di Roma, P.le Aldo Moro 2, I-00185, Rome, Italy}

\author{V.~I.~Tretyak }
\affiliation{Institute for Nuclear Research of NASU, 03028 Kyiv, Ukraine }

\author{V.~I.~Umatov }
\affiliation{National Research Centre Kurchatov Institute, Institute of Theoretical and Experimental Physics, 117218 Moscow, Russia }

\author{L.~Vagneron }
\affiliation{Univ Lyon, Universit\'{e} Lyon 1, CNRS/IN2P3, IP2I-Lyon, F-69622, Villeurbanne, France }

\author{M.~Vel\'azquez }
\affiliation{Universit\'e Grenoble Alpes, CNRS, Grenoble INP, SIMAP, 38420 Saint Martin d'H\`Weieres, France }

\author{B.~Welliver }
\affiliation{ University of California, Berkeley, California 94720, USA }

\author{L.~Winslow }
\affiliation{Massachusetts Institute of Technology, Cambridge, MA 02139, USA }

\author{M.~Xue }
\affiliation{Department of Modern Physics, University of Science and Technology of China, Hefei 230027, PR China }

\author{E.~Yakushev }
\affiliation{Laboratory of Nuclear Problems, JINR, 141980 Dubna, Moscow region, Russia }

\author{M.~Zarytskyy}
\affiliation{Institute for Nuclear Research of NASU, 03028 Kyiv, Ukraine }

\author{A.~S.~Zolotarova }
\affiliation{Universit\'e Paris-Saclay, CNRS/IN2P3, IJCLab, 91405 Orsay, France }

\begin{abstract}
    
The CUPID-Mo experiment, located at Laboratoire Souterrain de Modane (France), was a demonstrator experiment for CUPID. It consisted of an array of 20 Li$_2^{100}$MoO$_4$ (LMO) calorimeters each equipped with a Ge light detector (LD) for particle identification. In this work, we present the result of a search for two-neutrino and neutrinoless double beta decays of $^{100}$Mo to the first 0$^+$ and $2^+$ excited states of $^{100}$Ru using the full CUPID-Mo exposure (2.71 kg$\times$yr of LMO).  
We measure the half-life of $2\nu\beta\beta$ decay to the $0^{+}_1$ state as $T_{1/2}^{2\nu \rightarrow 0_1^+}=\Thalftwonu$. The bolometric technique enables measurement of the electron energies as well as the gamma rays from nuclear de-excitation and this allows us to set new limits on the two-neutrino decay to the $2_1^+$ state of $T^{2\nu \rightarrow 2_1^+}_{1/2}>\Limtwoplus$ and on the neutrinoless modes of $T_{1/2}^{0\nu\rightarrow 2_1^+}>\Tzeronutwo$, $T_{1/2}^{0\nu\rightarrow 0_1^+}>\Tzeronuzero$.  
Information on the electrons spectral shape is obtained which allows us to make the first comparison of the single state (SSD) and higher state (HSD) $2\nu\beta\beta$ decay models for the $0_1^+$ excited state of $^{100}$Ru.

\end{abstract}

\maketitle

\section{Introduction}
Two-neutrino double beta decay ($2\nu\beta\beta)$ is an allowed Standard Model process which occurs in some even-even nuclei for which single beta decays are energetically forbidden or heavily disfavored due to large changes in angular momentum~\cite{Saakyan:2013,Barabash:2020}. These decays have been observed to the $0^+$ ground states ($0^+_{\text{g.s.}}$) in eleven nuclei and to the first zero-plus excited state ($0_1^+$) for two nuclei, $^{100}$Mo and $^{150}$Nd~\cite{Saakyan:2013,Barabash:2020,Barabash:2017}. In addition, if neutrinos are Majorana particles then an additional decay mode becomes allowed: neutrinoless double beta decay ($0\nu\beta\beta$)~\cite{Goswami2015, PhysRevD.25.2951, Furry:1939}. Whereas 2\vbb \ decay conserves lepton number, 0\vbb decay would violate this symmetry~\cite{Bilenky2015, Goswami2015, Dolinski:2019a} and provide a possible path to explain the predominance of matter over antimatter in the universe~\cite{Dolinski:2019a, Goswami2015, FUKUGITA198645, DAVIDSON2008105,Deppisch:2017}.
\\ \indent
One of the most promising experimental techniques to search for $0\nu\beta\beta$ decay are cryogenic calorimeters. These detectors provide an excellent energy resolution, high detection efficiency and are scalable to tonne scale arrays, such as the CUORE experiment (Cryogenic Underground Observatory for Rare Events) \cite{CUORE1ton}. The CUPID experiment \cite{CUPIDInterestGroup:2019inu} (CUORE Upgrade with Particle IDentification) is an upgrade for CUORE which will use a scintillating bolometer technology to discriminate $\beta/\gamma$ and $\alpha$ particles. CUPID-Mo \cite{Armengaud:2020} was a demonstrator experiment for CUPID aiming to validate the technology of lithium molybdate (LMO) scintillating calorimeters.
\\ \indent
Multiple models exist to describe 0\vbb decay~\cite{Bilenky2015, Dolinski:2019a, Deppisch:2012, Rodejohann:2012, PhysRevD.68.034016, Atre2009, Blennow2010, MITRA201226, Cirigliano2018}, however the minimal extension to the Standard Model needed to explain 0\vbb decay is the light Majorana neutrino exchange mechanism~\cite{Benato:2015}, in which 0\vbb decay is mediated by a Majorana neutrino. In this framework, the decay rate of 0\vbb decay is related to the effective Majorana neutrino mass $\langle m_{\beta\beta}\rangle$ (a weighted sum of the three neutrino masses) by:
\begin{equation}
    (T_{1/2}^{0\nu})^{-1}=G_{0\nu}g_A^4|M_{0\nu}|^2\frac{\langle m_{\beta\beta}\rangle^2}{m_e^2},
\end{equation}
where $T_{1/2}^{0\nu}$ is the 0\vbb decay half-life, $G_{0\nu}$ is the phase space factor, $M_{0\nu}$ is the nuclear matrix element (NME), $m_e$ is the electron mass and $g_A$ is the axial vector coupling constant.
The phase space factor $G_{0\nu}$ can be computed accurately, but the NME is the result of complicated many-body nuclear physics calculations \cite{Engel}. Several models with different approximations are used, but these only agree to within a factor of a few on the value of the NME (cf.~\cite{Rath:2013,Simkovic:2013,Vaquero:2013,Barea:2015}). At present the upper limit on the effective Majorana mass, $\langle m_{\beta\beta}\rangle$, ranges from 60 to 600~\,meV~\cite{Dolinski:2019a,Agostini:2020,Gando:2016,CUORE1ton,Azzolini:2018dyb}. Despite its low exposure, CUPID-Mo has set the most stringent limit on $0\nu\beta\beta$ decay in $^{100}$Mo, with $T_{1/2}^{0\nu}>1.8\times 10^{24} \ \mathrm{yr}$ or $\langle m_{\beta\beta}\rangle<(0.28$\,--\,$0.49)\,\mathrm{eV}$ \cite{CUPIDMo:2022}.

For $2\nu\beta\beta$ decay, the half-life is related to the NME as:
\begin{equation}
  (T_{1/2}^{2\nu})^{-1}=G_{2\nu}|M^{\text{eff}}_{2\nu}|^2,
\end{equation}
where $G_{2\nu}$ is the phase space factor and and $M^{\text{eff}}_{2\nu}$ is the dimensionless effective nuclear matrix element including the factor $g_A^2$, $M^{\text{eff}}_2\nu=g_A^2\times M_{2\nu}$ \cite{Barabash:2020}.
Measurements of $T_{1/2}^{2\nu}$ for both ground states (g.s.) and excited states (\es) of a daughter nucleus provide experimental data which can be used to validate the methods to calculate the NME $M_{2\nu}$, and by extension, provide better confidence in the calculation of $M_{0\nu}$.
\\ \indent
Whilst the phase space, and therefore decay probability, is lower for decays to \es compared to transitions to the g.s., the presence of monochromatic $\gamma$'s can lead to a very clear detection signature. In the case of a positive observation of $0\nu\beta\beta$ decay, measurements of the $0\nu\beta\beta$ decay to \es will be a useful tool to understand the decay mechanism~\cite{Barabash:2017}. Additionally, whilst the $0\nu\beta\beta$ decay to $0_1^+$ state is expected to be suppressed relative to the decay to $0^+_{\text{g.s.}}$ state, it can still improve the overall sensitivity to $\langle m_{\beta\beta} \rangle$. 
Finally, a $2\nu\beta\beta$ decay to $2_1^+$ \es is expected to be strongly suppressed by angular momentum conservation. However, this mode is expected to be much more likely in a framework with Bosonic neutrinos~\cite{DOLGOV20051, Barabash:2007} and can be used to test this model.
\\ \indent
\begin{figure}
    \centering
    \includegraphics[width=0.45\textwidth]{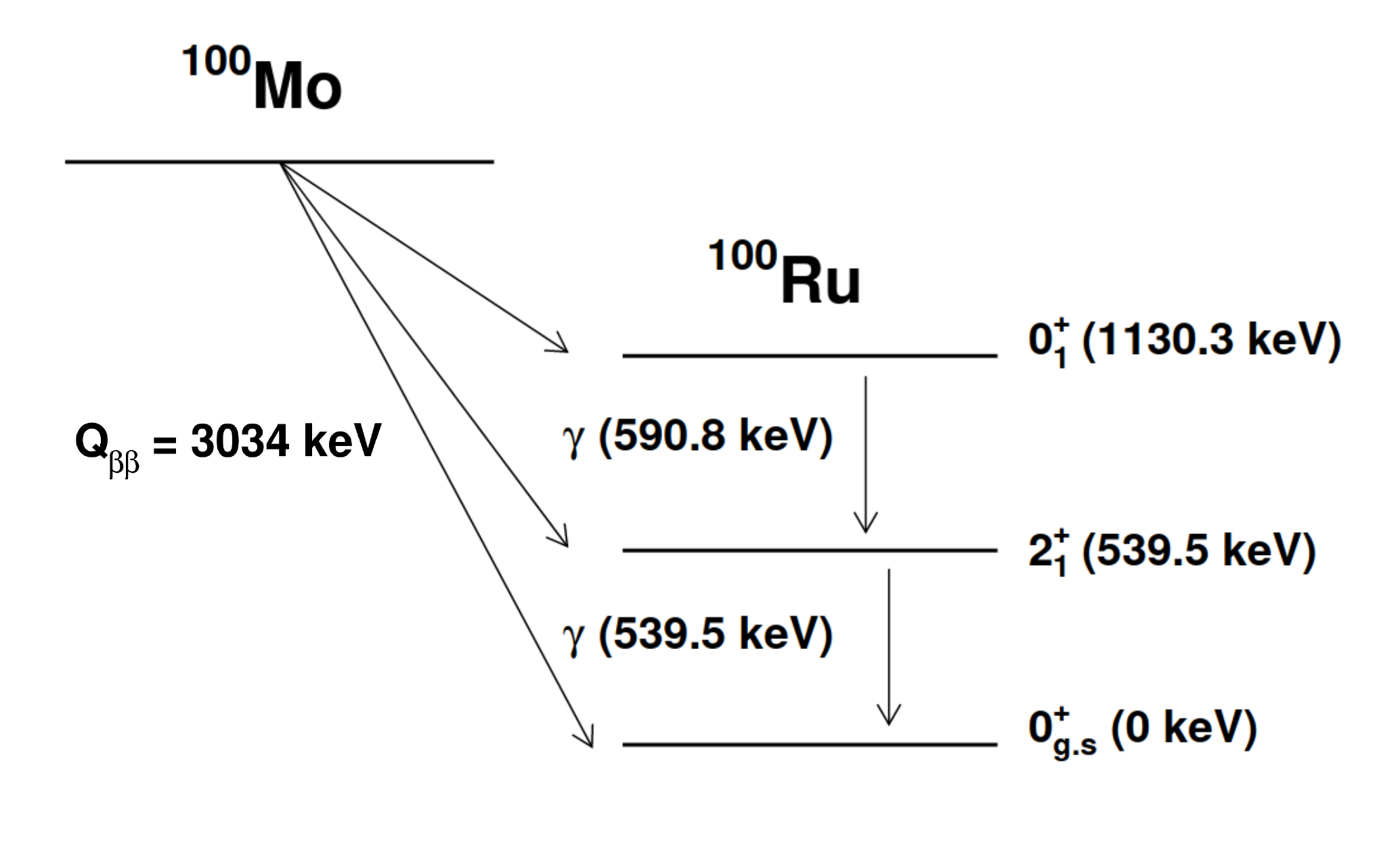}
    \caption{Transition scheme for the decay of $^{100}$Mo to $^{100}$Ru showing the double beta decays to the $0_1^+$ (1130.3 keV) and $2_1^+$ (539.5 keV) states. The decay to the $0_1^+$ state proceeds via the $2_1^+$ state with $\sim100\%$ probability, so both a 539.5 keV and 590.8 keV $\gamma$ are emitted in every decay, while only the $539.5$ keV $\gamma$ is emitted in the double beta decay to the $2_1^+$ state.}
    \label{fig:DecayScheme}
\end{figure}
We can exploit the specific combination of energy depositions associated with these events to measure \bb decays to \es As shown in Figure~\ref{fig:DecayScheme}, the two $\beta$'s are accompanied by one or more $\gamma$ particles. These have a much longer range ($\sim$\,$10 \ \mathrm{cm}$ \cite{NIST_Xray} for 1 MeV $\gamma$) compared to $\beta^{-}$ ($\sim$\,2 mm \cite{NIST_electron}) in LMO. A $\beta\beta$ \es event therefore has a high probability to induce a multi-detector signature with the $\gamma$’s escaping the detector where the decay took place and being measured in neighboring detectors. The additional information provided by these multi-detector events allows us to significantly reduce the background rate while the $\gamma$ peaks provide a clean signature to robustly measure the decay rate \cite{Barabash:1990}. \\ \indent
Since decays to higher \es are generally disfavored by the lower phase space due to lower Q-value (and also by angular momentum conservation for $2^+$ decays) we focus on the lowest energy \es: the first zero plus ($0_1^+$) and two plus ($2_1^+$) \es of $^{100}$Ru at 1130.3 keV and 539.5 keV (see Figure \ref{fig:DecayScheme}). 
\\ \indent
The $2\nu\beta\beta$ decay of $^{100}$Mo to $0_1^+$ \es has been measured in several experiments \cite{Barabash:1995,Barabash:1999,NEMO_ES,Kidd,armonia,NEMO3-ES.2014}, mostly using high purity Germanium (HPGe) counters (cf. \cite{Barabash:2017,universe_2020,Barabash:2020} for a recent review) with an external Mo (enriched) source and also NEMO-3 using a tracking detector and external sources. For the HPGe measurements given that the source and detector did not coincide, only $\gamma$'s were detected. The $2\nu\beta\beta$ decay to $2_1^+$ \es has not been observed and only lower limits on the half-life are available~\cite{NEMO3-ES.2014, NEMO_ES}. Placing constraints on this decay is complicated by the overlap with the decays to the $0_1^+$ state since both decays include a 539.5 keV $\gamma$ emission. By using a setup where the source is embedded in the detector, such as cryogenic calorimeters, we are able to measure the electron energies as well as the $\gamma$'s. This provides additional information to separate the $2\nu\beta\beta$ decays to the $2_1^+$ and $0_1^+$ \es, and to validate that the decay is indeed $2\nu\beta\beta \rightarrow 0_1^+$ and not a background process. This type of detector also allows us to effectively distinguish the neutrinoless and two-neutrino decay modes, which is not possible with HPGe detectors using an external $\beta\beta$ source. 

\begin{figure*}
    \centering
    \includegraphics[width=\textwidth]{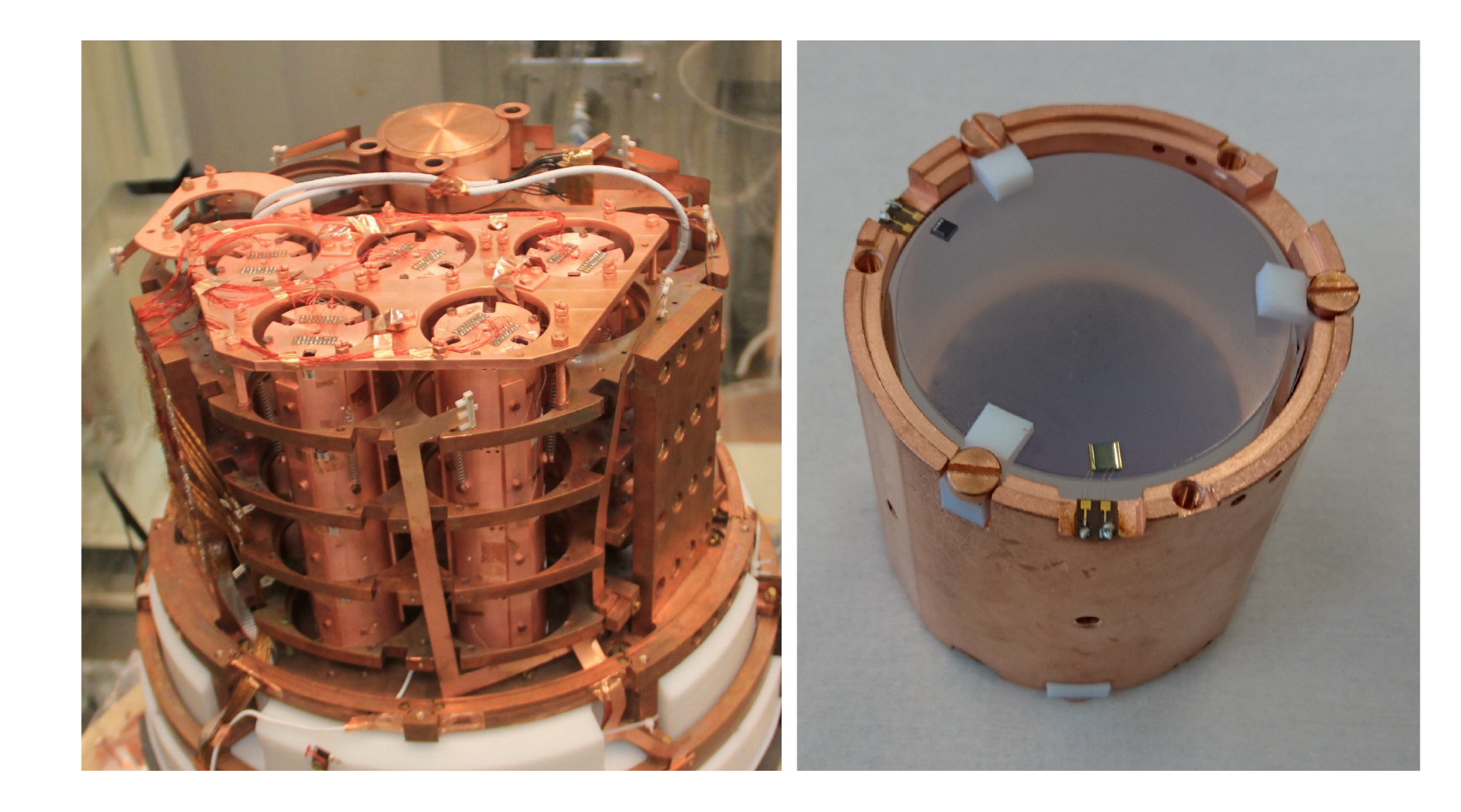}
    \caption{CUPID-Mo experimental geometry showing ({\it Left}) the CUPID-Mo towers mounted in the EDELWEISS cryostat and single detector module ({\it Right}) with LMO crystal, PTFE clamps, NTD thermistor, Si heater and Cu holder. The Ge LD is seen through the LMO crystal. }
    \label{Experiment_fig}
 
\end{figure*}

In this work, we present an analysis of $2\nu\beta\beta/0\nu\beta\beta \rightarrow 0_1^+/2_1^+$ \es which exploits the measurement of both $\beta$ and $\gamma$ energies. In Section \ref{Experiment} we describe the CUPID-Mo experiment; we introduce the search technique and {\sc Geant4} Monte Carlo simulations in Section \ref{Analysis}. In Section \ref{Data}, an overview of the experimental data and treatment of multi-site events used for this analysis are given. We describe our Bayesian analysis framework in Section \ref{Bayesian}, our treatment of systematic uncertainties in Section \ref{Syst}, and the obtained results in Section \ref{Results}.  We conclude in Section \ref{Discussion} by comparing our measurement of $2\nu\beta\beta \rightarrow 0_1^+$ state to theoretical calculations and we discuss the prospects to further investigate the decay mechanism as well as sensitivity to models beyond the light Majorana neutrino exchange.

\section{Experiment}
\label{Experiment}
CUPID-Mo consisted of an array of twenty 
$\sim$\,200\,g LMO detectors (Figure~\ref{Experiment_fig}) enriched to $\sim$\,97\,\% in $^{100}$Mo, with a total mass of 4.16~kg of LMO~\cite{Armengaud:2020} and 2.26~kg of $^{100}$Mo. These detectors were operated as cryogenic calorimeters at $\sim$\,20\,mK, in the EDELWEISS \cite{Armengaud:2017b} cryostat at the Laboratoire Souterrain de Modane (LSM), France. This technique allows for a very high detection efficiency ($\sim$\,75\,\% for $0\nu\beta\beta$ decay to g.s.) due to the detector containing the source, and excellent energy resolution $(\sigma \sim 0.1\% )$ at 3 MeV~\cite{Armengaud:2020}. 
In addition, CUPID-Mo employed a dual readout with twenty cryogenic light detectors (LDs) consisting of Ge wafers also operated as calorimeters. These allow for discrimination between $\alpha$'s and $\beta/\gamma$'s since LMO is a scintillating crystal and the amount of light produced for $\beta/\gamma$ is about five times higher compared to that produced by $\alpha$ particles of the same energy~\cite{Armengaud:2020,Armengaud:2017}.

Each LMO crystal is assembled into an independent detector module with a Ge LD, NTD-Ge \cite{Haller:1984} thermistors on both LMO and LD, Si heater, copper holder, reflective foil (3M Vikuiti\textsuperscript{\texttrademark}, which guides the light to the LDs), and PTFE clamps (shown in Figure~\ref{Experiment_fig}). While this design allows for a modular pre-assembly and is compatible with the operation of multiple payloads in the EDELWEISS cryogenic system, it results in a relatively high copper/LMO mass ratio of $\sim 1.5$. 
The individual detector modules are arranged in an array of five towers (see Figure~\ref{Experiment_fig}) so that each LMO detector faces two LDs (apart from those on the top floor which only have a lower light detector).

CUPID-Mo took data between 2019 and 2020, collecting a total exposure of 2.71 kg$\times$yr of LMO. The data-taking was organized into 9 datasets, periods of stable operations lasting around a month. As in \cite{Armengaud:2020c}, three short periods of data-taking which could not be calibrated to the same accuracy were discarded.

\section{Search Technique and Simulations}
\label{Analysis}

In $\beta\beta$ decays to \es the electrons are accompanied by $\gamma$'s so they often reconstruct in multi-site or coincident events where multiple detectors are triggered simultaneously. 
The particular set of energies obtained allows us to make a pure selection of \es decay events with very low background.
We define the multiplicity, \m, as the number of crystals subject to simultaneous energy depositions above the analysis energy threshold, chosen as 40\,keV, well above the detector trigger thresholds and within a time window of $\pm$ 10\,ms well above the detector time resolution \cite{CUPIDMo:2022}.

For an event which is a time coincidence between several LMO detectors, our minimal experimental observable information is the energy deposited in each detector. We define energies $E_1,E_2,E_3,..,E_{\mathcal{M}}$ as the energy deposited in each detector, sorted so that:
\begin{equation}
    E_1>E_2>E_3...>E_{\M}.
\end{equation}
In principle, the most sensitive approach exploiting all the information contained in the data, would be a fit directly to this multidimensional energy spectrum. However, this analysis has significant complexity due to the difficulty of quantifying the background shape in this high dimensional space.

An alternative approach that has been used by the CUORE-0, CUORE, and CUPID-0 experiments \cite{Adams:2021,Alduino:2018_1,Azzolini:2018oph} is to select categories of signal signatures consistent with being from an \es decay. For each category only the energy variable containing the $\gamma$ (or two electrons - $\beta\beta$ for $0\nu\beta\beta$ decay) peak is fitted (the peak energy $E_p$). The other energy variables are referred to as ``projected out" energies. This technique has the advantage that the decay rate can be extracted as a set of one dimensional fits to a peak over a background, which is a simple and robust technique.
We use this approach in our analysis, however we modify this technique by dividing up categories based on their projected out energy variables. Since the decays to different \es significantly overlap this allows us to better discriminate between them. Furthermore, it avoids integrating together regions with different background levels.

 
\subsection{MC simulations}
We use dedicated Monte Carlo (MC) simulations to identify the most promising categories of events to include and optimize the associated energy cuts. The objective is to identify categories which possess a high signal to background ratio, to define the energy variable to project onto, to select the energy boundaries for each signal category, and to assess the containment efficiency therein. A {\sc Geant4}~\cite{Allison:2016} model has been created which implements the CUPID-Mo detector into the existing EDELWEISS MC software package~\cite{Armengaud:2013}. Special care is taken to implement the detector structure as accurately as possible, in particular accounting for the individual dimensions of each LMO crystal, and the passive holder material close to the detectors. We simulate both $2\nu\beta\beta$ decays and $0\nu\beta\beta$ decays to the $0_1^+$ and $2_1^+$ \es. 
This accounts for the angular correlation between 540 keV and 591 keV $\gamma$'s from the $0_1^+$ ($0^+_1-2^+_1-0^+_{\text{g.s}}$) cascade as \cite{Hamilton:1940}:
\begin{equation}
   \rho(\theta) = 1-3\cos^2{(\theta)}+4\cos^4{(\theta)}.
\end{equation}
Here $\rho(\theta)$ is the probability for $\gamma$'s to be emitted at angle $\theta$ from each other. Particles are then propagated through the experimental geometry using the Livermore low energy physics list~\cite{livermore}, applicable down to very low energy (250 eV). For the $2\nu\beta\beta$ decay to $0_1^+$ \es simulations are performed using both the single state dominance (SSD) and higher state dominance (HSD) models~\cite{Simkovic:2000}. Currently, no experimental data favors one model over another for the $0_1^+$ state, however the SSD model is strongly favored for the ground state decay \cite{Arnold:2019,Armengaud:2020b} and so we use it as our default. The difference to the HSD model is then treated as a systematic uncertainty.
We combine the MC simulations with inputs from experimental data to reproduce the detector response of each individual detector, accounting for:
\begin{itemize}
    \item The energy threshold (set at 40 keV)~\cite{CUPIDMo:2022};
    \item The multiplicity of events;
    \item Energy resolution of each detector-dataset pair (energy dependent);
    \item Rejected periods of detector operations due to periods of cryogenic instability (i.e. due to environmental disturbances);
    \item Scintillation light, light detector energy resolution, and cuts based on the LDs (see Section \ref{Data});
\end{itemize}
\begin{figure}
    \centering
    \includegraphics[width=\columnwidth]{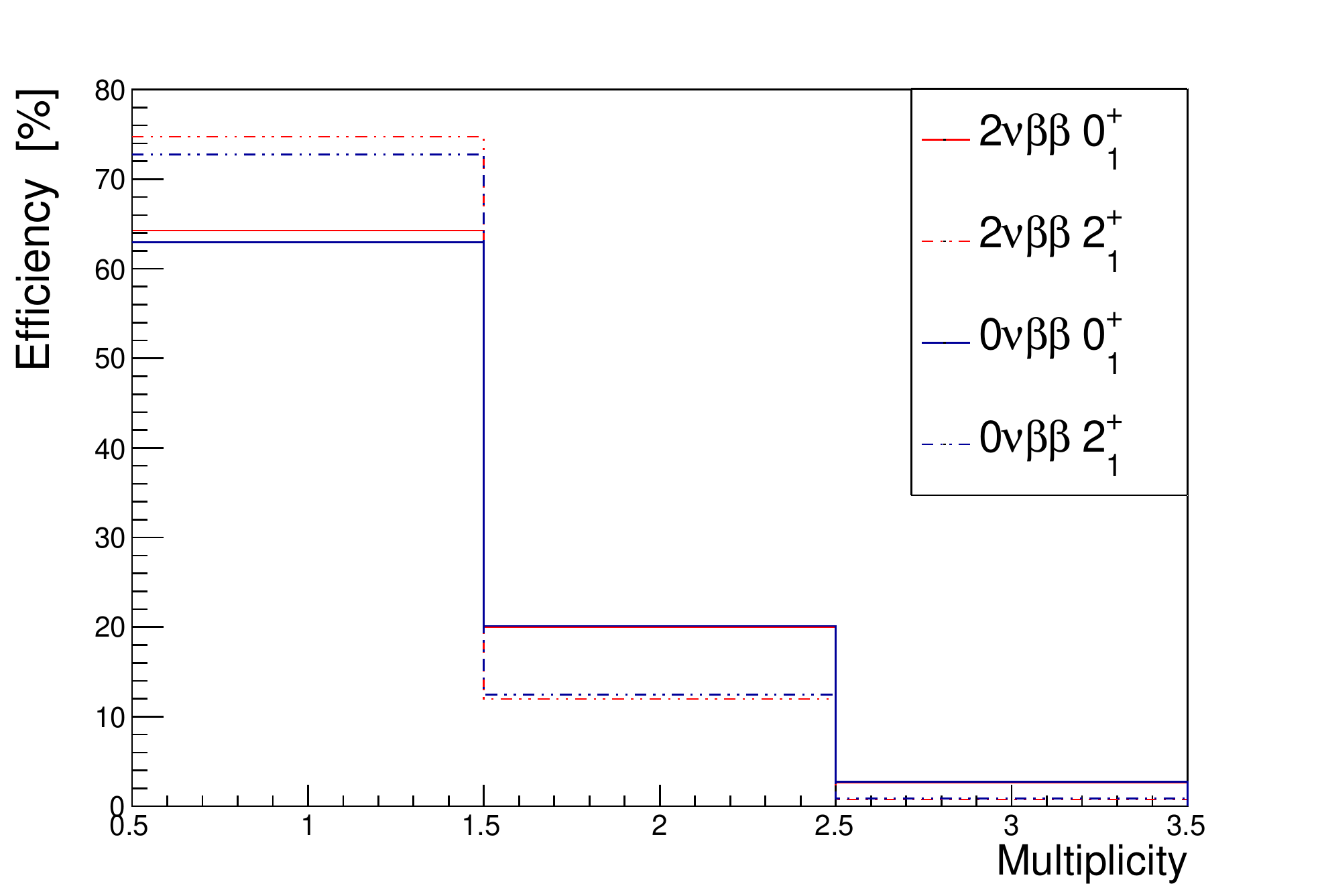}
    \caption{Distribution of multiplicities (MC simulation) for each \es~decay. Due to the small crystal array and significant amount of non-sensitive holder material, both decay modes to the ($0_1^+,2_1^+$) state are dominated by single crystals hits, (10\,--\,20\%) of events cause energy deposits in two crystals, and $\mathcal{O}$(1\%) in three or more.}
    \label{multis}
\end{figure}
\begin{figure*}
    \centering
    \includegraphics[width=\columnwidth]{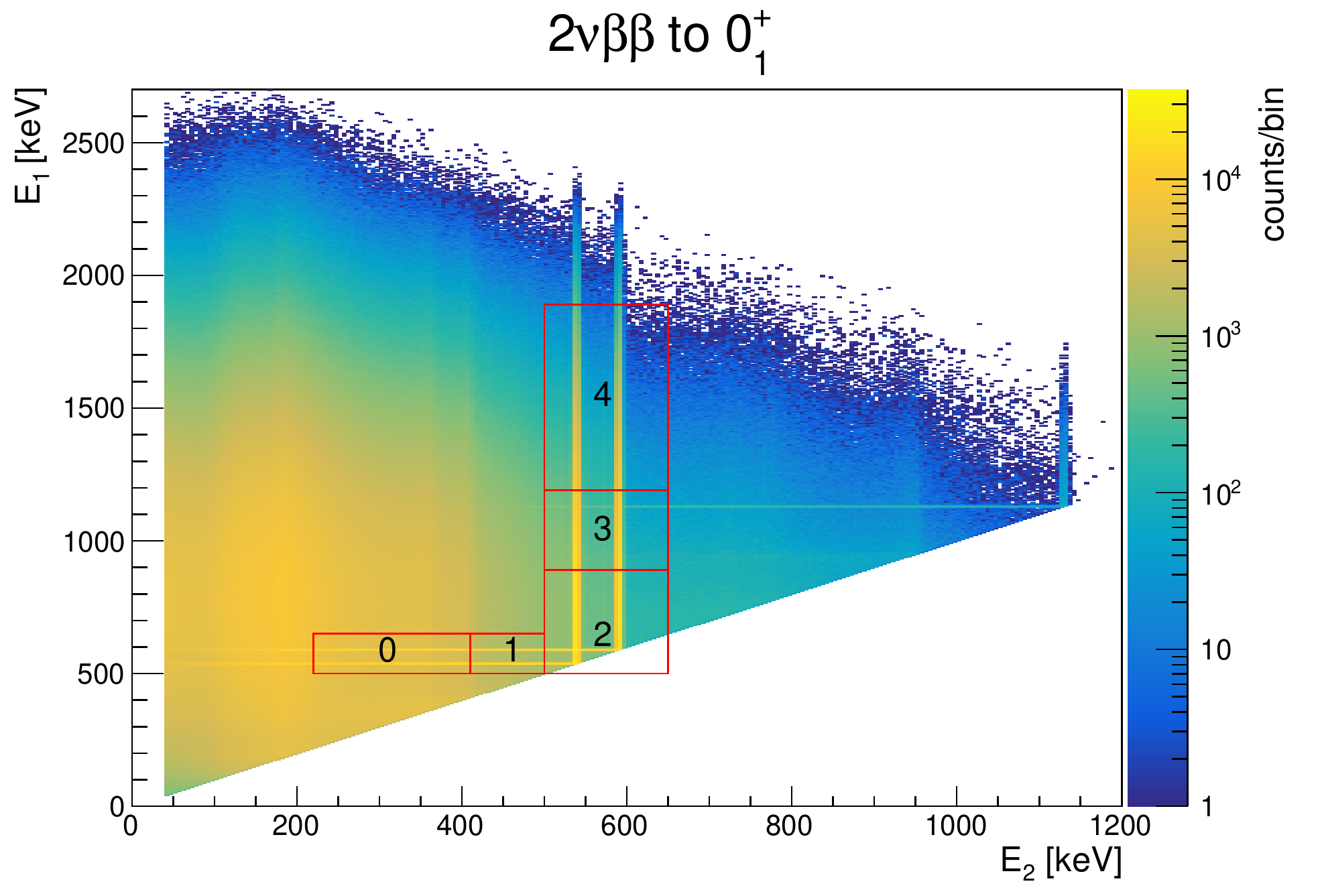}
        \includegraphics[width=\columnwidth]{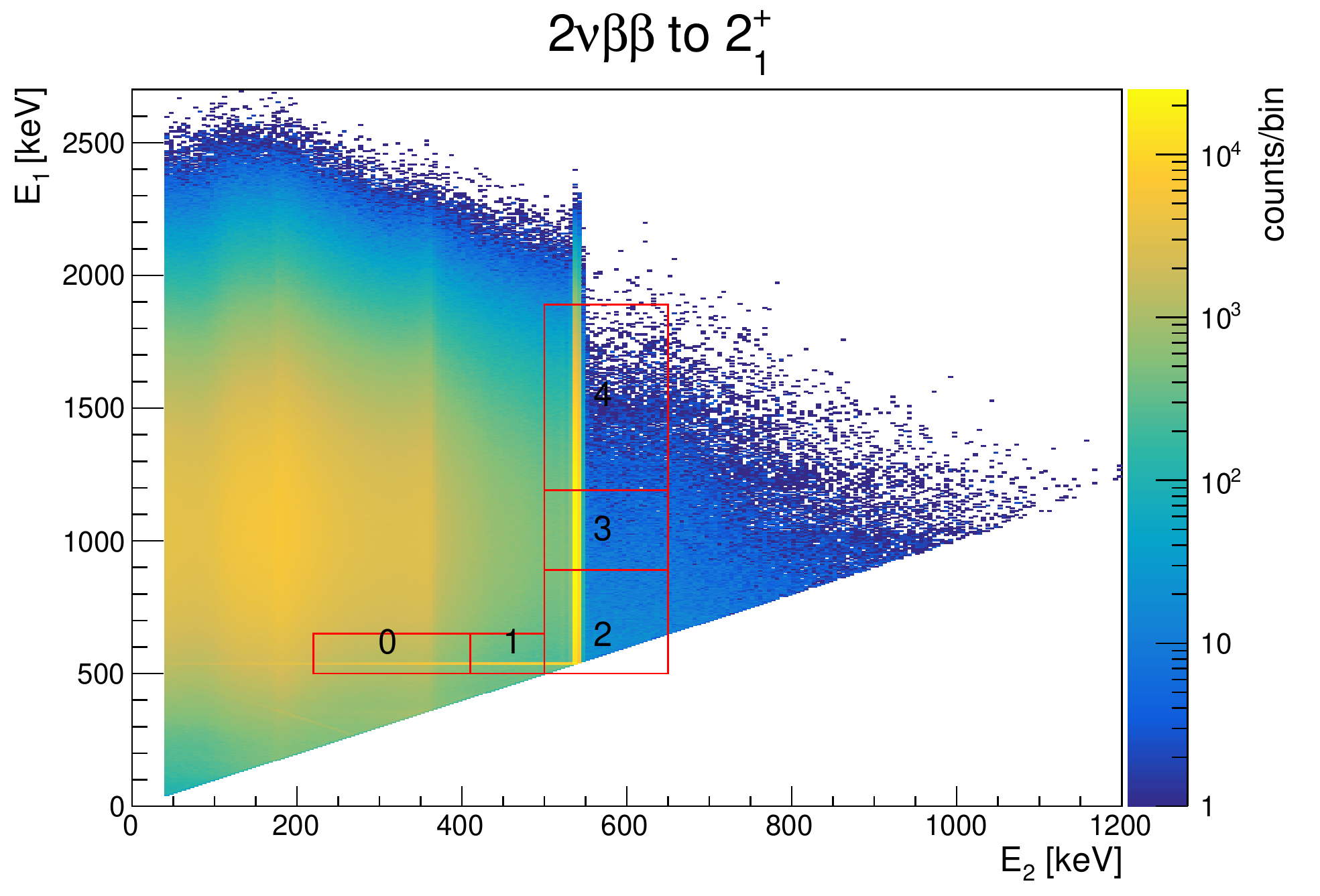}
    \includegraphics[width=\columnwidth]{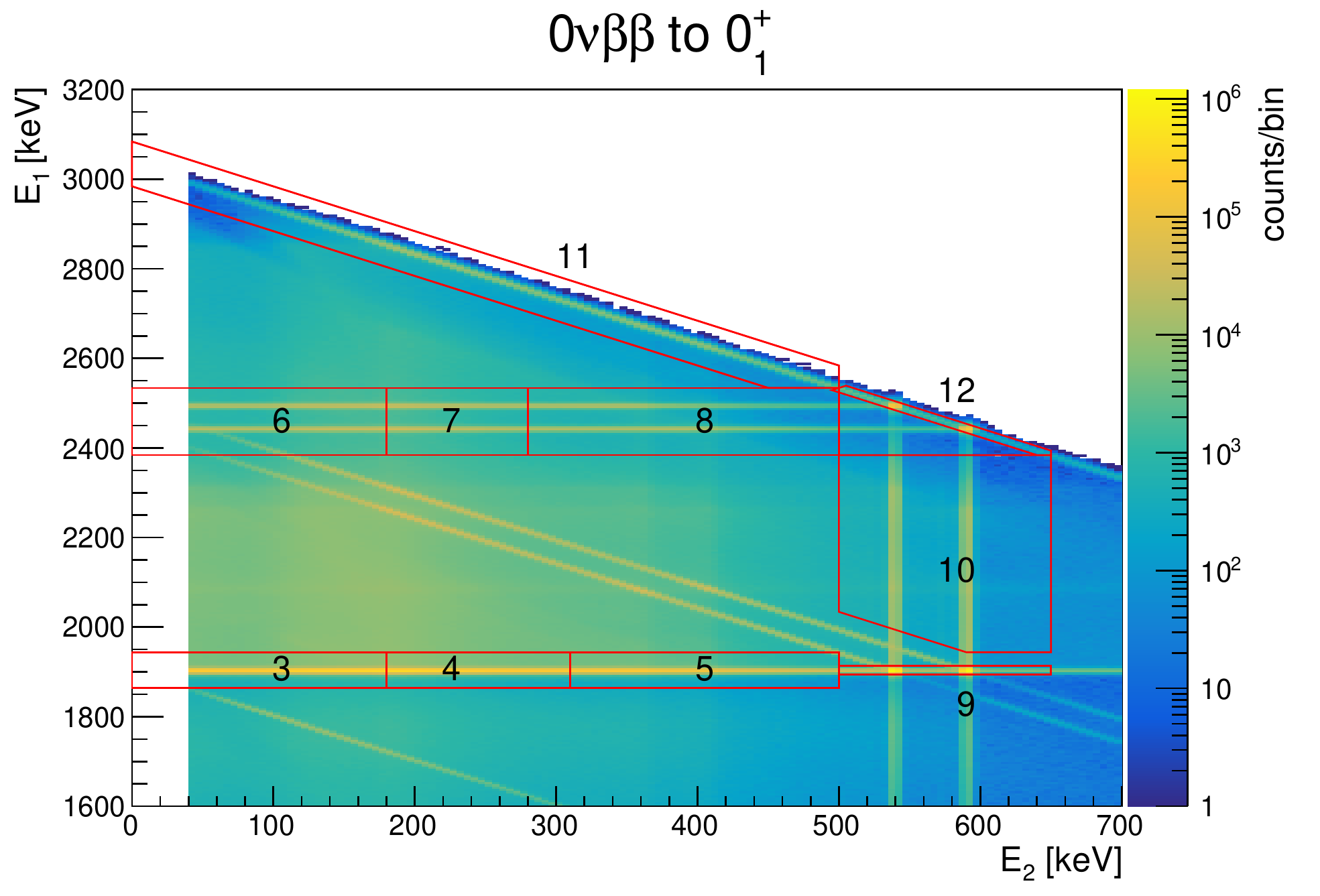}
    \includegraphics[width=\columnwidth]{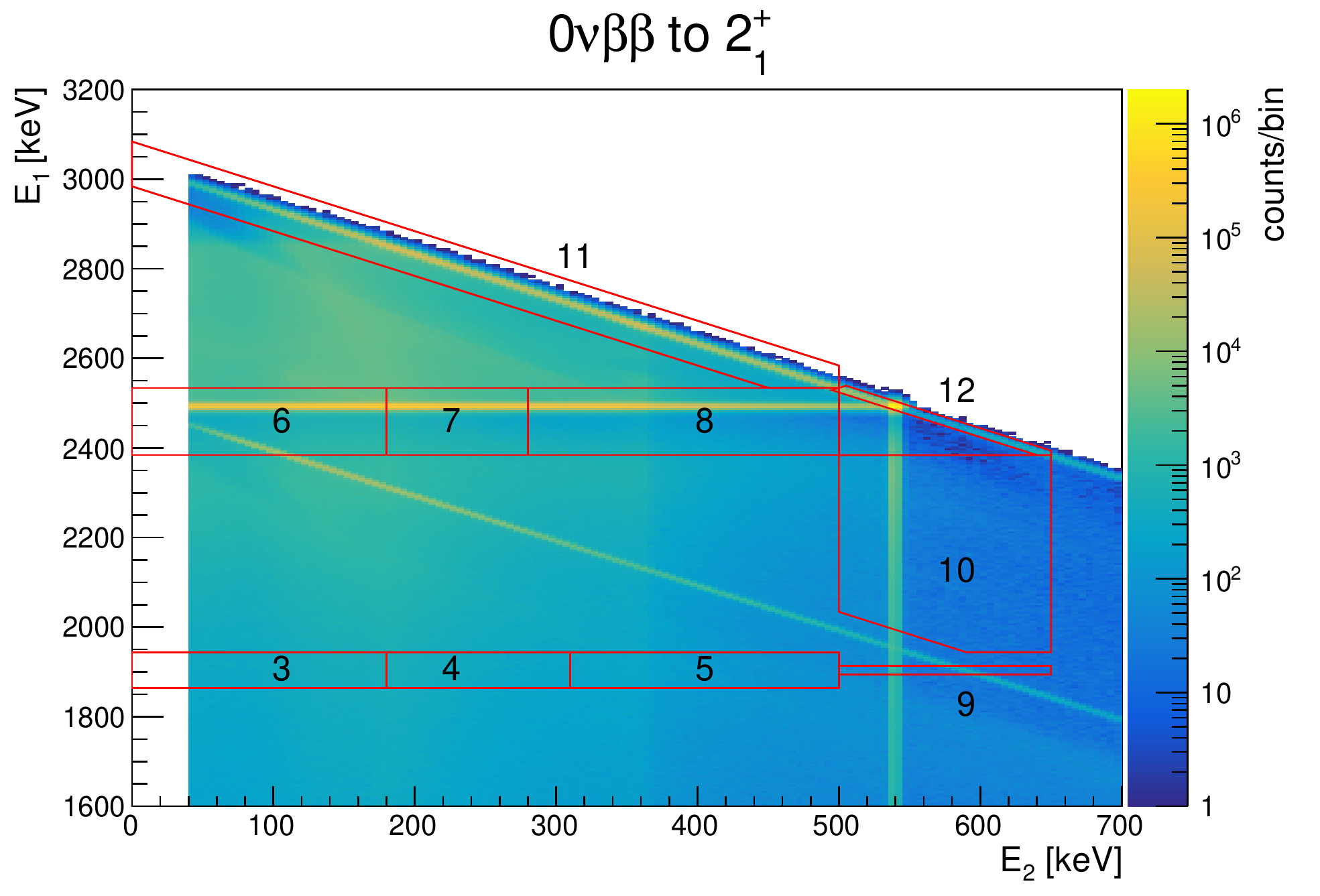}

    \caption{Monte Carlo simulated two-dimensional $\mathcal{M}_2$ energy distributions for decays to \es  {\it Top left}: $2\nu\beta\beta$ decay to $0_1^+$. {\it Top Right}: $2\nu\beta\beta$ decay to $2_1^+$. {\it Bottom Left}: $0\nu\beta\beta$ decay to $0_1^+$. {\it Bottom Right}: $0\nu\beta\beta$ decay to $2_1^+$. We show red boxes to highlight the $\mathcal{M}_2$ signal categories used in the analysis (see Tables \ref{tab:2nu_sigs}, \ref{tab:0nu_sigs}) and we label the associated category numbers. }
    \label{2Dist}
\end{figure*}
In our MC simulations, 
 we observe that most excited states events are in $\M_{1-2}$ with a small contribution of $\mathcal{M}_3$ (Figure \ref{multis}). The large fraction of $\M_1$ events is caused by the significant volume of passive material (Cu housing) and a relatively small modular detector array. We exclude \mone 2\vbb decay events as the continuous energy spectrum of the $\beta$'s does not produce a peak  to perform a fit. For $0\nu\beta\beta$ decays, there are mono-energetic peaks in $\mathcal{M}_1$, albeit with a high background. We therefore focus on $\M_2$ events which have the highest sensitivity, and we also include some categories of $\mathcal{M}_3$ events.

\subsection{$2\nu\beta\beta$ decay categories}
We first describe categories for $2\nu\beta\beta$ decay analysis.
For the analysis of $\mathcal{M}_2$ events the energies can be visualized as two-dimensional histograms. We plot these two-dimensional energy spectra for both \es decays ($2\nu\beta\beta$ and $0\nu\beta\beta$ decay) to $0_1^+$ and $2_1^+$ \es in Figure~\ref{2Dist}. When selecting categories of events to use in our analysis, we require at least one peak to perform a fit. This selection is equivalent to looking for lines or points on the two-dimensional plane.

For $2\nu\beta\beta$ decays in MC simulation we observe a very clear signature of events where either $E_1$ or $E_2$ has a $\gamma$ energy (either $539.5$ or $590.8$ keV) as shown in Figure \ref{2Dist} (top). These events can be interpreted as the $\beta\beta$ being contained in one detector, giving rise to a continuous distribution, and a $\gamma$ that is fully contained in a second LMO detector. In the case of the decay to the $0_1^+$ state, the second $\gamma$ either escapes the detector or is contained in the crystal with the $\beta\beta$'s.
We notice that the $2_{1}^{+}$ and $0_{1}^{+}$ decays (partially) overlap and the vertical/horizontal lines cover a wide range of projected out ($\beta\beta$) energies. 
This motivates us to divide the observed signal distribution into five slices, two for the horizontal and three for the vertical lines since the vertical lines cover a larger range in energy. The boundaries of this categories are defined to maximise the experimental sensitivity using MC simulations as explained in Appendix \ref{ap:opt_cat}. This leads to the energy categories highlighted in Figure \ref{2Dist} (top) and listed in Table \ref{tab:2nu_sigs} (index 0 to 4). 
\\ \indent
To identify $\mathcal{M}_3$ categories we note that in MC simulations generally the electrons carry the largest kinetic energy and are thus likely to be contained in the $E_1$ detector due to the relatively low $\gamma$ energies. We hence focus our search on events 
with $\gamma$'s in the $E_2$ and $E_3$ variables. In particular, this leads to two categories, one where the $\gamma$ energy is split between $E_2,E_3$ and another where one $\gamma$ is fully contained in $E_2$ with a Compton scatter event in $E_3$.
\\ \indent
Combined with the $\mathcal{M}_2$ categories this leads to a total of 7 independent categories used for the $2\nu\beta\beta$ decay analysis. 

\begin{table*}[htpb!]
    \centering
    \caption{Categories for $2\nu\beta\beta$ decay to $0_1^+$, $2_1^+$ \es , these are given a name describing the multiplicity, peak energy variable and then an index (if necessary to distinguish between categories). The table lists the energy cuts for the variable that is projected out, while the projected energy range for each category is 500\,--\,650 keV. For every category the $\gamma$'s are either 540 or 591 keV.}
    \begin{tabular}{ccccc}
     Cat. & Multiplicity & Peak Variable &  Energy cuts    & Interpretation  \\
      \hline \hline 
    0     & 2  & $E_1$  & $E_2\in[220,410]$ & $\gamma$ in $E_1$, $\beta\beta$ in $E_2$\\
    1 &2 &  $E_1$  &$E_2\in [410,500]$ &$\gamma$ in $E_1$, $\beta\beta$ in $E_2$\\
    2 & 2 & $E_2$ & $E_1\in [500,890]$  & $\gamma$ in $E_2$, $\beta\beta$ in $E_1$\\
        3 & 2& $E_2$  & $E_1\in [890,1190]$ &$\gamma$ in $E_2$, $\beta\beta$ in $E_1$\\
        4 & 2 & $E_2$  & $E_1\in [1190,1880]$ &$\gamma$ in $E_2$, $\beta\beta$ in $E_1$\\
   5 & 3 & $E_2$  & - &$\gamma$ in $E_2$, $\beta\beta$ in $E_1$, Compt. in $E_3$ \\
   6 & 3& $E_2+E_3$  & $E_2<500$ &  $\gamma$ in $E_2+E_3$ (shared Compt.), $\beta\beta$ in $E_1$
    \end{tabular}
    \label{tab:2nu_sigs}
\end{table*}

\subsection{$0\nu\beta\beta$ decay categories}
Unlike for $2\nu\beta\beta$ decay, in $0\nu\beta\beta$ decay the electrons have a fixed total energy so we are also able to search for a peak from the summed electron energy. Therefore a different set of categories are necessary, however the same strategy can be used. 
For $\mathcal{M}_1$ events we include three categories corresponding to the $\beta\beta$ and possibly $\gamma$'s being contained in the detector where the decay took place.  
For $\mathcal{M}_2$ events, we show the two-dimensional distributions in Figure \ref{2Dist} (bottom two figures). We observe a variety of features, vertical, horizontal diagonal lines and points.  Each of these require a different set of energy cuts to contain the signal, as shown in Table \ref{tab:0nu_sigs}. These cuts are carefully constructed to ensure that none of the categories overlap. Some categories feature a mono-energetic peak in both energy variables, for these we still choose one as the peak variable and use a range of $\pm 10$ keV for the other, projected energy, much larger than the energy resolution. 
We employ a similar strategy to the $2\nu\beta\beta$ decay categories, dividing the horizontal lines, events with a $\beta\beta$ peak ($1904, 2444, 2494 $~keV) in $E_1$ and a Compton scatter in $E_2$, into 3 categories each to account for the large changes in background level. This is not necessary for other categories due to the low background, or a small range in the projected out energy.
We neglect some possible categories which have very small containment efficiency (events with $E_2>600$ keV), that overlap significantly with $2\nu\beta\beta$ decay (events with $E_2\sim 500$\,--\,$650$ keV, $E_1<1904$ keV) or with $^{214}$Bi background, diagonal lines with energy summing to $\sim$\,2400 keV.
This leads to 10 $\mathcal{M}_2$ $0\nu\beta\beta$ decay categories which are most easily interpreted geometrically (see the boxes in Figure \ref{2Dist}).


As in $2\nu\beta\beta$ decay analysis we also consider $\mathcal{M}_3$ events focusing on those where the electrons are contained in $E_1$. We choose four categories: For events with $E_1\sim 1904$ keV, we split the data into $E_2 \in [500,650]$ keV, $E_2+E_3\in [500,650]$ keV or neither of these two. For events with $E_1\sim 2400$ keV the background is expected to be low so we just include one category.
This procedure leads to 17 experimental signatures as shown in Table
\ref{tab:0nu_sigs}.

\begin{table*}[]
    \centering
    \caption{Categories for $0\nu\beta\beta$ to excited states, for each category we give an index, a name, the projected energy variable and its energy window, the cuts used to define the category and a physical interpretation. Most of these cuts are more easily visualized in Figure \ref{2Dist}. }
    \small
    \begin{tabular}{cccccc}
      Cat. & Multiplicity & Peak Variable & Range [keV] & Energy cuts [keV]   & Interpretation  \\
      \hline \hline 
    0     & 1 & $E_1$ & $1864$--$1944$ &  & $\beta\beta$ in $E_1$, $\gamma$ escape\\
    1 & 1 &  $E_1$ & $2384$--$2534$ & &$\beta\beta$ in $E_1$, $\gamma$ escape\\
    2 & 1 & $E_1$ & $2984$--$3084$ &  & $\beta\beta$ and $\gamma$ in $E_1$\\
    3 & 2& $E_1$ &$1864$--$1944$ & $E_2<180$ & $\beta\beta$ in $E_1$, Compt. $\gamma$ in $E_2$ \\
    4 & 2 &  $E_1$ & $1864$--$1944$ & $E_2\in [180,310]$ & $\beta\beta$ in $E_1$, Compt. $\gamma$ in $E_2$ \\
    5 & 2 & $E_1$&$1864$--$1944$ & $E_2\in [310,500]$ & $\beta\beta$ in $E_1$, Compt. $\gamma$ in $E_2$ \\
    6 & 2 & $E_1$ & $2384$--$2534$ & $E_2<180$ & $\beta\beta$ in $E_1$, Compt. $\gamma$ in $E_2$ \\
    7 & 2 &  $E_1$ &$2384$--$2534$ & $E_2\in [180,280]$ & $\beta\beta$ in $E_1$, Compt. $\gamma$ in $E_2$ \\
    8 & 2 & $E_1$ & $2384$--$2534$ & $E_2\in [280,500]$ & $\beta\beta$ in $E_1$, Compt. $\gamma$ in $E_2$ \\
    9 & 2 & $E_2$ & $500$--$650$ & $E_1\in [1894,2904]$ & $\beta\beta$ in $E_1$, $\gamma $ in $E_2$ \\
    10 & 2& $E_2$ & $500$--$650$ & $E_1\in [1944,2384]$, $E_1+E_2>2534$ & $\beta\beta$ + Compt. $E_1$, $\gamma$ in $E_2$ \\
    11 & 2 & $E_1$+$E_2$ & $2984$--$3084$ & $E_1>2534$, $E_2<500$ & $\beta\beta$ + Compt. $E_1$, Compt. $E_2$ \\
    12 & 2 & $E_1-E_2$ & $1734$--$2034$ & $E_2<650$, $E_1>2384$, $E_1$+$E_2\in [3024,3044]$ & $\beta\beta$ in $E_1$, $\gamma$ in $E_2$ \\
    13 & 3 & $E_2+E_3$ & $500$--$650$ & $E_1\in [1894,1914]$, $E_2<500$ & $\beta\beta$ $E_1$, Compt. $E_2/E_3$ \\
    14 & 3 &$E_2$ & $500$--$650$ &$E_1\in [1894,1914]$ & $\beta\beta$ in $E_1$, $\gamma$ in $E_2$, Compt. $E_3$ \\
    15 & 3 & $E_1$ & $1854$--$1954$ & $E_2\not\in [500,650]$, $E_2+E_3\not\in [500,650]$ &$\beta\beta$ in $E_1$ and Compt. $E_2/E_3$ \\
    16 & 3 & $E_1$ & $2384$--$2534$ &  & $\beta\beta$ in $E_1$
    \end{tabular} 
    \label{tab:0nu_sigs}
\end{table*}
\subsection{Determination of signal shape and efficiency}
\label{sec:sigshape}
\begin{figure}
    \centering
    \includegraphics[width=0.45\textwidth]{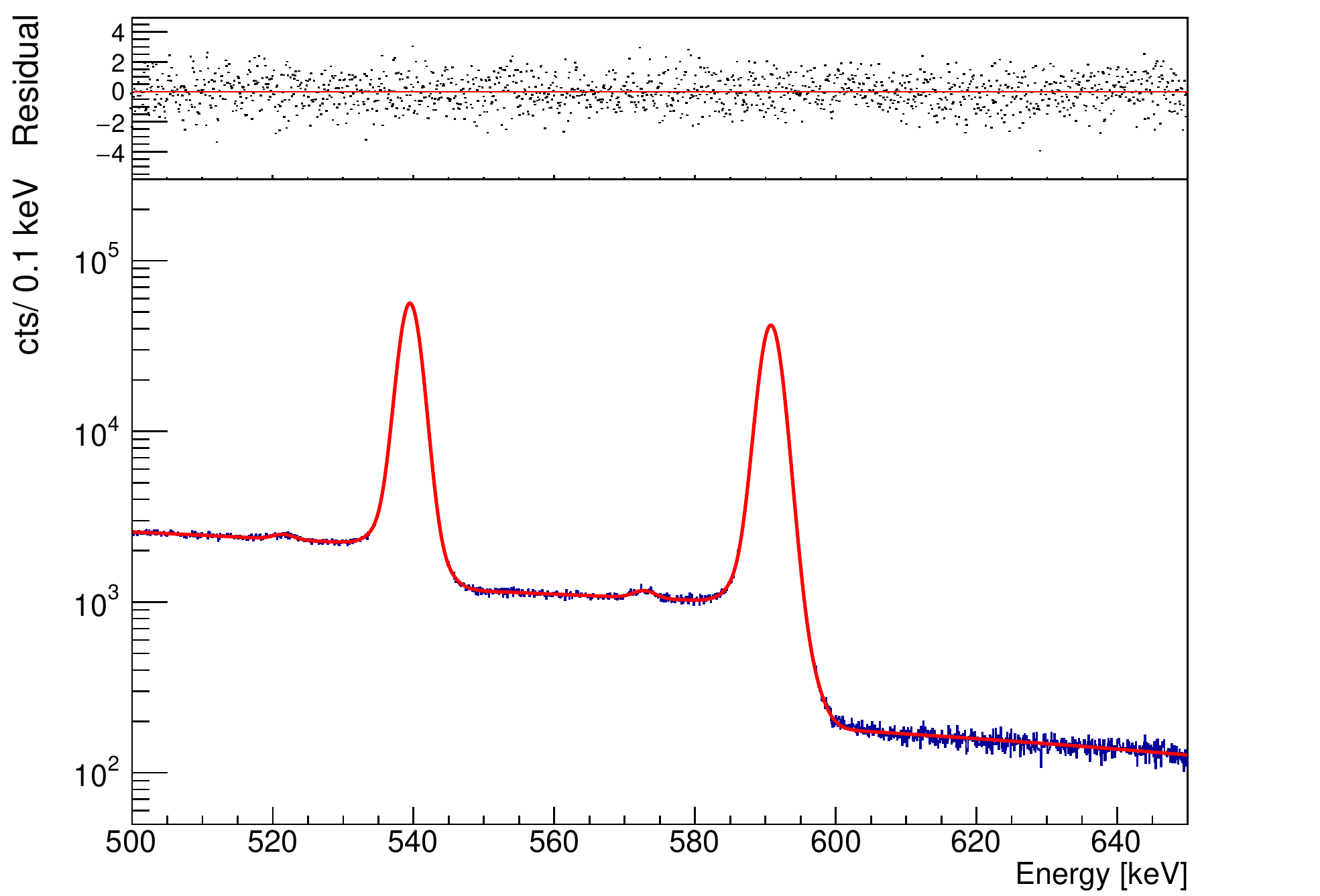}
    \caption{Parameterization of $2\nu\beta\beta$ to $0_1^+$ category 2 signal. The model consists of two photo-peaks, with their respective $^{100}$Mo X-ray escape peaks, two smeared step functions modeling Compton interactions and a linear background. The residual defined as $(\text{value}-\text{model})/\text{error}$ is shown in the upper figure. }
    \label{params1}
        \includegraphics[width=0.45\textwidth]{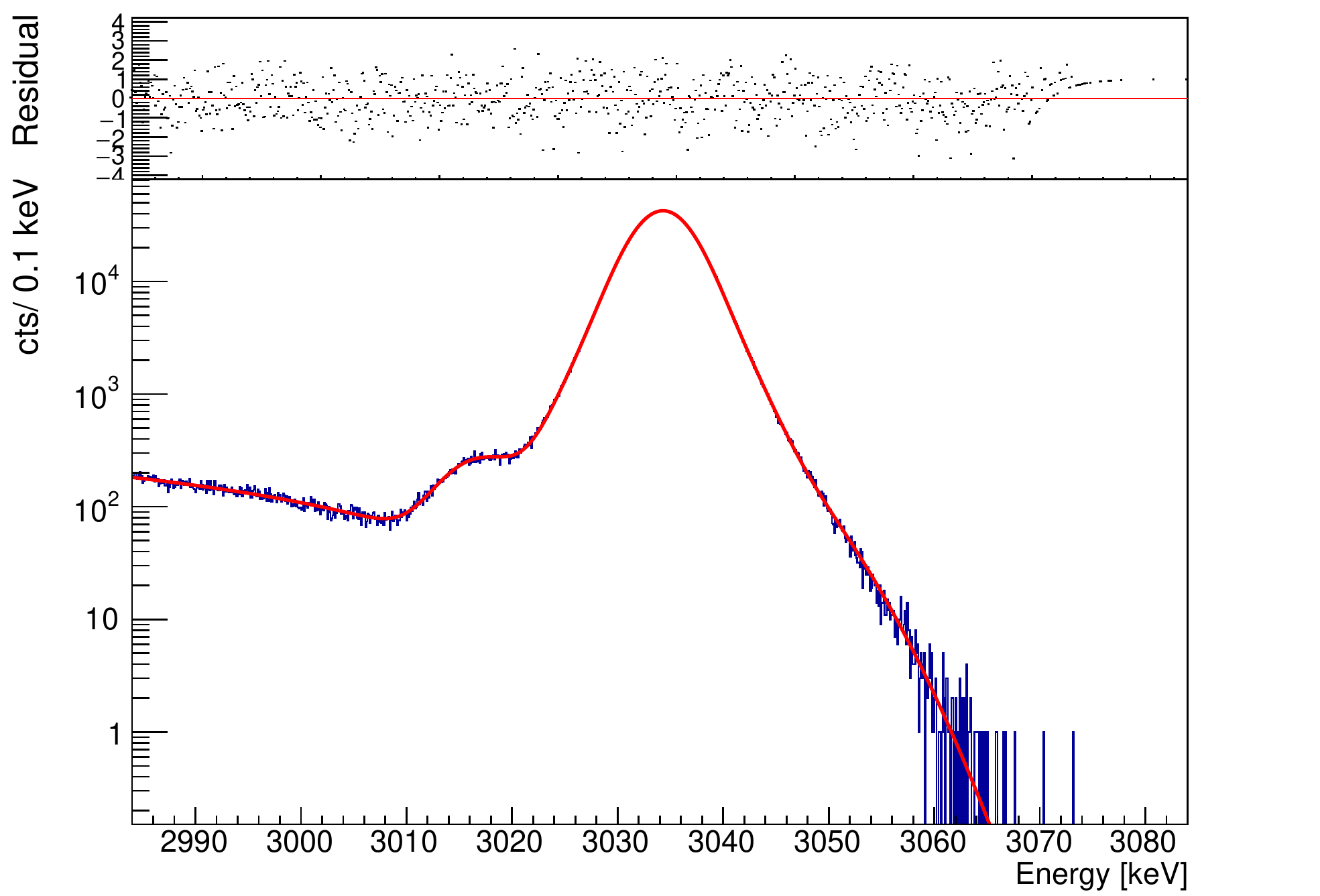}
    \caption{Parameterization of $0\nu\beta\beta$ decay to $0_1^+$ category 2 signal. The model contains a photo-peak, X-ray escape peak, and linear smeared step functions. The residual defined as $(\text{value}-\text{model})/\text{error}$ is shown in the upper figure.  }
    \label{params2}
\end{figure}

We use MC simulations to determine the signal shapes and efficiencies for our Bayesian analysis. For each category we make the energy cuts as described in Tables \ref{tab:2nu_sigs}, \ref{tab:0nu_sigs}.
We also apply a set of selection cuts to the data and MC simulations to remove events that likely arise from a known $\gamma$-ray background peak. The cuts are chosen to correspond to roughly $\pm \Delta E (\text{FWHM})$ with the values shown in Table \ref{tab:excl_cuts}.
\begin{table}[htpb]
    \centering
    \caption{Table showing the $\gamma$ exclusion cuts used to remove events likely originating from a background $\gamma$ ray.}
    \begin{tabular}{ccc}
       Variable  & Cut [keV] & Decay\\ \hline \hline
      $E_1$   &  $1333\pm 5$  & $^{60}$Co\\
      $E_1$   &  $1173\pm 5$ & $^{60}$Co\\
      $E_1+E_2$   &  $1333\pm 5$ & $^{60}$Co\\
      $E_1+E_2$   &  $1173\pm 5$ & $^{60}$Co\\
$E_1+E_2$ & $1461 \pm 5$ & $^{40}$K\\
$E_1$ & $2615 \pm 7$ & $^{208}$Tl\\
$E_1+E_2$ & $2615\pm 7$ & $^{208}$Tl
    \end{tabular}
    \label{tab:excl_cuts}
\end{table}
After the projection onto the peak energy variable we extract one-dimensional histograms for each decay mode ($0_1^+$ or $2_1^+$) in each category from MC simulations.
\\ \indent
From these histograms we obtain the containment efficiency $\varepsilon_{\text{cont},r}^{0+/2+}$, or the probability for a decay to reconstruct in category $r$. We also compute the total containment efficiency (summed over all categories) as $4.4/2.0 \ \%$ for $2\nu\beta\beta$ to $0_1^+/2_1^+$ \es and $46/57 \ \%$ respectively for $0\nu\beta\beta$ decay.
\\ \indent
We fit these histograms to phenomenological functions to extract analytic signal shapes to use in our analysis. These functions are described in more detail in Appendix \ref{ap:functions}, the photo-peak is approximated via three Gaussian's to account for the slight non-Gaussianity of the line-shape (due to each detector-dataset pair having a different resolution), this is discussed more in Section \ref{sec:reso}.
Each function is normalized to unity and then we model the MC simulated spectrum as:
\begin{equation}
    f(E)= N_{\text{tot}}\sum_{i=1}^N p_i\cdot f_i(E),
\end{equation}
where the sum runs over the normalized functions $f_i$, $p_i$ is the probability for an event to be in a given spectral feature $i$ of the signal distribution for the considered category, for example in one of the gamma peaks or in the continuum, and $N_{\text{tot}}$ is the total number of events. 
We show two example fits (both $0_1^+$ signal) in Figure \ref{params1}, $2\nu\beta\beta$ category 2 and Figure \ref{params2}, $0\nu\beta\beta$ category 2. We see that these models describe the MC simulations very well, in particular the slightly non-Gaussian photo-peak shape. 
These functions encode most of our systematic uncertainties, in particular the resolution and peak position of Gaussian and normalization of the peaks which can then be treated as nuisance parameters of the final fit.

\section{Data analysis}
\label{Data}
For this search we use the full CUPID-Mo data corresponding to 2.71 kg$\times$yr exposure of LMO. We process our data using a C++ software, {\tt Diana} \cite{Alduino:2016,Azzolini2018b}, developed by the Cuoricino, CUORE, and CUPID-0 collaborations and further developed by CUPID-Mo. Most of the data processing steps are the same as described in \cite{Armengaud:2020c,CUPIDMo:2022}. An optimal trigger is used to identify physics events and an optimal filter which maximizes signal-to-noise ratio is used to compute amplitudes (both LD and LMO detectors) \cite{Gatti:1986}. Thermal gain stabilization and calibration are performed using data collected with a Th/U calibration source. For each LMO detector we associate ''side-channels", as the LDs which face this crystal.
For $\mathcal{M}_1$ data (3 $0\nu\beta\beta$ decay categories) we use the same selection cuts as in \cite{CUPIDMo:2022}: a principal component analysis (PCA) based pulse shape discrimination (PSD) cut~\cite{Huang:2020}, a normalized light distance cut, delayed coincidence (DC) cuts to remove $^{222}$Rn, $^{228}$Th decay chain events and, muon veto anti-coincidence.
\\ \indent
However, for coincidence events ($\mathcal{M}>1$) it is beneficial to modify several of these steps. Random coincidences between non-causally related events (hereafter {\it accidentals}) could provide a large possible background. Similar to CUORE \cite{Alduino:2016} and as described in \cite{CUPIDMo:2022} we adjust the time difference of events to account for the characteristic detector response based time offsets between detector pairs. This allows us to reduce the time window used to define coincidences to $10\,\mathrm{ms}$ (from 100 ms in previous analyses \cite{Armengaud:2020c}) and thus reduce the accidental coincidence background.
\\ \indent
In order to select a clean sample of higher multiplicity \es candidate events, we make use of the dual read-out and implement a light yield based cut for $\M>1$ events. Unlike the $\M_1$ analysis where this cut primarily rejects $\alpha$ backgrounds, it is designed here to tag and reject events where an energetic electron escapes an LMO crystal, punches through an LD, and is stopped in the adjacent crystal within a tower. In addition, this cut very efficiently tags events from a $^{60}$Co contamination identified on one of the LDs where the $\beta$ particle typically leaves significant excess energy in the LD.
Developing a light cut to remove these events is less straightforward than for $\mathcal{M}_1$ data, since a coincidence between two vertically adjacent crystals is accompanied by scintillation light signal that can be absorbed in the common, intermediate LD. For each event we compute the expected scintillation light deposited on the LD based on the values observed in $\mathcal{M}_1$ events (mainly $2\nu\beta\beta$ decays to g.s.). This accounts for these multiple contributions to light yield. We then normalize the light energies using a conservative estimate of the LD resolution ($\sigma^{\text{LD}}_{i,s}$) measured in high energy background data ($\M_2$ total energy $>1500$~keV) as:
\begin{equation}
    n_{i,s}= \frac{E^{\text{LD}}_{i,s}-E^{\text{LD,exp}}_{i,s}}{\sigma^{\text{LD}}_{i,s}},
\end{equation}
where $i$ refers to the pulse index (one for $E_1$ etc.), $s$ is the side channel (either 0 or 1), $E_{i,s}^{\text{LD}}$ is the measured LD energy, while $E_{i,s}^{\text{LD,exp}}$ is the expected LD energy for $\gamma/\beta$ like event and $\sigma_{i,s}^{\text{LD}}$ is the predicted energy resolution of the LD.
We then place a cut of $|n_{i,s}|<10$ for each LD and side channel in an $\mathcal{M}>1$ event. We place a cut of $-10<n_{i,s}<3$ for the LD with the $^{60}$Co contamination.
This contamination can also lead to background events in other LMO detectors; to remove this we make a global LD anti-coincidence cut. We remove any LMO event (excluding those who directly face this LD) with a trigger on this LD with LD energy $>5$ keV within a $\pm 5$~ms window.
\\ \indent We also adjust the PCA based PSD cuts \cite{Armengaud:2020c,CUPIDMo:2022} to place a cut on the shape of each pulse in a higher multiplicity event. We require that each pulse has a normalized PCA reconstruction error of less than 23 median absolute deviations. This was optimized by comparing the efficiency of $\mathcal{M}_2$ events $\varepsilon_s$, obtained from events summing to the $^{40}$K 1461 keV peak and estimating the background efficiency, $\varepsilon_b$, in a side-band $E_2\in(450$\,--\,$520)$~keV. This sideband, which is only used in optimizing the PCA cut, approximates the background for the dominant $2\nu\beta\beta$ decay categories (2, 3 and 4) whilst not including any signal peaks. We then maximize a figure of merit $\varepsilon_s/\sqrt{\varepsilon_b}$ which is  proportional to the experimental sensitivity. 
For $\mathcal{M}_1$ data a cut of $<9$ on the PCA normalized reconstruction error is used (as in \cite{CUPIDMo:2022}).
\\ \indent Finally, we employ a data blinding by adding simulated MC events directly to the data files. The rate is chosen by sampling randomly from a uniform distribution with range, $ \Gamma \in \Gamma_{\text{measured}}\cdot (2,10)$,
where $\Gamma_{\text{measured}}$ is the current leading limit or measurement \cite{Barabash:2020} and $(2,10)$, is a uniform distribution between 2 and 10 chosen to ensure the injected signal is significantly larger than any possible signal in the data. The injected rate is hidden during the analysis of the blinded data. The blinded data are used to optimize and test the Bayesian fitting routine and prevent biasing our results.

\subsection{Energy resolution and linearity}
\label{sec:reso}
We determine the response of our detector to a monochromatic energy deposit using both calibration sources and $\gamma$ peaks from natural radioactivity using the same procedure described in detail in \cite{CUPIDMo:2022}. 
In particular, a simultaneous fit, over each detector in each dataset, of the 2615 keV peak in calibration data is used to extract the resolution of each detector-dataset pair at 2615 keV $\sigma_{\text{chan,dataset}}^{2615}$.
\\ \indent
We model the line-shape of our peaks in $\mathcal{M}_1$ physics data as an exposure weighted sum of Gaussians for each channel-dataset pair. The individual Gaussians have a common mean $\mu$ but a resolution which is a product of $\sigma_{\text{chan,dataset}}^{2615}$ and a global scaling $R(E)$.
We fit this line-shape model to our peaks to extract $R(E)$ and $\mu(E)$ for each peak in physics data and parameterize the energy dependence as $R(E)=\sqrt{r_0^2+r_1E}$ and $\mu(E)=b_0+b_1E+b_2E^2$. The resulting parametrization is used in the MC simulations to model the detector response. In particular, we account for the uncertainty on the parameters of these functions as systematic uncertainties (Section \ref{Syst}).
\subsection{Cut efficiencies}
\label{sec:effs}
To measure $2\nu\beta\beta/0\nu\beta\beta$ decay rates it is necessary to determine the analysis efficiency, or probability that a signal event will pass all selection cuts. We employ the same strategy that was utilised in the \mone 0\nbb analysis~\cite{CUPIDMo:2022} in order to compute these. We evaluate the pileup efficiency, or probability that a signal event will not have another trigger in the same waveform using random triggers. Several other cuts (multiplicity selection, muon veto cut, and LD anti-coincidence) induce dead times in one or more detectors. For these we evaluate the efficiencies using $^{210}$Po Q-value peak events by counting the number of events passing the cuts with energy in $(5407 \pm 50)$~keV. These are a proxy for physical $\mathcal{M}_1$ events due to the high energy and modular structure meaning $\alpha$ particles can only deposit energy in one crystal. \\ \indent Finally, for our pulse shape and light yield cuts the efficiency is evaluated using $\gamma$ peaks in $\mathcal{M}_1$ and summed $\mathcal{M}_2$ energy which are a clean sample of signal like events. For each prominent $\gamma$ peak, we fit the energy distribution of events passing and failing each cut to a Gaussian plus linear background. From the measured number of events $N_{\text{pass}},N_{\text{fail}}$ (events in the Gaussian peak only for the two fits) we compute the efficiency $\varepsilon= N_{\text{pass}}/(N_{\text{pass}}+N_{\text{fail}})$. We estimate numerically the uncertainty on $\varepsilon$ by sampling from the measured uncertainties on $N_{\text{pass}},N_{\text{fail}}$. We observe an energy independent efficiency for all of our cuts  (shown in col. 2 of Table \ref{tab:cut_eff}). 
The energy range for the $\mathcal{M}_1$ 3034 keV category of $0\nu\beta\beta$ decay lies outside the range of our $\gamma$ peaks. Similar to \cite{CUPIDMo:2022} we use a 1st order polynomial to extrapolate the PCA and light yield cut efficiency to 3034 keV, accounting for the possibility of a slight energy dependence of the cut (shown in col. 3 of Table \ref{tab:cut_eff}). We summarise the cut efficiencies in Table \ref{tab:cut_eff}, and use them as nuisance parameters with Gaussian priors in our Bayesian analysis.
\begin{table*}
    \centering
        \caption{Efficiency for analysis cuts, the energy independent efficiencies are computed assuming the cut efficiency does not depend on energy. For the extrapolated efficiency in $\mathcal{M}_1$ data at 3034 keV, we fit the efficiency as a function of energy to a first order polynomial and extrapolate this to 3034 keV, this is only relevant for the $\M_1$ signatures with 3034 keV energy.  }

    \begin{tabular}{cccc}
   Cut      & Efficiency Energy Independent [\%] & Efficiency Extrapolated [\%] & Method of Evaluation \\ \hline \hline
    Pileup     & $95.7\pm 1.0$ & - & Noise events\\
    Multiplicity & $99.55\pm 0.07$ & - & $^{210}$Po \\
    LD Anti-coincidence & $99.976\pm 0.017$ & - & $^{210}$Po \\
    Delayed coincidence & $99.16\pm 0.01$ & - & $^{210}$Po \\
    Muon Veto & $99.62\pm 0.07$ & - & $^{210}$Po \\
    Light Cut ($\mathcal{M}_1$) &$99.4\pm 0.4$ & $99.7\pm 0.8$ & Gamma Peaks ($\mathcal{M}_1$) \\
    Light Cut $\mathcal{M}>1$ &$97.7\pm1.8$ & -  & Gamma Peaks ($\mathcal{M}_2$) \\
    PCA $<23$ ($\mathcal{M}_1$) & $99.2\pm 0.3$ & - & Gamma Peaks ($\mathcal{M}_1$) \\
    PCA $<9$ $(\mathcal{M}1)$  & $95.2\pm 0.5$ & $94.3\pm 1.5$ & Gamma Peaks $(\mathcal{M}_1)$
    \end{tabular}
    \label{tab:cut_eff}
\end{table*}

\section{Bayesian analysis}
\label{Bayesian}
\begin{figure}
    \centering
    \includegraphics[width=0.47\textwidth]{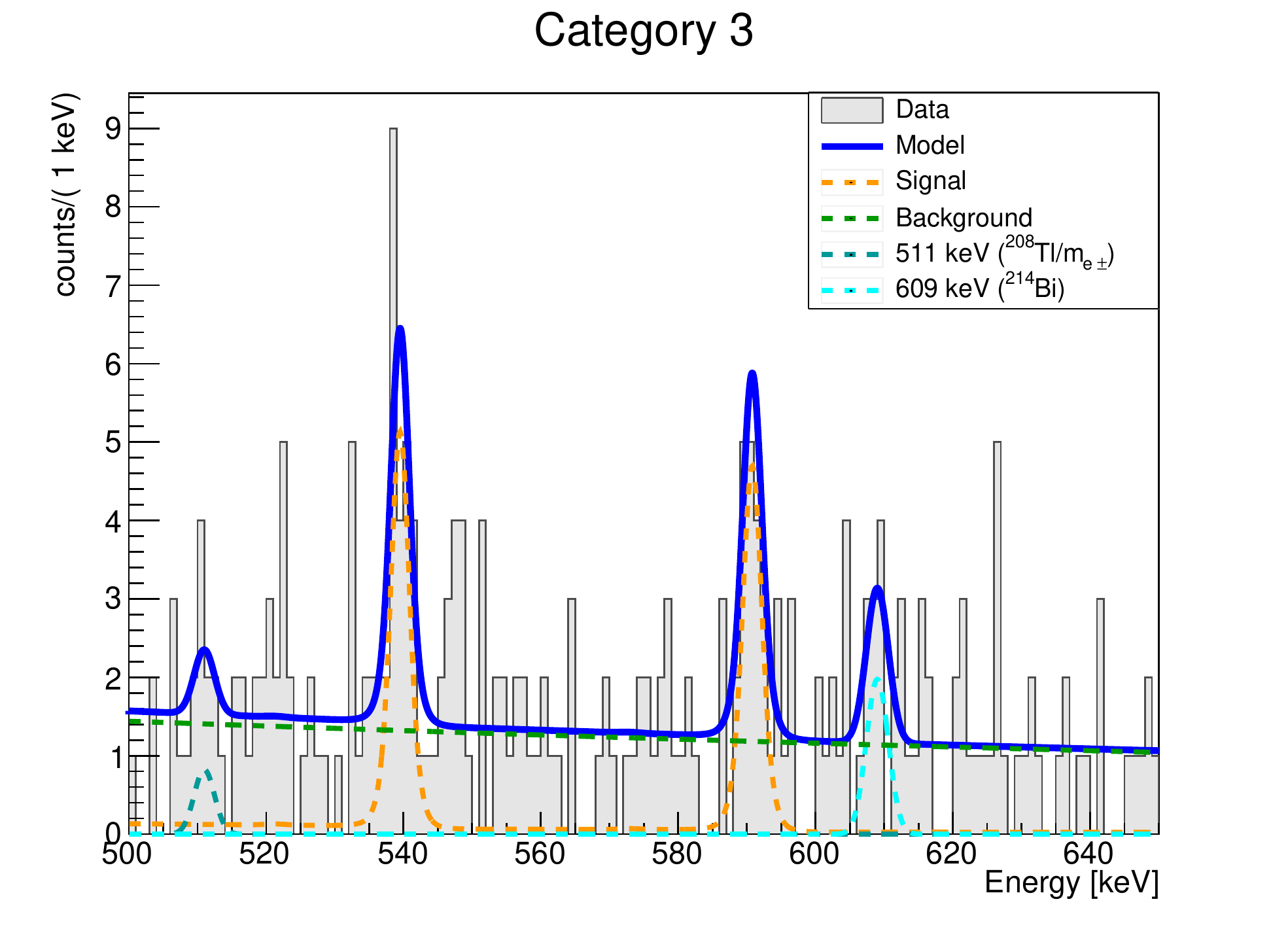}

    \caption{Fit of the $2\nu\beta\beta$ decay category 3. The experimental data are binned for visualization and we show the best fit reproduction of our model and the contributions from the signal peaks, background peaks and the exponential background.}
    \label{fig:one_fit}
\end{figure}

We perform a Bayesian analysis to extract decay rates. In particular, we use an extended unbinned maximum likelihood fit implemented using the Markov Chain Monte Carlo (MCMC) of the Bayesian Analysis Toolkit (BAT) \cite{Caldwell:2009} for all categories except for two $0\nu\beta\beta$ decay categories, for which we use a binned fit due to their exceptionally high statistics. 
We use a fit of the summed data of all 19 detectors and 9 datasets. 
\\ \indent
Two separate fits are created, one to extract the decay rates of $2\nu\beta\beta$ decay and another for the $0\nu\beta\beta$ decay. Both fits are implemented in the same framework.
In each category (index $r$) we model our experimental data as: 
\begin{align}
    f_r(E) &= \varepsilon_{a,r}n_{2^+}\cdot f_{2^+,r}(E)+\varepsilon_{a,r}n_{0^+}\cdot f_{0^+,r}(E) \\ &+ n_{r,\text{bkg}} \cdot f_{r,\text{bkg}}(E) +\sum_{p=1}^5 n_{r,p} f_p(E) \nonumber
\end{align}
where
\begin{itemize}
    \item $n_{0^+/2^+}$ are the observed number of counts to $0_1^+/2_1^+$ excited states.
    \item $f_{0^+/2^+,r}(E)$ are the phenomenological functions describing the signal shape of category $r$, normalized to unity over the sum of all categories. 
    \item $\varepsilon_{a,r}$ is the analysis efficiency for category $r$ (described in Section \ref{Syst}).
    \item $f_{r,\text{bkg}}(E)$ is a function describing the background in category $r$, either exponential or flat for categories with low statistics.
    \item $n_{r,p}$ are the number of background counts from known $\gamma$ lines (a subset of 511, 583, 609, 2448 and 2505~keV) with index $p$ and $f_{p}(E)$ is the model of this spectral shape, a single Gaussian.

\end{itemize}
We show in Fig. \ref{fig:one_fit} the fit of the $2\nu\beta\beta$ decay category 3 and the contribution from the exponential background, signal and background peaks.
Our likelihood function is:
\begin{align}
    \text{log}(\mathcal{L})=&\sum_{n=1}^N\text{log
}( f_{r(n)}(E_n))\\ +&\sum_{r=1}^{C} -\lambda_r+n_r\text{log}(\lambda_r)-\text{log}(n_r!) \nonumber\\
+&\sum_{b=1}^{B} \sum_{i=1}^{N_{\text{bin}}} - \lambda_i + n_{i,b}\cdot \text{log(}\lambda_{i,b})-\text{log}(n_{i,b}!), \nonumber
\end{align}
where the first sum is over the events in the experimental data (unbinned categories), $r(n)$ is the category of event $n$, $C$ is the number of categories, $\lambda_r$ is the total predicted number of events in category $r$, while $n_r$ is the observed number of events.  The last sums are a binned likelihood where the sum $b$ runs over the $B$ categories which use a binned likelihood. These are the categories 0 and 1 for the $0\nu\beta\beta$ decay where a large number of events make an unbinned fit computationally unfeasible. $\lambda_i$ is the expectation value for bin $i$ and $n_{i,b}$ is the number of events in experimental data. We use 0.5 keV bins for these fits, much smaller than the energy resolution over the full spectral range.
\\ \indent We use BAT to sample the full posterior distribution of the parameters of interest $\vec{\theta}= (n_{0_1^+},n_{2_1^+})$ and nuisance parameters $\vec{\nu}$, which include all of the background components (both the exponential/flat background and the $\gamma$ peaks) and also parameters of systematic uncertainties (see Section \ref{Syst})
\begin{equation}
    p(\vec{\theta},\vec{\nu}|\mathcal{D}) = \frac{P(\mathcal{D}|\vec{\theta},\vec{\nu})P(\vec{\theta},\vec{\nu})}{P(\mathcal{D})},
\end{equation}
where $\mathcal{D}$ represents the data.
We use uninformative (flat) priors on all background model components and parameters of interest $n_{2^+/0^+}$.
We define observables for the decay rates of $0_1^+/2_1^+$ decay ($\Gamma_{0^+/2^+}$)
\begin{align}
    \Gamma_{0^+/2^+}& = \frac{n_{0^+/2^+}\cdot W}{\varepsilon_{\text{cont},0^+/2^+}\cdot\eta \cdot Mt \cdot N_a\cdot 1000},
\end{align}
where $W$ is the molecular mass for enriched Li$_2\,^{100}$MoO$_4$, $N_a$ is the Avogadro constant, $Mt$ is the exposure (in $\mathrm{kg}\times \mathrm{yr}$), $\eta$ is the isotopic abundance of $^{100}$Mo and, $\varepsilon_{\text{cont},0^+/2^+}$ is the containment efficiency for $0_1^+/2_1^+$ signal, i.e. the total probability a simulated event is in one of the categories. The half-life is then given by $\ln{(2)}/\Gamma$.
For each step of the Markov chains the values of $\Gamma_{0^+/2^+}$ are stored and used to compute the marginalized posterior distributions.
We include all of our systematic uncertainties as nuisance parameters in our analysis as described in Section \ref{Syst}.

\section{Systematic uncertainties}
\label{Syst}
\subsection{Energy resolution and bias}
We account for the uncertainty in the energy resolution of our peaks. Both the terms $r_1$ and $r_0$ in the resolution scaling function ($R(E)$, see Section \ref{sec:reso}) are given Gaussian priors based on the measured uncertainty from the fit to $\gamma$ peak resolution. A multivariate prior is not necessary since the correlation is very low ($-0.2$\%) because the $r_0$ term is fixed very well by noise events (random triggers used to estimate energy resolution at 0 keV). At each stage in the Markov chain this resolution is applied to all the signal components and background peaks, by adjusting the resolution of the functions $f_{2^+/0^+,r}(E),f_{\text{r,bkg}}(E)$. 

We also account for the uncertainty in the energy scale. Our energy bias is parameterized as a second order polynomial
as described earlier in Section \ref{sec:reso}. We adjust the mean position of all peaks in our signal and background template functions ($f_{0^+/2^+},f_{\text{bkg}}$) by this bias. 
We add a nuisance parameter to the model accounting for the uncertainty on the bias, which is given a Gaussian prior with $0.07$ keV $\sigma$, and vary the position of each peak (in a correlated way) by this amount. For the $0\nu\beta\beta$ decay analysis the peaks cover a wide range in energy and we take a conservative approach varying the position of all the peaks by the largest uncertainty of $0.3 \ \mathrm{keV}$ (from 3034 keV). 
\subsection{Analysis efficiency}
We account for the uncertainty in the analysis efficiency, or the probability that a signal like event would pass all cuts. These cut efficiencies are included as nuisance parameters in our model which are given Gaussian priors based on the estimated uncertainty (see Section \ref{sec:effs}). A nuisance parameter is included for each cut, including two separate parameters for the constant and extrapolated light cut, and PCA$<9$ cut. We then predict the efficiency in category $r$ as:
\begin{equation}
    \varepsilon_r = \prod_{c=1}^{11} \varepsilon_{c}^{p(\mathcal{M}(r),c)},
\end{equation}
where the product $c$ runs over all cuts (both energy independent and extrapolated, see Table \ref{tab:cut_eff}) and $p(\mathcal{M}(r),c)$ is a power which represents how many times a cut was applied to an event. For example, for PSD the cut is applied for both pulses so the efficiency is raised to the power of the multiplicity.
\subsection{Containment Efficiency}
\label{sec:cont}
The final systematic uncertainty we account for is the containment efficiency from MC simulations. In particular, this is related to the accuracy of our {\sc Geant4} MC simulations which can broadly be divided into two parts: the accuracy of the simulated geometry and the implemented physics process. For the simulated geometry we vary the amounts of the three main materials in the experimental setup (see Figure \ref{Experiment_fig}):
\begin{itemize}
    \item The dimensions of the LMO crystals (and therefore density since the mass is very well known);
    \item The thickness of the copper holders;
    \item The thickness of the Ge LDs.
\end{itemize}
We run a set of simulations varying the LMO dimensions (by $\pm 100/200 \ \mathrm{\mu m}$ or around $\pm$0.2/0.4\%), the Cu holder thickness (by $\pm 100/200 \ \mathrm{\mu m}$ or $\pm$2.5/5\%) and the Ge LD thickness (by $\pm 10/20 \ \mathrm{\mu m}$  or around $\pm$6/12\%). We also account for the possible inaccuracy of {\sc Geant4} physics cross sections by running MC simulations varying both the Compton effect and photo-effect cross sections by $\pm 5$\% which is a conservative estimate of {\sc Geant4} cross section accuracy \cite{Weidenspointner:2013,Cirrone:2010}. For each systematic effect (Cu thickness, LMO dimension, Ge thickness, Compton or photo-effect cross section scale) we introduce a parameter $\nu_i$. For every effect, decay, category (index $r$) and peak (index $p$) we parameterize the variation in peak containment efficiency. This is the product of the overall containment efficiency and the Gaussian peak probability (see Section \ref{sec:sigshape}). We use first order polynomial fits to obtain $d\varepsilon_{\text{cont},r,p}/d\nu_i$. We show the results of all these tests in Appendix \ref{ap:cont}. We observe that the most significant effects are the photo-effect cross section (higher scale tends to increase containment) and the Compton cross section (higher tends to decrease containment). The geometrical uncertainties provide a much smaller effect. Effects where the variation in containment efficiency is consistent with 0 (within $2\sigma$) are not included. Based on the parameters $\vec{\nu}$ we predict the efficiency for category $r$ and peak $p$ as:
\begin{equation}
    \varepsilon_{\text{cont},r,p}=\varepsilon^0_{\text{cont},r,p}+\sum_{c=1}^5\nu_c\cdot d\varepsilon_{\text{cont},r,p}/d(\nu_c),
\end{equation} 
where the sum $c$ runs over systematic effects, and $\varepsilon^0_{\text{cont},r,p}$ is the default efficiency (for the MC simulation without varying any parameters). In our fit, the nuisance parameters accounting for uncertainties of copper thickness and LMO dimension are given Gaussian priors with a standard deviation of 100$ \ \mathrm{\mu m}$; for the Ge thickness we use 20$\, \mathrm{\mu m}$, and we use $ \ 5\%$ for the Compton and photo-effect physics cross sections. At each step of the Markov chain the amplitude of each peak is adjusted according to these parameters. The normalization of the fit functions are adjusted accordingly and we recompute the containment efficiency for both $0_1^+/2_1^+$ signals.
\subsection{Choice of $2\nu\beta\beta$ decay to $0_1^+$ state model}
The final systematic uncertainty we consider is the $2\nu\beta\beta$ to $0_1^+$ state decay model. By default we use the SSD model, however we also run MC simulations and a fit using the HSD model. We then consider the difference between these two results as a systematic on the $2\nu\beta\beta$ decay to $0_1^+$ state decay rate.
\section{Results}
\label{Results}
\subsection{Toy Monte-Carlo sensitivity and bias}
\begin{figure} 
    \centering
    \includegraphics[width=0.5\textwidth]{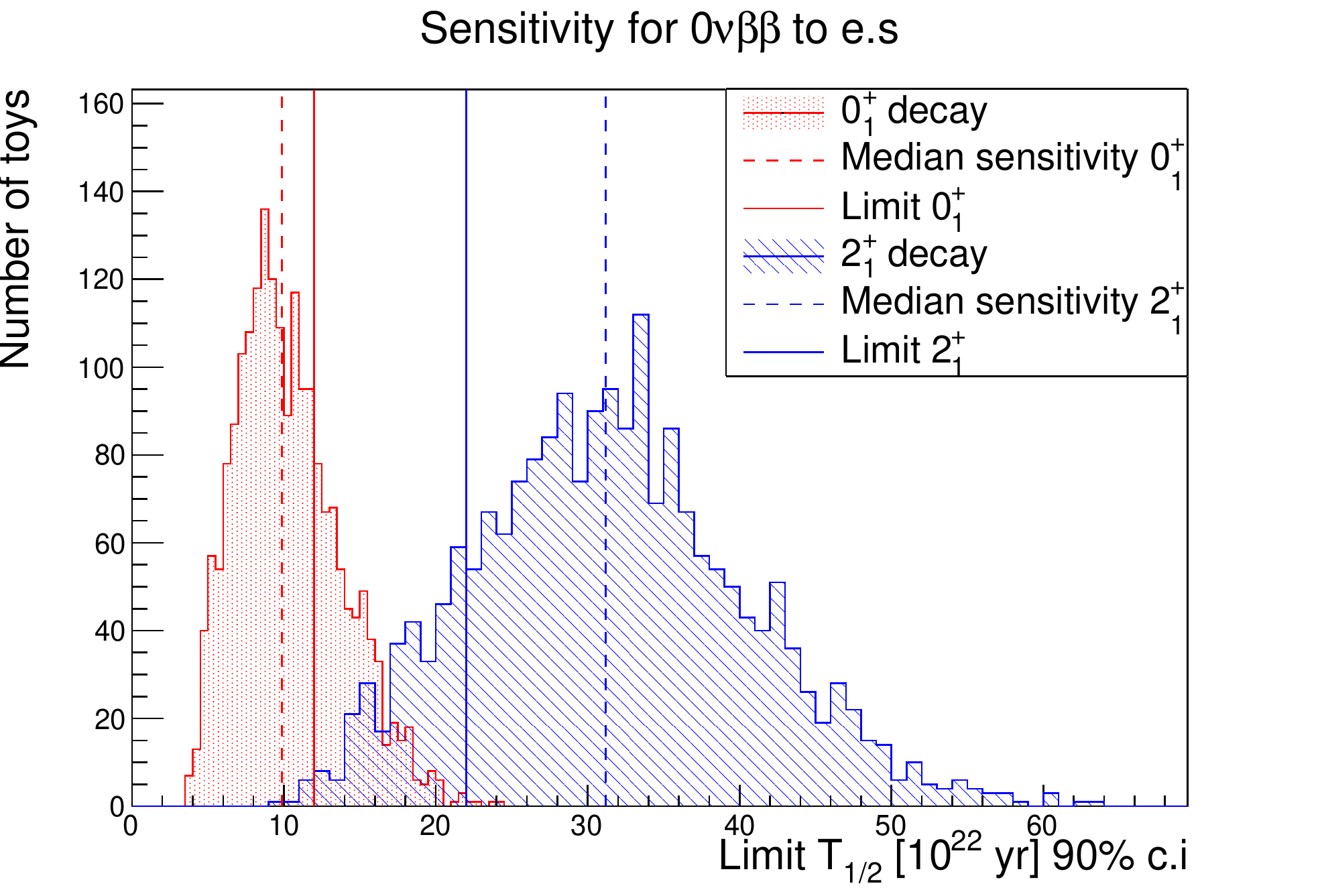}
    \caption{Sensitivity for $0\nu\beta\beta$ decay to both $0_1^+$ and $2_1^+$ excited states obtained from an ensemble of 2000 toy experiments.  We show the median sensitivity (dashed line) and the limit obtained on real data (solid lines). In both cases the limits on real data are consistent with those expected from toy experiments.} 
    \label{sens_0nu}
        \includegraphics[width=0.5\textwidth]{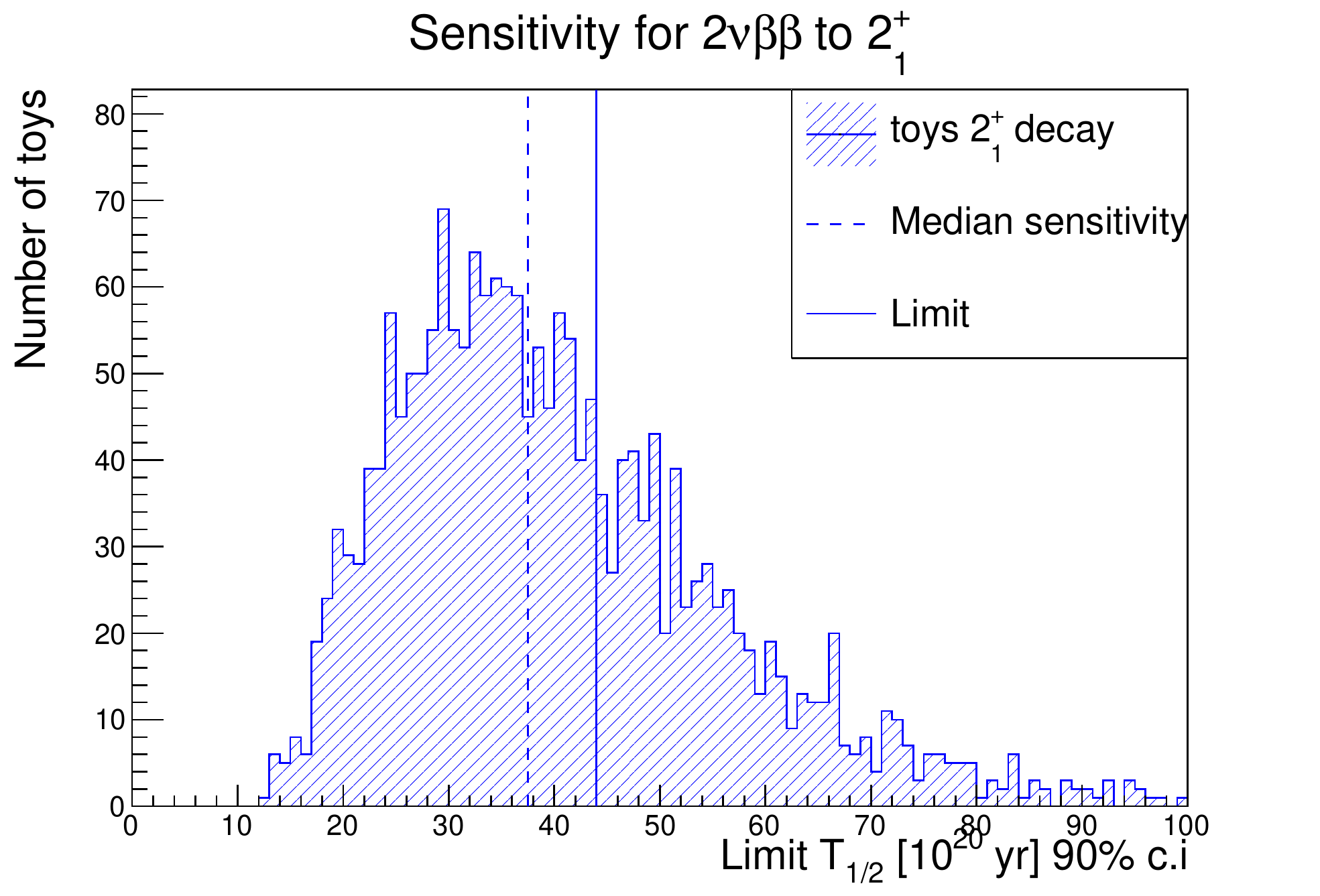}
    \caption{Sensitivity for $2\nu\beta\beta$ decay to $2_1^+$ excited state obtained from an ensemble of 2000 toy experiments. As with $0\nu\beta\beta$ decay we show the median sensitivity (dashed line) and the limit obtained on real data (solid line). Again the limit on real data is consistent with the toy experiments.}
    \label{sens_2nu}

\end{figure}

\begin{figure}
    \centering
    \includegraphics[width=0.5\textwidth]{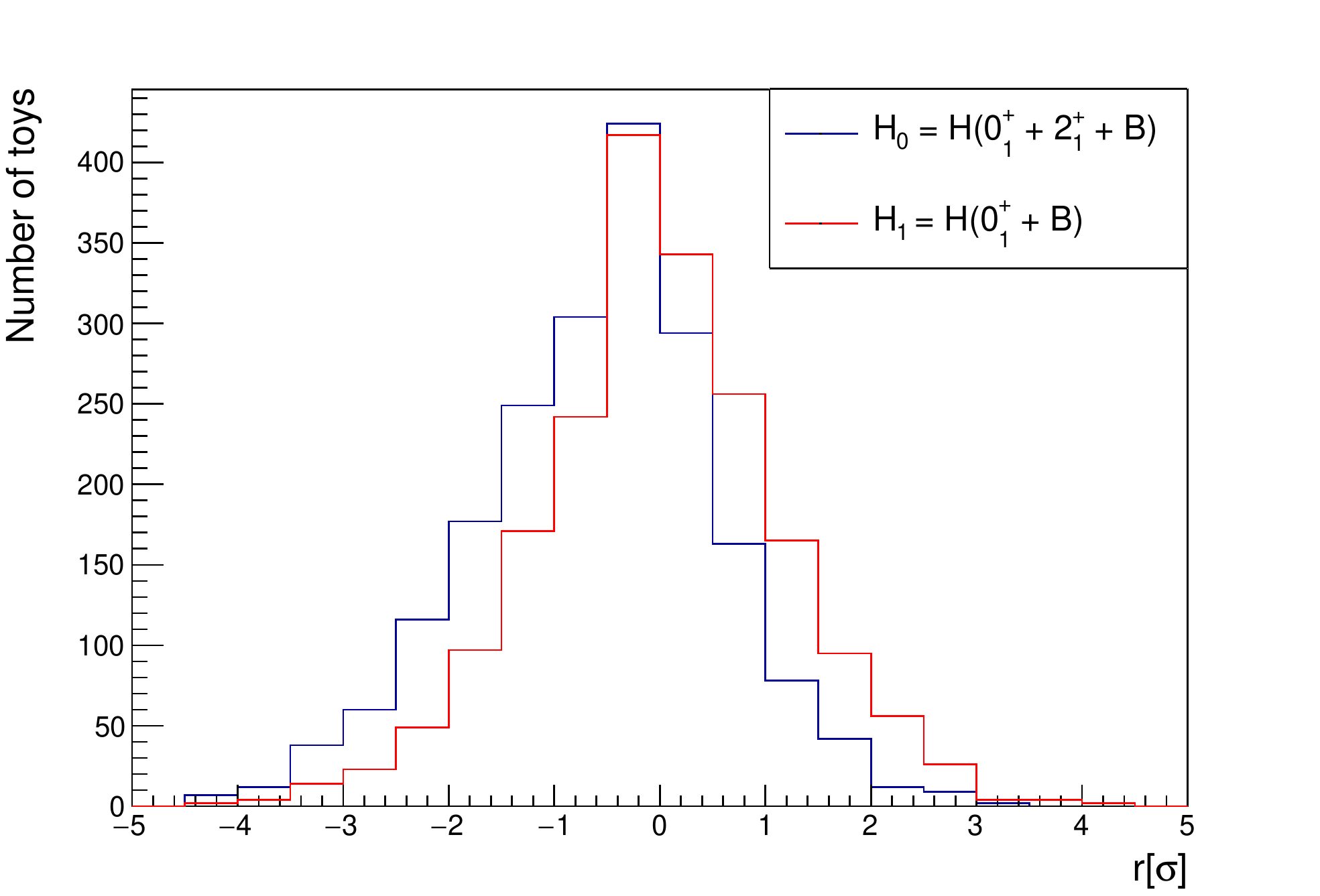}
    \caption{The distribution of $r_i$ (see Eq. \ref{r}) for both $H_0$ and $H_1$ models for our fits to toy data. We observe a clear bias for the $H_0$ model while the $H_1$ model is unbiased.}
    \label{fig:biastoys}
\end{figure}

To validate our fitting routine and predict the median exclusion sensitivity we use an ensemble of pseudo-experiments (toy MC). We generate 2000 toy datasets for both $0\nu\beta\beta$  and $2\nu\beta\beta$ decay analysis based on the background parameters from the fit to data (see Sections \ref{2nu_fit_b},\ref{0nu_fit}). We assume the $2\nu\beta\beta \rightarrow 0_1^+$ signal from the best fit to data and zero signal of $2\nu\beta\beta \rightarrow2_1^+$ or $0\nu\beta\beta$ decay modes. We sample from a Poisson distribution for the number of counts of each component in each toy experiment. We fit each dataset to determine the distribution of possible limits and therefore the median sensitivity for each decay. This is shown in Figures \ref{sens_0nu} and \ref{sens_2nu}. We extract median exclusion sensitivities, $\hat{T}$, at 90\% credibility interval (c.i.) for the three decays which have not yet been measured of:
\begin{align}
    \hat{T}_{1/2}^{2\nu\rightarrow 2_1^+}&=\Senstwonu, \\
        \hat{T}_{1/2}^{0\nu \rightarrow 2_1^+}&=\Senstwo, \\
                \hat{T}_{1/2}^{0\nu\rightarrow 0_1^+}&=\Senszero.
\end{align}
We also use these toys to verify the fitting procedure is not biased. 
We fit our toy datasets for $2\nu\beta\beta$ decay analysis with both a model including the $2_1^+$ signal called $H_0=H(0_1^++2_1^++B)$ and one including only $0_1^+$ signal, $H_1=H(0_1^++B)$. We compute 
\begin{equation}
\label{r}
    r_i= \frac{n_{0^+}^i-\bar{n_{0^+}}}{\sigma(n_{0^+})},
\end{equation}where $\bar{n_{0^+}}$ is the marginalized mode from the fit to real data used to generate the toys, $n_{0^+}^i$ is the marginalized mode in toy $i$, and $\sigma(n_{0^+})$ is the estimated error (central 68\% c.i.). The distribution of $r_i$ for both models is shown in Fig. \ref{fig:biastoys}. We observe a clear bias in $r_i$ for the $H_0$ model with a mean of $-0.58\sigma$, while $r_i$ is unbiased for the $H_1$ model (mean of $-0.005\sigma$). Hence, we decide to use the $H_1$ model for the case that no evidence of a decay to $2_1^+$ state is found. This bias is due to non-negative Bayesian priors on the rate of $2_1^+$ decay which allow the rate of $2_1^+$ to fluctuate only up. Since the $0_1^+$ and $2_1^+$ rates are anti-correlated this causes the $0_1^+$ rate to be biased.
\begin{figure*}
    \centering
        \includegraphics[width=0.45\textwidth]{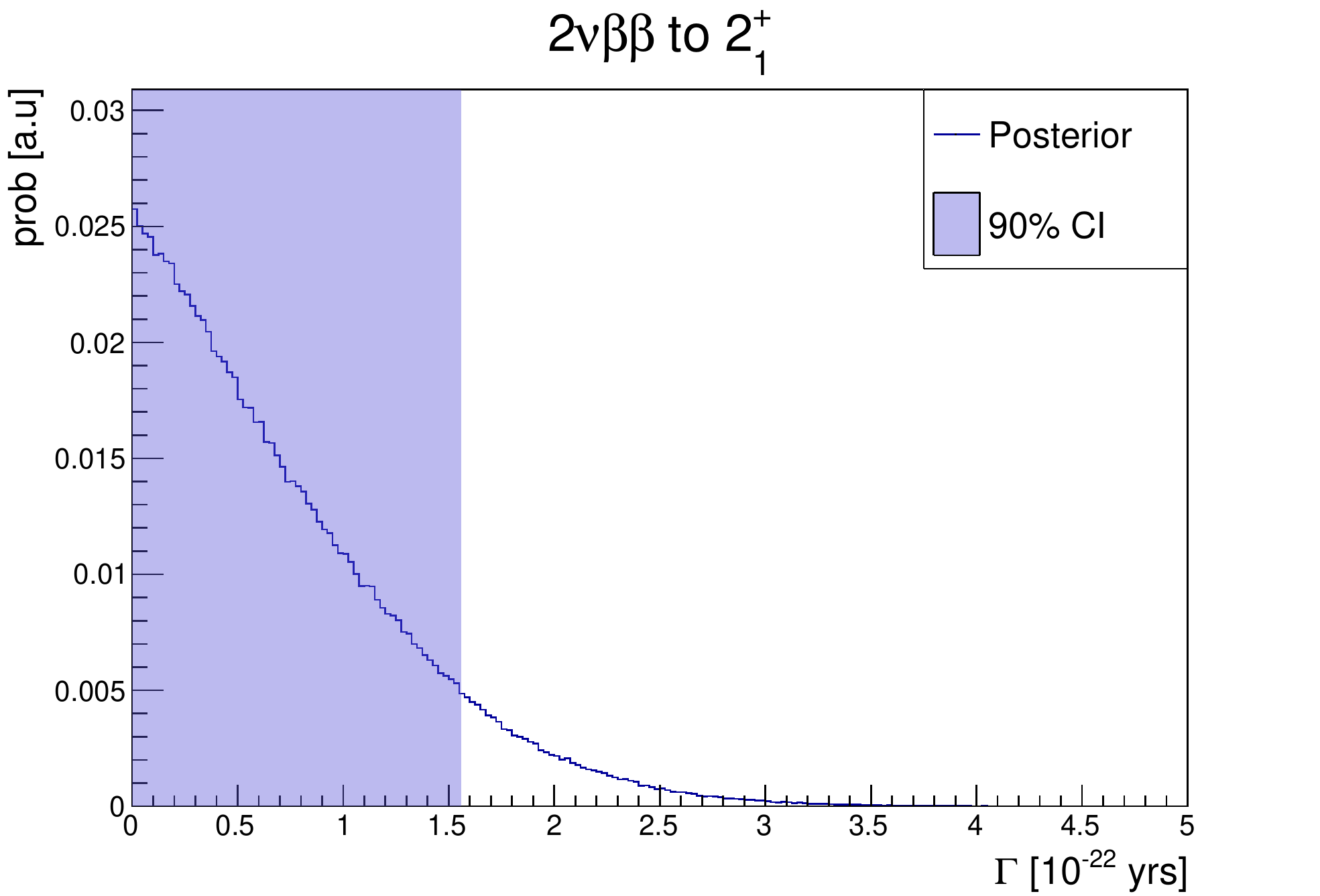}
    \includegraphics[width=0.45\textwidth]{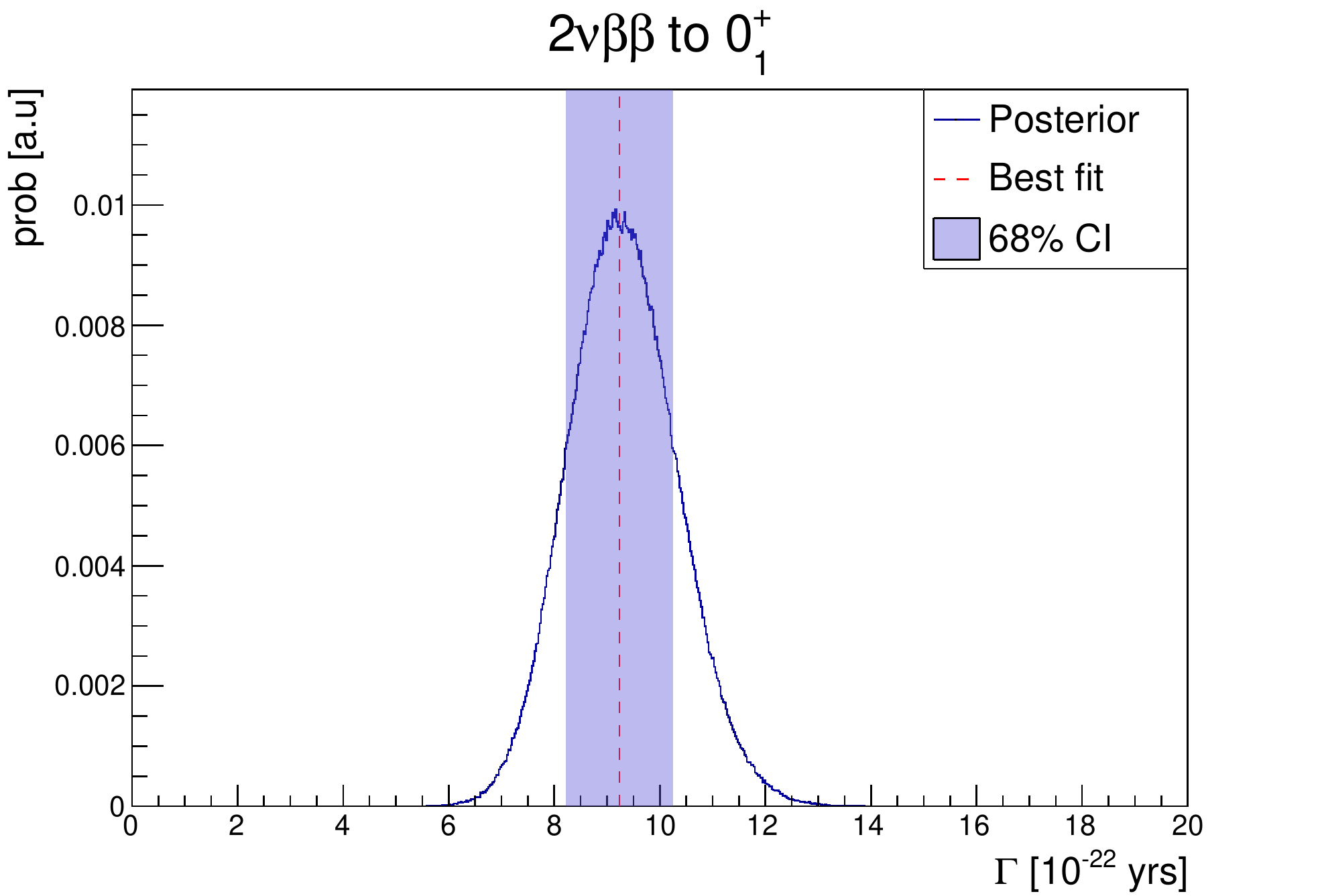}
            \includegraphics[width=0.45\textwidth]{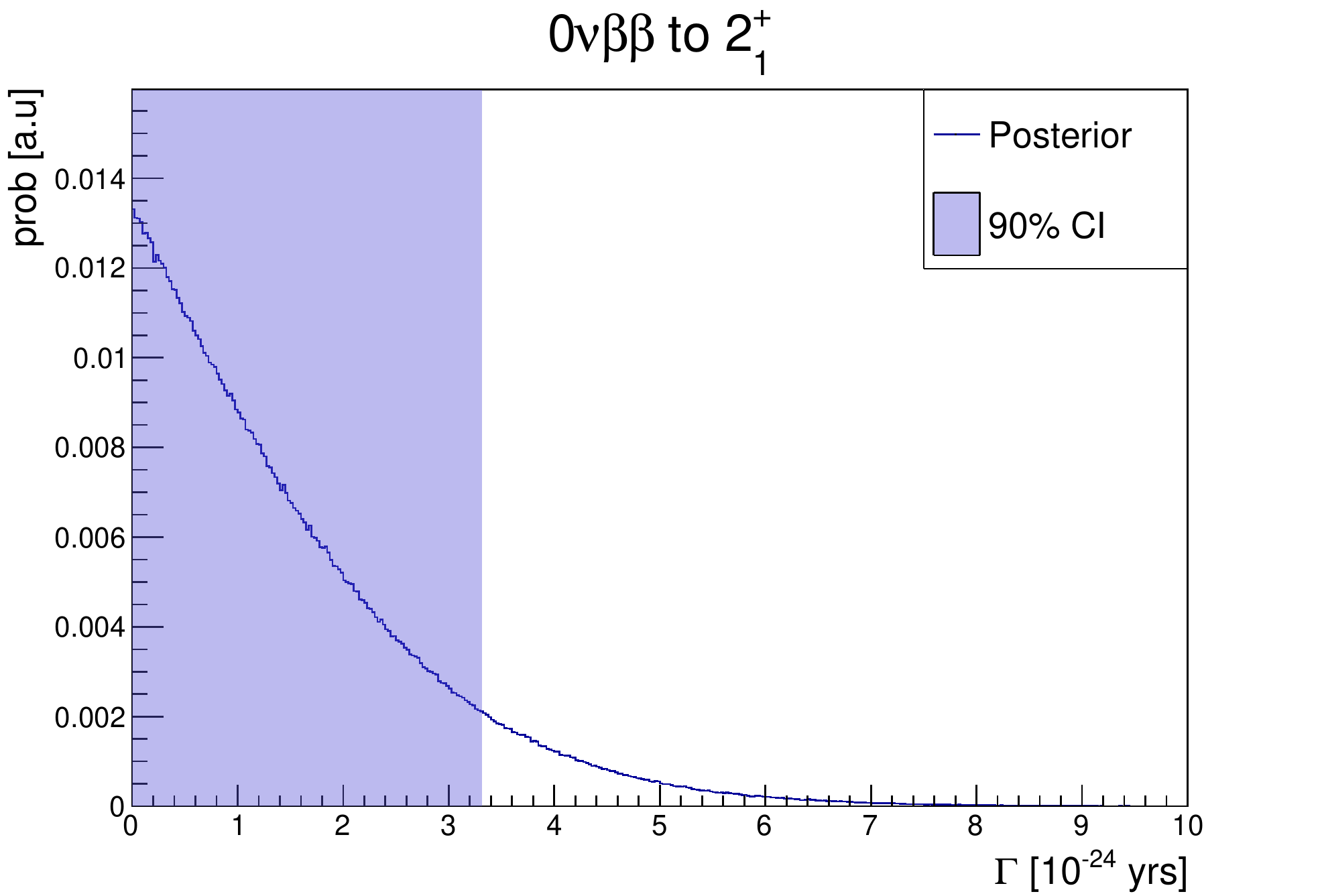}
    \includegraphics[width=0.45\textwidth]{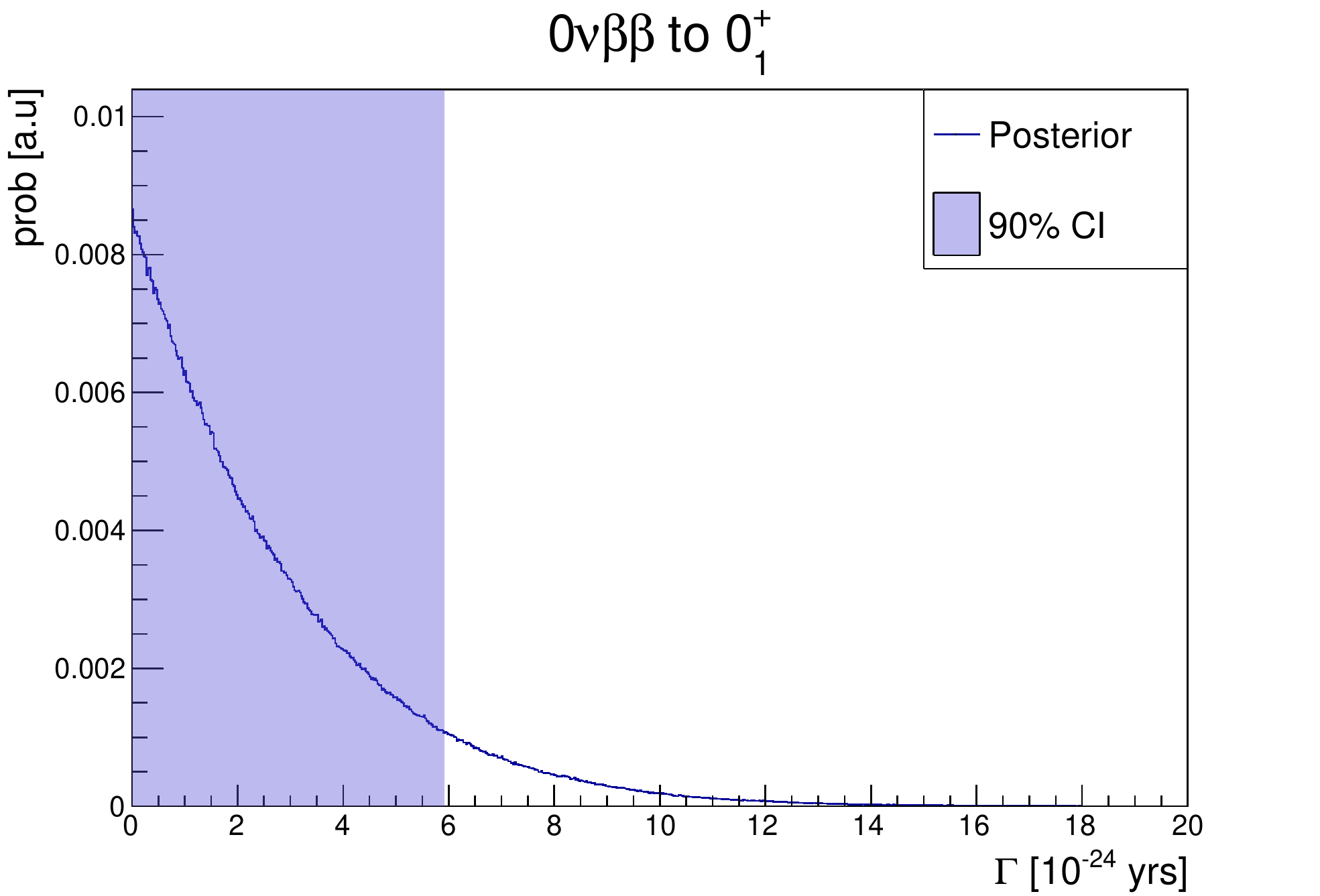}

    \caption{Posterior probability distributions for the decay rates, $\Gamma^{2\nu}$ to $2_1^+$ state (top left) and $0_1^+$ state (top right). The mode for $2_1^+$ decay is at zero rate and so we find no evidence for this decay and we extract the measurement of $0_1^+$ mode using the $H(0_1^+ +B )$ model. Bottom left/right are the posteriors of $\Gamma^{0\nu}$ to $2_1^+$ and $0_1^+$ state respectively, again showing no evidence of the decay.}
    \label{fig:posterior}
\end{figure*}

\subsection{Fit of $2\nu\beta\beta$ with $H_0=H(0_{1}^+2_1^++B)$ model}
\label{2nu_fit_a}
We run the $2\nu\beta\beta$ decay fit with both $2\nu\beta\beta$ decay to $2_1^+$ and $0_1^+$ \es signals and background components, i.e. the $H_0$ model. We can use this fit to determine if we observe evidence for the decay to $2_1^+$ state and to set a limit if no evidence is found.  We consider first a fit including all possible background peaks, the {\it maximal model}. We then repeat the fit removing any background peak where the central 68\% confidence interval contains 0 counts and replacing the exponential background with a constant if the slope is compatible with 0 (within 1 $\sigma$). We call this {\it the minimal model} and it is this model we use for statistical inference. We show the marginalized posterior distribution of the decay rate $\Gamma$ for $2\nu\beta\beta$ decay to $2_1^+$ state (including all systematics) in Fig. \ref{fig:posterior} (left). We see that the mode is at 0 rate and therefore find no evidence for $2\nu\beta\beta$ to $2_1^+$ state.
We correspondingly set a limit (including all systematics) of:
\begin{equation}
    T_{1/2}^{2\nu\rightarrow 2_1^+}>\Limtwoplus
    \end{equation}
This is the most stringent constraint on this process in $^{100}$Mo, an improvement of $\sim 80\%$ over the previous constraint, $2.5\times 10^{21}\,\mathrm{yr}$ \cite{NEMO3-ES.2014}.
\subsubsection{Fit of $2\nu\beta\beta$ with $H_1=H(0_{1}^++B)$ model}
\label{2nu_fit_b}

\begin{figure*}
    \centering
\end{figure*}
Since we find no evidence of the $2\nu\beta\beta$ decay to $2_1^+$ state and toy experiments show that including this parameter in our fit would bias our measurement, we run the $2\nu\beta\beta$ decay $H_1=H(0_1^++B)$ fit without the $2_1^+$ decay contribution. We find that this model is able to describe our data in all seven categories very well (see Fig. \ref{fig:one_fit} for an example and all fits in Appendix \ref{ap:fits}). We verify the goodness of fit using our ensemble of toy experiments. 
For each fit we extract the {\it global mode} or the set of parameters $\vec{\theta}$ with the maximum probability. We use the posterior probability of these parameters:
\begin{equation}
    k = p(\vec{\theta}|\mathcal{D}),
\end{equation}
as a test statistic to quantify the quality of the fit by comparing the probability distribution obtained in toy experiments $p(k^{'})$ to the fit on real data.
We extract the $p$-value:
\begin{equation}
    p=\int_{0}^{k} p(k^{'}) dk^{'}= \twonuprob
\end{equation}
indicating that the model describes the data well. \\ \indent
We extract the posterior distribution on $\Gamma^{2\nu\rightarrow0_1^+}$ from the fit on the data and  we extract the central 68.3\% c.i. on the decay rate as:
\begin{equation}
    \Gamma_{2\nu \rightarrow 0_1^+} = \gammazeroSSD.
\end{equation}
As the systematics are allowed to float freely in our fit this error is the sum of the statistical and systematic components. 
To evaluate the statistical error, we repeat the fit fixing all nuisance parameters connected to systematics, obtaining:
\begin{equation}
    \Gamma_{2\nu \rightarrow 0_1^+} = \gammazerostat.
\end{equation}
A difference in the error compared to the statistical plus systematic fit is only observed after the first digit.
We also run a fit using the HSD model, which leads to:
\begin{equation}
    \Gamma_{2\nu \rightarrow 0_1^+}^{\text{HSD}}=\gammazero,
\end{equation}
a $4\%$ decrease in the decay rate.
Under the assumption that statistical and systematic errors add in quadrature we obtain:
\begin{equation}
        \Gamma_{2\nu \rightarrow 0_1^+} = \gammazerostatandsyst,
\end{equation}
or converting to half life:
\begin{equation}
    T_{1/2}^{2\nu \rightarrow 0_{1}^{+}}=\Thalftwonu.
\end{equation}
This is a new independent measurement of this decay, with total uncertainty consistent with the previous leading measurement \cite{NEMO3-ES.2014}. 
\subsubsection{$0\nu\beta\beta$ decay fit}
\label{0nu_fit}
From the fit for $0\nu\beta\beta$ decay we find no evidence of either the decay to $0_1^+/2_1^+$ excited states. The best fit reproductions of the experimental data are shown in Appendix \ref{ap:fits}. The posterior distributions of $\Gamma^{0\nu \rightarrow 2_1^+/0_1^+}$ are shown in Fig. \ref{fig:posterior} (bottom left and right), this leads to limits (both at 90\% c.i. including all systematic uncertainties) of:
\begin{align}
    T_{1/2}^{0\nu\rightarrow2_1^+}>\Tzeronutwo, \\
    T_{1/2}^{0\nu\rightarrow0_1^+}>\Tzeronuzero.
\end{align}
These are new leading limits on these processes, a factor of 1.3 stronger than previous limits from \cite{NEMO_ES} in both cases, despite a factor of $\sim 6$ lower exposure of $^{100}$Mo.
\subsubsection{$2\nu\beta\beta$ decay spectral shape}
The CUPID-Mo source equals detector geometry also allows us to investigate the $2\nu\beta\beta\rightarrow 0_{1}^{+}$ spectral shape.
This provides a concrete demonstration of a method for this analysis which can be applied to future experiments with large statistics. Our analysis described in Section \ref{Analysis} does not reconstruct the $\beta\beta$ spectral shape directly, so we develop a procedure to extract this shape. For simplicity this analysis is only applied to $\mathcal{M}_2$ data which is the main contribution to the sensitivity.\\

We select all events that have either $E_1$ or $E_2$ in the range $[539.5-1.75,539.5+1.75]$ keV and $[590.8-1.75,590.8+1.75]$ keV. This choice of the interval is the FWHM energy resolution, roughly optimal for maximising the sensitivity $S/\sqrt{B}$. We note there is some ambiguity in this selection when $E_1 \sim E_2$ (when the $\beta\beta$ has the same energy as the $\gamma$) however the excellent CUPID-Mo energy resolution makes this negligible. We then define $E_{\beta\beta,540},E_{\beta\beta,591}$ as the energy which does not contain the peak for both 540 and 591 keV peaks respectively and we construct histograms of these energies. This procedure is applied to both the data, signal MC simulations, and to the preliminary background model reconstruction of the data.

\begin{figure*}
    \centering
    \includegraphics[width=0.48\textwidth]{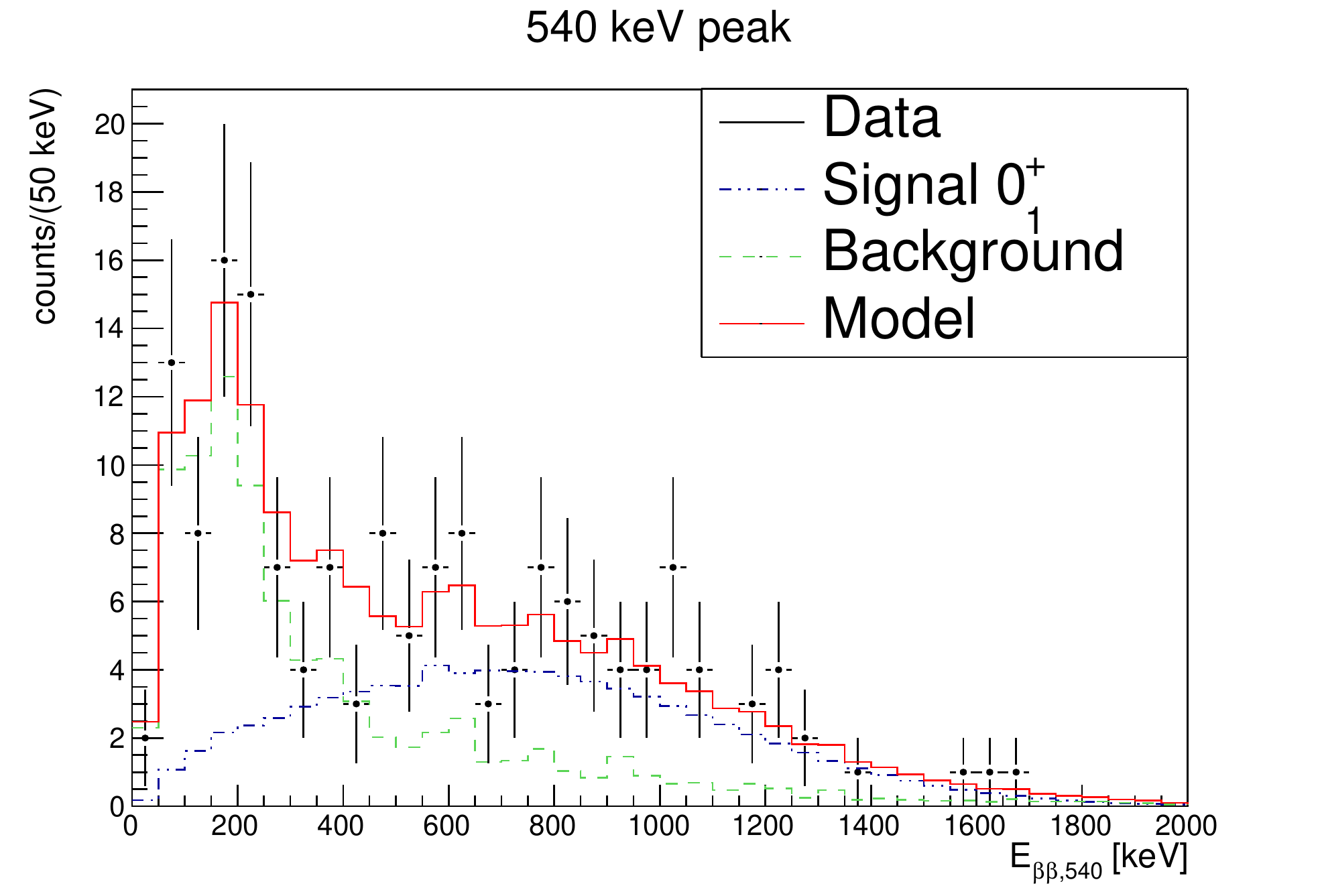}
        \includegraphics[width=0.48\textwidth]{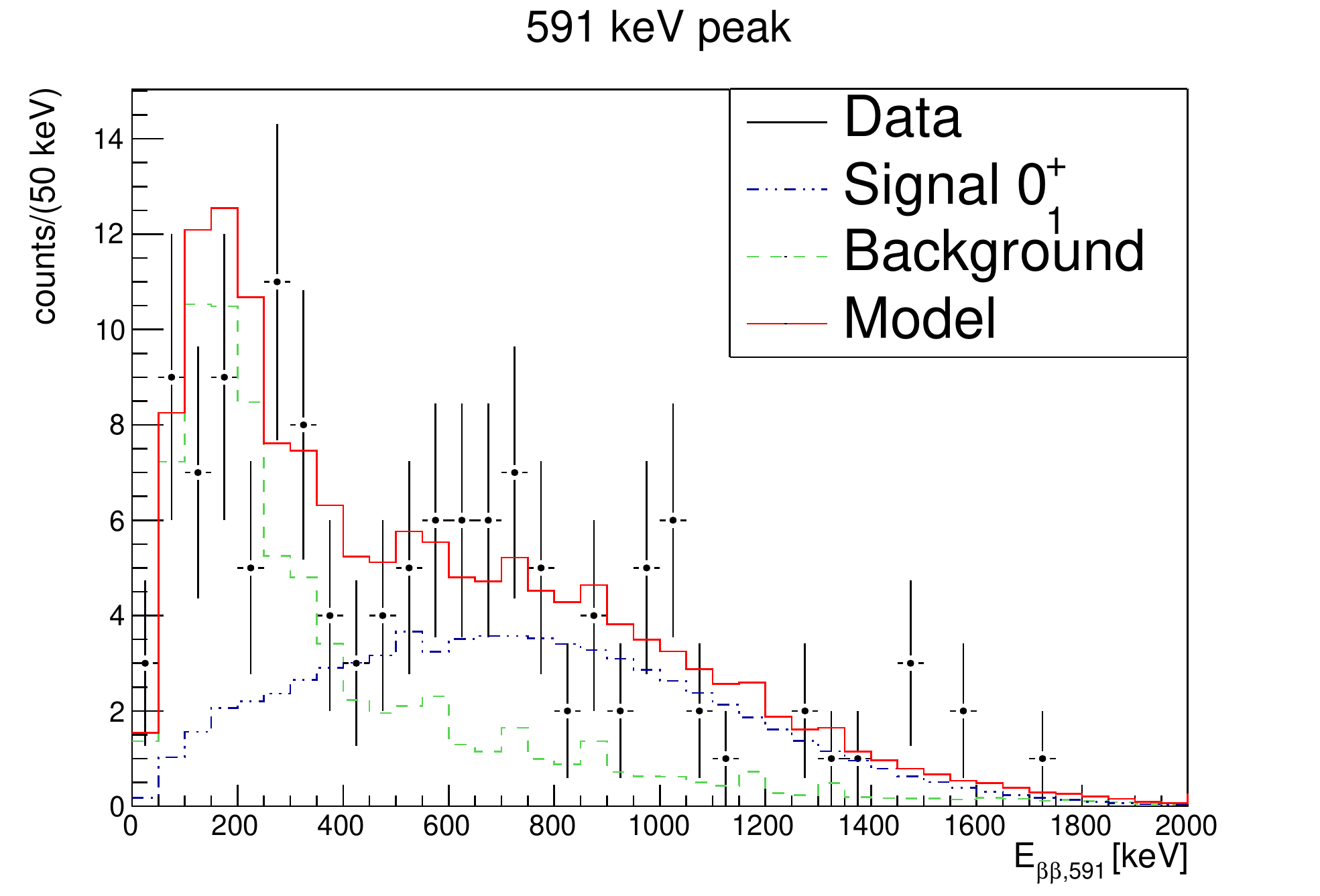}
    \caption{Plots showing the $2\nu\beta\beta$ decay to \es $\beta\beta$ spectral shape for 540 keV peak (\textit{left}) and 591 keV peak (\textit{right}). We overlay the best fit prediction. We find that in the region $>600$ keV, the spectrum is dominated by $2\nu\beta\beta \rightarrow 0_1^+$.}
    \label{beta_plot}
\end{figure*}

To compare the SSD and HSD models we use a simultaneous binned Bayesian likelihood fit of the two $\beta\beta$ spectra. We use three components; background, $0_1^+$ signal and $2_1^+$ signal and 50 keV bins. Using the SSD model, this fit reconstructs the half-life as:
\begin{align}
    T_{1/2}^{2\nu \rightarrow 0_1^+}&=\SpectralThalf, \\
    T_{1/2}^{2\nu \rightarrow 2_1^+}&> \SpectralLimit.
\end{align}
These are consistent with the values from Sections \ref{2nu_fit_a},\ref{2nu_fit_b} and should be considered a cross check of this more robust analysis which does not depend on the quality of the background model. We show this fit reproduction in Fig. \ref{beta_plot}. Repeating the fit using MC simulations obtained with the HSD model, we extract the evidence for both models $p(\mathcal{D}|\text{SSD/HSD})$ and therefore probabilities for the HSD or SSD models of:
\begin{equation}
  p(\text{SSD}|\mathcal{D})= \pSSD, \ p(\text{HSD}|\mathcal{D})=\pHSD.
\end{equation}
This indicates that the CUPID-Mo data is not able to differentiate between the two models. However, this analysis provides a method which could be used to differentiate these two models in a future experiment with larger statistics.
\section{Discussion}
\label{Discussion}
\subsubsection{Matrix element for $2\nu\beta\beta$ decay to $0_1^+$}
From our measurement of $2\nu\beta\beta \rightarrow 0_1^+$ excited state we can measure experimentally the nuclear matrix element for this process based on:
\begin{equation}
    \Gamma^{2\nu}/\ln{(2)}= G_{2\nu}\cdot |M^{\text{eff}}_{2\nu}|^2.
\end{equation}
The phase space factor is given by $(60.55$\,--\,$65.18) \times 10^{-21} \ \mathrm{1/yr}$ depending on whether the HSD or SSD model is used \cite{Kotila:2012}. Therefore the matrix element (assuming the SSD model) is given by:
\begin{align}
    M^{\text{eff}}_{2\nu \rightarrow 0_1^+}&=\sqrt{\frac{\Gamma}{(\ln{(2)}\cdot G)}}\pm \frac{1}{2\sqrt{\ln{(2)}G}}\frac{\Delta \Gamma}{\sqrt{\Gamma}} \\ &= \Mtwonu. \nonumber
\end{align}
We compare the theoretical predictions of dimensionless $M^{\text{eff}}_{2\nu}=g_A^2 M_{2\nu}$ calculated assuming and unquenched value of $g_A=1.27$. These are 0.395, 0.595, 0.185 for the shell model \cite{Coraggio:2022vgy}, microscopic interacting Boson model \cite{BareaIBM2}, and quasi-particle random phase approximation \cite{Jouni2015}, respectively. 
This shows the decay rate is quenched relative to theoretical predictions.
\subsubsection{Effective Majorana neutrino mass, $m_{\beta\beta}$ }
The limits on $0\nu\beta\beta \rightarrow 0_1^+$, despite the lower phase space, can be used to set a limit on $m_{\beta\beta}$. We use the phase space factor ($3.162\times 10^{-15}\, \mathrm{yr^{-1}}$ from \cite{Kotila:2012}) and the NMEs from \cite{Coraggio:2022vgy,BareaIBM2} to obtain:
\begin{equation}
    m_{\beta\beta}< (11-15) \,\mathrm{eV},
\end{equation}
depending on the NMEs used. Whilst this is still several orders of magnitude above the constrains from the g.s. decay this can be improved significantly in future for a large experiment with minimized dead material. 
\subsubsection{Bosonic neutrinos}
Double beta decays to $J\neq 0$ excited states can be sensitive to a Bosonic contribution to the neutrino wave-function \cite{Barabash:2007}. In particular, under the SSD hypothesis, the predicted half-lives for $2\nu\beta\beta$ to the $2_1^+$ \es are $2.4\times 10^{22}, \ 1.7\times 10^{23}\, \mathrm{yr}$ for Bosonic and Fermionic neutrinos respectively. The current limit of $4.4\times 10^{21}\,\mathrm{yr}$ is still below these predictions. 
\subsubsection{SSD vs HSD models}
Both NEMO-3 and CUPID-Mo have demonstrated that SSD hypothesis describes the experimental data very well for the $2\nu\beta\beta $ to $0^+_{\text{g.s.}}$ decay in $^{100}$Mo \cite{Arnold:2019,Armengaud:2020b}. However, for the decay to $0_1^+$ state the mechanism (SSD or HSD/closure approximation) is not known. CUPID-Mo is only the second experiment to reconstruct the $2\nu\beta\beta$ to $0_1^+$ $\beta\beta$ spectral shape. However, the present level of statistics is insufficient to distinguish the two modes. The excellent energy resolution and low background rates mean this will be possible in a future experiment. In this paper, we demonstrate a method to quantify numerically this and obtained the first result on the compatibility of these two models with data.

\section{Conclusion}

In this paper we have presented a new analysis of $0\nu\beta\beta$ and $2\nu\beta\beta$ transitions of $^{100}$Mo to the first two ($2_1^+/0_1^+$) excited states of $^{100}$Ru using the full exposure of CUPID-Mo.
This analysis exploits the information available for a source equals detector experiment, where both $\beta$ and $\gamma$'s can be measured.
A measurement of $^{100}$Mo $T_{1/2}^{2\nu\rightarrow 0_1^+}$ was obtained:
\begin{equation}
    T_{1/2}^{2\nu\rightarrow 0_1^+}=\Thalftwonu.
\end{equation}
For the other three decay modes, no evidence was found and we extract the limits:
\begin{align}
    T_{1/2}^{2\nu\rightarrow 2_1^+}&>\Limtwoplus,\\
    T_{1/2}^{0\nu\rightarrow 0_1^+}&>\Tzeronuzero,\\
    T_{1/2}^{0\nu\rightarrow 2_1^+}&>\Tzeronutwo.
\end{align}
These are the leading limits on these processes. The sensitivity was limited by the small size of the array and large amount of dead material which results in a low containment efficiency. Future experiments such as CUPID \cite{CUPIDInterestGroup:2019inu} or CROSS \cite{CROSS:2019xov} will feature a more tightly packed array, much lower amounts of dead material, and a much larger exposure. This will lead to a significantly improved sensitivity with a precision measurement of $2\nu\beta\beta$ decay to $0_1^+$ state, distinction between the SSD and HSD mechanisms of the decay and the possibility to measure the $2\nu\beta\beta$ decay to $2_1^+$ state. 
\section{Acknowledgements}
This work has been partially performed in the framework of the LUMINEU program, a project funded by the Agence Nationale de la Recherche (ANR, France). The help of the technical staff of the Laboratoire Souterrain de Modane and of the other participant laboratories is gratefully acknowledged. 
We thank the mechanical workshops of CEA/SPEC for their valuable contribution in the detector conception and of LAL (now IJCLab) for the detector holders fabrication. 
F.A. Danevich, V.V. Kobychev, V.I. Tretyak and M.M. Zarytskyy were supported in part by the National Research Foundation of Ukraine Grant No. 2020.02/0011. O.G. Polischuk was supported in part by the project “Investigations of rare nuclear processes” of the program of the National Academy of Sciences of Ukraine “Laboratory of young scientists”. A.S. Barabash, S.I. Konovalov, I.M. Makarov, V.N. Shlegel and V.I. Umatov were supported by the Russian Science Foundation under grant No. 18-12-00003. We acknowledge the support of the P2IO LabEx (ANR-10-LABX0038) in the framework “Investissements d’Avenir” (ANR-11-IDEX-0003-01 – Project “BSM-nu”) managed by the Agence Nationale de la Recherche (ANR), France. 
Additionally the work is supported by the Istituto Nazionale di Fisica Nucleare (INFN) and by the EU Horizon2020 research and innovation program under the Marie Sklodowska-Curie Grant Agreement No. 754496. This work is also based on support by the US Department of Energy (DOE) Office of Science under Contract Nos. DE-AC02-05CH11231, and by the DOE Office of Science, Office of Nuclear Physics under Contract Nos. DE-FG02-08ER41551, DE-SC0011091; by the France-Berkeley Fund, the MISTI-France fund and  by the Chateau-briand Fellowship of the Office for Science \& Technology of the Embassy of France in the United States. J. Kotila is supported by Academy of Finland (Grant Nos. 3314733, 320062, 345869)
This research used resources of the National Energy Research Scientific Computing Center (NERSC).
This work makes use of the {\tt Diana} data analysis software which has been developed by the Cuoricino, CUORE, LUCIFER, and CUPID-0 Collaborations. \\ \indent
Russian and Ukrainian scientists have given and give crucial contributions to CUPID-Mo. For this reason, the CUPID-Mo collaboration is particularly sensitive to the current situation in Ukraine. The position of the collaboration leadership on this matter, approved by majority, is expressed at \url{https://cupid-mo.mit.edu/collaboration#statement} . Majority of the work described here was completed before February 24, 2022.

\bibliography{Biblio_PRC.bib}

\begin{thebibliography}{69}%
\makeatletter
\providecommand \@ifxundefined [1]{%
 \@ifx{#1\undefined}
}%
\providecommand \@ifnum [1]{%
 \ifnum #1\expandafter \@firstoftwo
 \else \expandafter \@secondoftwo
 \fi
}%
\providecommand \@ifx [1]{%
 \ifx #1\expandafter \@firstoftwo
 \else \expandafter \@secondoftwo
 \fi
}%
\providecommand \natexlab [1]{#1}%
\providecommand \enquote  [1]{``#1''}%
\providecommand \bibnamefont  [1]{#1}%
\providecommand \bibfnamefont [1]{#1}%
\providecommand \citenamefont [1]{#1}%
\providecommand \href@noop [0]{\@secondoftwo}%
\providecommand \href [0]{\begingroup \@sanitize@url \@href}%
\providecommand \@href[1]{\@@startlink{#1}\@@href}%
\providecommand \@@href[1]{\endgroup#1\@@endlink}%
\providecommand \@sanitize@url [0]{\catcode `\\12\catcode `\$12\catcode
  `\&12\catcode `\#12\catcode `\^12\catcode `\_12\catcode `\%12\relax}%
\providecommand \@@startlink[1]{}%
\providecommand \@@endlink[0]{}%
\providecommand \url  [0]{\begingroup\@sanitize@url \@url }%
\providecommand \@url [1]{\endgroup\@href {#1}{\urlprefix }}%
\providecommand \urlprefix  [0]{URL }%
\providecommand \Eprint [0]{\href }%
\providecommand \doibase [0]{https://doi.org/}%
\providecommand \selectlanguage [0]{\@gobble}%
\providecommand \bibinfo  [0]{\@secondoftwo}%
\providecommand \bibfield  [0]{\@secondoftwo}%
\providecommand \translation [1]{[#1]}%
\providecommand \BibitemOpen [0]{}%
\providecommand \bibitemStop [0]{}%
\providecommand \bibitemNoStop [0]{.\EOS\space}%
\providecommand \EOS [0]{\spacefactor3000\relax}%
\providecommand \BibitemShut  [1]{\csname bibitem#1\endcsname}%
\let\auto@bib@innerbib\@empty
\bibitem [{\citenamefont {Saakyan}(2013)}]{Saakyan:2013}%
  \BibitemOpen
  \bibfield  {author} {\bibinfo {author} {\bibfnamefont {R.}~\bibnamefont
  {Saakyan}},\ }\bibfield  {title} {\bibinfo {title} {{Two-Neutrino Double-Beta
  Decay}},\ }\href {https://doi.org/10.1146/annurev-nucl-102711-094904}
  {\bibfield  {journal} {\bibinfo  {journal} {Annu. Rev. Nucl. Part. Sci.}\
  }\textbf {\bibinfo {volume} {63}},\ \bibinfo {pages} {503} (\bibinfo {year}
  {2013})}\BibitemShut {NoStop}%
\bibitem [{\citenamefont {Barabash}(2020)}]{Barabash:2020}%
  \BibitemOpen
  \bibfield  {author} {\bibinfo {author} {\bibfnamefont {A.~S.}\ \bibnamefont
  {Barabash}},\ }\bibfield  {title} {\bibinfo {title} {{Precise Half-Life
  Values for Two-Neutrino Double-$\beta$ Decay: 2020 Review}},\ }\href
  {https://doi.org/10.3390/universe6100159} {\bibfield  {journal} {\bibinfo
  {journal} {Universe}\ }\textbf {\bibinfo {volume} {6}},\ \bibinfo {pages}
  {159} (\bibinfo {year} {2020})}\BibitemShut {NoStop}%
\bibitem [{\citenamefont {Barabash}(2017)}]{Barabash:2017}%
  \BibitemOpen
  \bibfield  {author} {\bibinfo {author} {\bibfnamefont {A.~S.}\ \bibnamefont
  {Barabash}},\ }\bibfield  {title} {\bibinfo {title} {{Double beta decay to
  the excited states: Review}},\ }\href {https://doi.org/10.1063/1.5007627}
  {\bibfield  {journal} {\bibinfo  {journal} {AIP Conf. Proc.}\ }\textbf
  {\bibinfo {volume} {1894}},\ \bibinfo {pages} {020002} (\bibinfo {year}
  {2017})},\ \Eprint {https://arxiv.org/abs/1709.06890} {arXiv:1709.06890
  [nucl-ex]} \BibitemShut {NoStop}%
\bibitem [{\citenamefont {Dell'Oro}\ \emph {et~al.}(2016)\citenamefont
  {Dell'Oro}, \citenamefont {Marcocci}, \citenamefont {Viel},\ and\
  \citenamefont {Vissani}}]{Goswami2015}%
  \BibitemOpen
  \bibfield  {author} {\bibinfo {author} {\bibfnamefont {S.}~\bibnamefont
  {Dell'Oro}}, \bibinfo {author} {\bibfnamefont {S.}~\bibnamefont {Marcocci}},
  \bibinfo {author} {\bibfnamefont {M.}~\bibnamefont {Viel}},\ and\ \bibinfo
  {author} {\bibfnamefont {F.}~\bibnamefont {Vissani}},\ }\bibfield  {title}
  {\bibinfo {title} {{Neutrinoless Double Beta Decay: 2015 Review}},\
  }\href@noop {} {\bibfield  {journal} {\bibinfo  {journal} {Advances in High
  Energy Physics}\ }\textbf {\bibinfo {volume} {2016}},\ \bibinfo {pages}
  {2162659} (\bibinfo {year} {2016})}\BibitemShut {NoStop}%
\bibitem [{\citenamefont {Schechter}\ and\ \citenamefont
  {Valle}(1982)}]{PhysRevD.25.2951}%
  \BibitemOpen
  \bibfield  {author} {\bibinfo {author} {\bibfnamefont {J.}~\bibnamefont
  {Schechter}}\ and\ \bibinfo {author} {\bibfnamefont {J.~W.~F.}\ \bibnamefont
  {Valle}},\ }\bibfield  {title} {\bibinfo {title} {{Neutrinoless
  double-$\ensuremath{\beta}$ decay in
  SU(2)\ifmmode\times\else\texttimes\fi{}U(1) theories}},\ }\href@noop {}
  {\bibfield  {journal} {\bibinfo  {journal} {Phys. Rev. D}\ }\textbf {\bibinfo
  {volume} {25}},\ \bibinfo {pages} {2951} (\bibinfo {year}
  {1982})}\BibitemShut {NoStop}%
\bibitem [{\citenamefont {Furry}(1939)}]{Furry:1939}%
  \BibitemOpen
  \bibfield  {author} {\bibinfo {author} {\bibfnamefont {W.~H.}\ \bibnamefont
  {Furry}},\ }\bibfield  {title} {\bibinfo {title} {{On Transition
  Probabilities in Double Beta-Disintegration}},\ }\href@noop {} {\bibfield
  {journal} {\bibinfo  {journal} {Phys. Rev.}\ }\textbf {\bibinfo {volume}
  {56}},\ \bibinfo {pages} {1184} (\bibinfo {year} {1939})}\BibitemShut
  {NoStop}%
\bibitem [{\citenamefont {Bilenky}\ and\ \citenamefont
  {Giunti}(2015)}]{Bilenky2015}%
  \BibitemOpen
  \bibfield  {author} {\bibinfo {author} {\bibfnamefont {S.~M.}\ \bibnamefont
  {Bilenky}}\ and\ \bibinfo {author} {\bibfnamefont {C.}~\bibnamefont
  {Giunti}},\ }\bibfield  {title} {\bibinfo {title} {{Neutrinoless double-beta
  decay: A probe of physics beyond the Standard Model}},\ }\href@noop {}
  {\bibfield  {journal} {\bibinfo  {journal} {International Journal of Modern
  Physics A}\ }\textbf {\bibinfo {volume} {30}},\ \bibinfo {pages} {1530001}
  (\bibinfo {year} {2015})}\BibitemShut {NoStop}%
\bibitem [{\citenamefont {Dolinski}\ \emph {et~al.}(2019)\citenamefont
  {Dolinski}, \citenamefont {Poon},\ and\ \citenamefont
  {Rodejohann}}]{Dolinski:2019a}%
  \BibitemOpen
  \bibfield  {author} {\bibinfo {author} {\bibfnamefont {M.~J.}\ \bibnamefont
  {Dolinski}}, \bibinfo {author} {\bibfnamefont {A.~W.}\ \bibnamefont {Poon}},\
  and\ \bibinfo {author} {\bibfnamefont {W.}~\bibnamefont {Rodejohann}},\
  }\bibfield  {title} {\bibinfo {title} {{Neutrinoless Double-Beta Decay:
  Status and Prospects}},\ }\href@noop {} {\bibfield  {journal} {\bibinfo
  {journal} {Annu. Rev. Nucl. Part. Sci.}\ }\textbf {\bibinfo {volume} {69}},\
  \bibinfo {pages} {219} (\bibinfo {year} {2019})}\BibitemShut {NoStop}%
\bibitem [{\citenamefont {Fukugita}\ and\ \citenamefont
  {Yanagida}(1986)}]{FUKUGITA198645}%
  \BibitemOpen
  \bibfield  {author} {\bibinfo {author} {\bibfnamefont {M.}~\bibnamefont
  {Fukugita}}\ and\ \bibinfo {author} {\bibfnamefont {T.}~\bibnamefont
  {Yanagida}},\ }\bibfield  {title} {\bibinfo {title} {{Barygenesis without
  grand unification}},\ }\href@noop {} {\bibfield  {journal} {\bibinfo
  {journal} {Phys.~Lett.~B}\ }\textbf {\bibinfo {volume} {174}},\ \bibinfo
  {pages} {45 } (\bibinfo {year} {1986})}\BibitemShut {NoStop}%
\bibitem [{\citenamefont {Davidson}\ \emph {et~al.}(2008)\citenamefont
  {Davidson}, \citenamefont {Nardi},\ and\ \citenamefont
  {Nir}}]{DAVIDSON2008105}%
  \BibitemOpen
  \bibfield  {author} {\bibinfo {author} {\bibfnamefont {S.}~\bibnamefont
  {Davidson}}, \bibinfo {author} {\bibfnamefont {E.}~\bibnamefont {Nardi}},\
  and\ \bibinfo {author} {\bibfnamefont {Y.}~\bibnamefont {Nir}},\ }\bibfield
  {title} {\bibinfo {title} {{Leptogenesis}},\ }\href@noop {} {\bibfield
  {journal} {\bibinfo  {journal} {Physics Reports}\ }\textbf {\bibinfo {volume}
  {466}},\ \bibinfo {pages} {105} (\bibinfo {year} {2008})}\BibitemShut
  {NoStop}%
\bibitem [{\citenamefont {Deppisch}\ \emph {et~al.}(2018)\citenamefont
  {Deppisch} \emph {et~al.}}]{Deppisch:2017}%
  \BibitemOpen
  \bibfield  {author} {\bibinfo {author} {\bibfnamefont {F.~F.}\ \bibnamefont
  {Deppisch}} \emph {et~al.},\ }\bibfield  {title} {\bibinfo {title}
  {{Neutrinoless Double Beta Decay and the Baryon Asymmetry of the Universe}},\
  }\href {https://doi.org/10.1103/PhysRevD.98.055029} {\bibfield  {journal}
  {\bibinfo  {journal} {Phys. Rev. D}\ }\textbf {\bibinfo {volume} {98}},\
  \bibinfo {pages} {055029} (\bibinfo {year} {2018})},\ \Eprint
  {https://arxiv.org/abs/1711.10432} {arXiv:1711.10432 [hep-ph]} \BibitemShut
  {NoStop}%
\bibitem [{\citenamefont {Adams}\ \emph {et~al.}(2022)\citenamefont {Adams}
  \emph {et~al.}}]{CUORE1ton}%
  \BibitemOpen
  \bibfield  {author} {\bibinfo {author} {\bibfnamefont {D.~Q.}\ \bibnamefont
  {Adams}} \emph {et~al.} (\bibinfo {collaboration} {CUORE}),\ }\bibfield
  {title} {\bibinfo {title} {{Search for Majorana neutrinos exploiting
  millikelvin cryogenics with CUORE}},\ }\href
  {https://doi.org/10.1038/s41586-022-04497-4} {\bibfield  {journal} {\bibinfo
  {journal} {Nature}\ }\textbf {\bibinfo {volume} {604}},\ \bibinfo {pages}
  {53} (\bibinfo {year} {2022})},\ \Eprint {https://arxiv.org/abs/2104.06906}
  {arXiv:2104.06906 [nucl-ex]} \BibitemShut {NoStop}%
\bibitem [{\citenamefont {Armstrong}\ \emph {et~al.}(2019)\citenamefont
  {Armstrong} \emph {et~al.}}]{CUPIDInterestGroup:2019inu}%
  \BibitemOpen
  \bibfield  {author} {\bibinfo {author} {\bibfnamefont {W.~R.}\ \bibnamefont
  {Armstrong}} \emph {et~al.} (\bibinfo {collaboration} {CUPID Interest
  Group}),\ }\bibfield  {title} {\bibinfo {title} {{CUPID pre-CDR}},\
  }\href@noop {} {\bibfield  {journal} {\bibinfo  {journal}
  {{arXiv:1907.09376}}\ } (\bibinfo {year} {2019})}\BibitemShut {NoStop}%
\bibitem [{\citenamefont {Armengaud}\ \emph
  {et~al.}(2020{\natexlab{a}})\citenamefont {Armengaud} \emph
  {et~al.}}]{Armengaud:2020}%
  \BibitemOpen
  \bibfield  {author} {\bibinfo {author} {\bibfnamefont {E.}~\bibnamefont
  {Armengaud}} \emph {et~al.} (\bibinfo {collaboration} {{CUPID-Mo}}),\
  }\bibfield  {title} {\bibinfo {title} {{The CUPID-Mo experiment for
  neutrinoless double-beta decay: performance and prospects}},\ }\href@noop {}
  {\bibfield  {journal} {\bibinfo  {journal} {Eur. Phys. J. C}\ }\textbf
  {\bibinfo {volume} {80}},\ \bibinfo {pages} {44} (\bibinfo {year}
  {2020}{\natexlab{a}})}\BibitemShut {NoStop}%
\bibitem [{\citenamefont {Deppisch}\ \emph {et~al.}(2012)\citenamefont
  {Deppisch}, \citenamefont {Hirsch},\ and\ \citenamefont
  {P{\"a}s}}]{Deppisch:2012}%
  \BibitemOpen
  \bibfield  {author} {\bibinfo {author} {\bibfnamefont {F.~F.}\ \bibnamefont
  {Deppisch}}, \bibinfo {author} {\bibfnamefont {M.}~\bibnamefont {Hirsch}},\
  and\ \bibinfo {author} {\bibfnamefont {H.}~\bibnamefont {P{\"a}s}},\
  }\bibfield  {title} {\bibinfo {title} {{Neutrinoless double-beta decay and
  physics beyond the standard model}},\ }\href@noop {} {\bibfield  {journal}
  {\bibinfo  {journal} {J. Phys G: Nucl. Part. Phys.}\ }\textbf {\bibinfo
  {volume} {39}},\ \bibinfo {pages} {124007} (\bibinfo {year}
  {2012})}\BibitemShut {NoStop}%
\bibitem [{\citenamefont {Rodejohann}(2012)}]{Rodejohann:2012}%
  \BibitemOpen
  \bibfield  {author} {\bibinfo {author} {\bibfnamefont {W.}~\bibnamefont
  {Rodejohann}},\ }\bibfield  {title} {\bibinfo {title} {{Neutrinoless
  double-beta decay and neutrino physics}},\ }\href@noop {} {\bibfield
  {journal} {\bibinfo  {journal} {J. Phys G: Nucl. Part. Phys.}\ }\textbf
  {\bibinfo {volume} {39}},\ \bibinfo {pages} {124008} (\bibinfo {year}
  {2012})}\BibitemShut {NoStop}%
\bibitem [{\citenamefont {Pr\'ezeau}\ \emph {et~al.}(2003)\citenamefont
  {Pr\'ezeau}, \citenamefont {Ramsey-Musolf},\ and\ \citenamefont
  {Vogel}}]{PhysRevD.68.034016}%
  \BibitemOpen
  \bibfield  {author} {\bibinfo {author} {\bibfnamefont {G.}~\bibnamefont
  {Pr\'ezeau}}, \bibinfo {author} {\bibfnamefont {M.}~\bibnamefont
  {Ramsey-Musolf}},\ and\ \bibinfo {author} {\bibfnamefont {P.}~\bibnamefont
  {Vogel}},\ }\bibfield  {title} {\bibinfo {title} {{Neutrinoless double
  $\ensuremath{\beta}$ decay and effective field theory}},\ }\href@noop {}
  {\bibfield  {journal} {\bibinfo  {journal} {Phys. Rev. D}\ }\textbf {\bibinfo
  {volume} {68}},\ \bibinfo {pages} {034016} (\bibinfo {year}
  {2003})}\BibitemShut {NoStop}%
\bibitem [{\citenamefont {Atre}\ \emph {et~al.}(2009)\citenamefont {Atre},
  \citenamefont {Han}, \citenamefont {Pascoli},\ and\ \citenamefont
  {Zhang}}]{Atre2009}%
  \BibitemOpen
  \bibfield  {author} {\bibinfo {author} {\bibfnamefont {A.}~\bibnamefont
  {Atre}}, \bibinfo {author} {\bibfnamefont {T.}~\bibnamefont {Han}}, \bibinfo
  {author} {\bibfnamefont {S.}~\bibnamefont {Pascoli}},\ and\ \bibinfo {author}
  {\bibfnamefont {B.}~\bibnamefont {Zhang}},\ }\bibfield  {title} {\bibinfo
  {title} {{The search for heavy Majorana neutrinos}},\ }\href@noop {}
  {\bibfield  {journal} {\bibinfo  {journal} {Journal of High Energy Physics}\
  }\textbf {\bibinfo {volume} {2009}},\ \bibinfo {pages} {030} (\bibinfo {year}
  {2009})}\BibitemShut {NoStop}%
\bibitem [{\citenamefont {Blennow}\ \emph {et~al.}(2010)\citenamefont
  {Blennow}, \citenamefont {Fernandez-Martinez}, \citenamefont {Lopez-Pavon},\
  and\ \citenamefont {Men{\'e}ndez}}]{Blennow2010}%
  \BibitemOpen
  \bibfield  {author} {\bibinfo {author} {\bibfnamefont {M.}~\bibnamefont
  {Blennow}}, \bibinfo {author} {\bibfnamefont {E.}~\bibnamefont
  {Fernandez-Martinez}}, \bibinfo {author} {\bibfnamefont {J.}~\bibnamefont
  {Lopez-Pavon}},\ and\ \bibinfo {author} {\bibfnamefont {J.}~\bibnamefont
  {Men{\'e}ndez}},\ }\bibfield  {title} {\bibinfo {title} {{Neutrinoless double
  beta decay in seesaw models}},\ }\href@noop {} {\bibfield  {journal}
  {\bibinfo  {journal} {Journal of High Energy Physics}\ }\textbf {\bibinfo
  {volume} {2010}},\ \bibinfo {pages} {96} (\bibinfo {year}
  {2010})}\BibitemShut {NoStop}%
\bibitem [{\citenamefont {Mitra}\ \emph {et~al.}(2012)\citenamefont {Mitra},
  \citenamefont {Senjanovi{\'c}},\ and\ \citenamefont {Vissani}}]{MITRA201226}%
  \BibitemOpen
  \bibfield  {author} {\bibinfo {author} {\bibfnamefont {M.}~\bibnamefont
  {Mitra}}, \bibinfo {author} {\bibfnamefont {G.}~\bibnamefont
  {Senjanovi{\'c}}},\ and\ \bibinfo {author} {\bibfnamefont {F.}~\bibnamefont
  {Vissani}},\ }\bibfield  {title} {\bibinfo {title} {{Neutrinoless double beta
  decay and heavy sterile neutrinos}},\ }\href@noop {} {\bibfield  {journal}
  {\bibinfo  {journal} {Nuclear Physics B}\ }\textbf {\bibinfo {volume}
  {856}},\ \bibinfo {pages} {26} (\bibinfo {year} {2012})}\BibitemShut
  {NoStop}%
\bibitem [{\citenamefont {Cirigliano}\ \emph {et~al.}(2018)\citenamefont
  {Cirigliano}, \citenamefont {Dekens}, \citenamefont {de~Vries}, \citenamefont
  {Graesser},\ and\ \citenamefont {Mereghetti}}]{Cirigliano2018}%
  \BibitemOpen
  \bibfield  {author} {\bibinfo {author} {\bibfnamefont {V.}~\bibnamefont
  {Cirigliano}}, \bibinfo {author} {\bibfnamefont {W.}~\bibnamefont {Dekens}},
  \bibinfo {author} {\bibfnamefont {J.}~\bibnamefont {de~Vries}}, \bibinfo
  {author} {\bibfnamefont {M.~L.}\ \bibnamefont {Graesser}},\ and\ \bibinfo
  {author} {\bibfnamefont {E.}~\bibnamefont {Mereghetti}},\ }\bibfield  {title}
  {\bibinfo {title} {{A neutrinoless double beta decay master formula from
  effective field theory}},\ }\href@noop {} {\bibfield  {journal} {\bibinfo
  {journal} {Journal of High Energy Physics}\ }\textbf {\bibinfo {volume}
  {2018}},\ \bibinfo {pages} {97} (\bibinfo {year} {2018})}\BibitemShut
  {NoStop}%
\bibitem [{\citenamefont {Benato}(2015)}]{Benato:2015}%
  \BibitemOpen
  \bibfield  {author} {\bibinfo {author} {\bibfnamefont {G.}~\bibnamefont
  {Benato}},\ }\bibfield  {title} {\bibinfo {title} {{Effective Majorana Mass
  and Neutrinoless Double Beta Decay}},\ }\href
  {https://doi.org/10.1140/epjc/s10052-015-3802-1} {\bibfield  {journal}
  {\bibinfo  {journal} {Eur. Phys. J. C}\ }\textbf {\bibinfo {volume} {75}},\
  \bibinfo {pages} {563} (\bibinfo {year} {2015})},\ \Eprint
  {https://arxiv.org/abs/1510.01089} {arXiv:1510.01089 [hep-ph]} \BibitemShut
  {NoStop}%
\bibitem [{\citenamefont {Engel}\ and\ \citenamefont
  {Men\'endez}(2017)}]{Engel}%
  \BibitemOpen
  \bibfield  {author} {\bibinfo {author} {\bibfnamefont {J.}~\bibnamefont
  {Engel}}\ and\ \bibinfo {author} {\bibfnamefont {J.}~\bibnamefont
  {Men\'endez}},\ }\bibfield  {title} {\bibinfo {title} {{Status and Future of
  Nuclear Matrix Elements for Neutrinoless Double-Beta Decay: A Review}},\
  }\href {https://doi.org/10.1088/1361-6633/aa5bc5} {\bibfield  {journal}
  {\bibinfo  {journal} {Rept. Prog. Phys.}\ }\textbf {\bibinfo {volume} {80}},\
  \bibinfo {pages} {046301} (\bibinfo {year} {2017})},\ \Eprint
  {https://arxiv.org/abs/1610.06548} {arXiv:1610.06548 [nucl-th]} \BibitemShut
  {NoStop}%
\bibitem [{\citenamefont {Rath}\ \emph {et~al.}(2013)\citenamefont {Rath},
  \citenamefont {Chandra}, \citenamefont {Chaturvedi}, \citenamefont {Lohani},
  \citenamefont {Raina},\ and\ \citenamefont {Hirsch}}]{Rath:2013}%
  \BibitemOpen
  \bibfield  {author} {\bibinfo {author} {\bibfnamefont {P.~K.}\ \bibnamefont
  {Rath}}, \bibinfo {author} {\bibfnamefont {R.}~\bibnamefont {Chandra}},
  \bibinfo {author} {\bibfnamefont {K.}~\bibnamefont {Chaturvedi}}, \bibinfo
  {author} {\bibfnamefont {P.}~\bibnamefont {Lohani}}, \bibinfo {author}
  {\bibfnamefont {P.~K.}\ \bibnamefont {Raina}},\ and\ \bibinfo {author}
  {\bibfnamefont {J.~G.}\ \bibnamefont {Hirsch}},\ }\bibfield  {title}
  {\bibinfo {title} {{Neutrinoless \ensuremath{\beta}\ensuremath{\beta} decay
  transition matrix elements within mechanisms involving light Majorana
  neutrinos, classical Majorons, and sterile neutrinos}},\ }\href
  {https://doi.org/10.1103/PhysRevC.88.064322} {\bibfield  {journal} {\bibinfo
  {journal} {Phys. Rev. C}\ }\textbf {\bibinfo {volume} {88}},\ \bibinfo
  {pages} {064322} (\bibinfo {year} {2013})},\ \Eprint
  {https://arxiv.org/abs/1308.0460} {arXiv:1308.0460 [nucl-th]} \BibitemShut
  {NoStop}%
\bibitem [{\citenamefont {\v{S}imkovic}\ \emph {et~al.}(2013)\citenamefont
  {\v{S}imkovic}, \citenamefont {Rodin}, \citenamefont {Faessler},\ and\
  \citenamefont {Vogel}}]{Simkovic:2013}%
  \BibitemOpen
  \bibfield  {author} {\bibinfo {author} {\bibfnamefont {F.}~\bibnamefont
  {\v{S}imkovic}}, \bibinfo {author} {\bibfnamefont {V.}~\bibnamefont {Rodin}},
  \bibinfo {author} {\bibfnamefont {A.}~\bibnamefont {Faessler}},\ and\
  \bibinfo {author} {\bibfnamefont {P.}~\bibnamefont {Vogel}},\ }\bibfield
  {title} {\bibinfo {title}
  {{0\ensuremath{\nu}\ensuremath{\beta}\ensuremath{\beta} and
  2\ensuremath{\nu}\ensuremath{\beta}\ensuremath{\beta} nuclear matrix
  elements, quasiparticle random-phase approximation, and isospin symmetry
  restoration}},\ }\href {https://doi.org/10.1103/PhysRevC.87.045501}
  {\bibfield  {journal} {\bibinfo  {journal} {Phys. Rev. C}\ }\textbf {\bibinfo
  {volume} {87}},\ \bibinfo {pages} {045501} (\bibinfo {year} {2013})},\
  \Eprint {https://arxiv.org/abs/1302.1509} {arXiv:1302.1509 [nucl-th]}
  \BibitemShut {NoStop}%
\bibitem [{\citenamefont {Vaquero}\ \emph {et~al.}(2013)\citenamefont
  {Vaquero}, \citenamefont {Rodr{\'{i}}guez},\ and\ \citenamefont
  {Egido}}]{Vaquero:2013}%
  \BibitemOpen
  \bibfield  {author} {\bibinfo {author} {\bibfnamefont {N.~L.}\ \bibnamefont
  {Vaquero}}, \bibinfo {author} {\bibfnamefont {T.~R.}\ \bibnamefont
  {Rodr{\'{i}}guez}},\ and\ \bibinfo {author} {\bibfnamefont {J.~L.}\
  \bibnamefont {Egido}},\ }\bibfield  {title} {\bibinfo {title} {{Shape and
  pairing fluctuation effects on neutrinoless double beta decay nuclear matrix
  elements}},\ }\href@noop {} {\bibfield  {journal} {\bibinfo  {journal} {Phys.
  Rev. Lett.}\ }\textbf {\bibinfo {volume} {111}},\ \bibinfo {pages} {142501}
  (\bibinfo {year} {2013})}\BibitemShut {NoStop}%
\bibitem [{\citenamefont {Barea}\ \emph
  {et~al.}(2015{\natexlab{a}})\citenamefont {Barea}, \citenamefont {Kotila},\
  and\ \citenamefont {Iachello}}]{Barea:2015}%
  \BibitemOpen
  \bibfield  {author} {\bibinfo {author} {\bibfnamefont {J.}~\bibnamefont
  {Barea}}, \bibinfo {author} {\bibfnamefont {J.}~\bibnamefont {Kotila}},\ and\
  \bibinfo {author} {\bibfnamefont {F.}~\bibnamefont {Iachello}},\ }\bibfield
  {title} {\bibinfo {title} {{0$\nu\beta\beta$ and 2$\nu\beta\beta$ Nuclear
  Matrix Elements in the Interacting Boson Model With Isospin Restoration}},\
  }\href@noop {} {\bibfield  {journal} {\bibinfo  {journal} {Phys. Rev. C}\
  }\textbf {\bibinfo {volume} {91}},\ \bibinfo {pages} {034304} (\bibinfo
  {year} {2015}{\natexlab{a}})}\BibitemShut {NoStop}%
\bibitem [{\citenamefont {Agostini}\ \emph {et~al.}(2020)\citenamefont
  {Agostini} \emph {et~al.}}]{Agostini:2020}%
  \BibitemOpen
  \bibfield  {author} {\bibinfo {author} {\bibfnamefont {M.}~\bibnamefont
  {Agostini}} \emph {et~al.} (\bibinfo {collaboration} {GERDA}),\ }\bibfield
  {title} {\bibinfo {title} {{Final Results of GERDA on the Search for
  Neutrinoless Double-$\ensuremath{\beta}$ Decay}},\ }\href
  {https://doi.org/10.1103/PhysRevLett.125.252502} {\bibfield  {journal}
  {\bibinfo  {journal} {Phys. Rev. Lett.}\ }\textbf {\bibinfo {volume} {125}},\
  \bibinfo {pages} {252502} (\bibinfo {year} {2020})}\BibitemShut {NoStop}%
\bibitem [{\citenamefont {Gando}\ \emph {et~al.}(2016)\citenamefont {Gando}
  \emph {et~al.}}]{Gando:2016}%
  \BibitemOpen
  \bibfield  {author} {\bibinfo {author} {\bibfnamefont {A.}~\bibnamefont
  {Gando}} \emph {et~al.} (\bibinfo {collaboration} {KamLAND-Zen}),\ }\bibfield
   {title} {\bibinfo {title} {{Search for Majorana Neutrinos Near the Inverted
  Mass Hierarchy Region with KamLAND-Zen}},\ }\href@noop {} {\bibfield
  {journal} {\bibinfo  {journal} {Phys. Rev. Lett.}\ }\textbf {\bibinfo
  {volume} {117}},\ \bibinfo {pages} {082503} (\bibinfo {year}
  {2016})}\BibitemShut {NoStop}%
\bibitem [{\citenamefont {Azzolini}\ \emph
  {et~al.}(2018{\natexlab{a}})\citenamefont {Azzolini} \emph
  {et~al.}}]{Azzolini:2018dyb}%
  \BibitemOpen
  \bibfield  {author} {\bibinfo {author} {\bibfnamefont {O.}~\bibnamefont
  {Azzolini}} \emph {et~al.} (\bibinfo {collaboration} {CUPID-0}),\ }\bibfield
  {title} {\bibinfo {title} {{First Result on the Neutrinoless Double-$\beta$
  Decay of $^{82}$Se with CUPID-0}},\ }\href@noop {} {\bibfield  {journal}
  {\bibinfo  {journal} {Phys. Rev. Lett.}\ }\textbf {\bibinfo {volume} {120}},\
  \bibinfo {pages} {232502} (\bibinfo {year} {2018}{\natexlab{a}})}\BibitemShut
  {NoStop}%
\bibitem [{\citenamefont {Augier}\ \emph {et~al.}(2022)\citenamefont {Augier}
  \emph {et~al.}}]{CUPIDMo:2022}%
  \BibitemOpen
  \bibfield  {author} {\bibinfo {author} {\bibfnamefont {C.}~\bibnamefont
  {Augier}} \emph {et~al.},\ }\bibfield  {title} {\bibinfo {title} {{Final
  results on the $0\nu\beta\beta$ decay half-life limit of $^{100}$Mo from the
  CUPID-Mo experiment}},\ }\href@noop {} {\  (\bibinfo {year} {2022})},\
  \Eprint {https://arxiv.org/abs/2202.08716} {arXiv:2202.08716 [nucl-ex]}
  \BibitemShut {NoStop}%
\bibitem [{\citenamefont {Dolgov}\ and\ \citenamefont
  {Smirnov}(2005)}]{DOLGOV20051}%
  \BibitemOpen
  \bibfield  {author} {\bibinfo {author} {\bibfnamefont {A.}~\bibnamefont
  {Dolgov}}\ and\ \bibinfo {author} {\bibfnamefont {A.}~\bibnamefont
  {Smirnov}},\ }\bibfield  {title} {\bibinfo {title} {{Possible violation of
  the spin-statistics relation for neutrinos: Cosmological and astrophysical
  consequences}},\ }\href@noop {} {\bibfield  {journal} {\bibinfo  {journal}
  {Physics Letters B}\ }\textbf {\bibinfo {volume} {621}},\ \bibinfo {pages}
  {1} (\bibinfo {year} {2005})}\BibitemShut {NoStop}%
\bibitem [{\citenamefont {Barabash}\ \emph {et~al.}(2007)\citenamefont
  {Barabash}, \citenamefont {Dolgov}, \citenamefont {Dvornicky}, \citenamefont
  {Simkovic},\ and\ \citenamefont {Smirnov}}]{Barabash:2007}%
  \BibitemOpen
  \bibfield  {author} {\bibinfo {author} {\bibfnamefont {A.~S.}\ \bibnamefont
  {Barabash}}, \bibinfo {author} {\bibfnamefont {A.~D.}\ \bibnamefont
  {Dolgov}}, \bibinfo {author} {\bibfnamefont {R.}~\bibnamefont {Dvornicky}},
  \bibinfo {author} {\bibfnamefont {F.}~\bibnamefont {Simkovic}},\ and\
  \bibinfo {author} {\bibfnamefont {A.~Y.}\ \bibnamefont {Smirnov}},\
  }\bibfield  {title} {\bibinfo {title} {{Statistics of neutrinos and the
  double beta decay}},\ }\href
  {https://doi.org/10.1016/j.nuclphysb.2007.05.033} {\bibfield  {journal}
  {\bibinfo  {journal} {Nucl. Phys. B}\ }\textbf {\bibinfo {volume} {783}},\
  \bibinfo {pages} {90} (\bibinfo {year} {2007})},\ \Eprint
  {https://arxiv.org/abs/0704.2944} {arXiv:0704.2944 [hep-ph]} \BibitemShut
  {NoStop}%
\bibitem [{\citenamefont {Hubbell}\ and\ \citenamefont
  {Seltzer}(2004)}]{NIST_Xray}%
  \BibitemOpen
  \bibfield  {author} {\bibinfo {author} {\bibfnamefont {J.~H.}\ \bibnamefont
  {Hubbell}}\ and\ \bibinfo {author} {\bibfnamefont {S.~M.}\ \bibnamefont
  {Seltzer}},\ }\bibfield  {title} {\bibinfo {title} {{X-Ray mass attenuation
  coefficients}},\ }\href@noop {} {\bibfield  {journal} {\bibinfo  {journal}
  {NIST Standard Reference Database 126}\ } (\bibinfo {year}
  {2004})}\BibitemShut {NoStop}%
\bibitem [{\citenamefont {Berger}\ \emph {et~al.}(2017)\citenamefont {Berger}
  \emph {et~al.}}]{NIST_electron}%
  \BibitemOpen
  \bibfield  {author} {\bibinfo {author} {\bibfnamefont {M.}~\bibnamefont
  {Berger}} \emph {et~al.},\ }\bibfield  {title} {\bibinfo {title}
  {{Stopping-Power \& Range Tables for Electrons, Protons, and Helium Ions}},\
  }\bibfield  {journal} {\bibinfo  {journal} {NIST Standard Reference Database
  124}\ }\href {https://doi.org/{https://dx.doi.org/10.18434/T4NC7P}}
  {{https://dx.doi.org/10.18434/T4NC7P}} (\bibinfo {year} {2017})\BibitemShut
  {NoStop}%
\bibitem [{\citenamefont {Barabash}(1990)}]{Barabash:1990}%
  \BibitemOpen
  \bibfield  {author} {\bibinfo {author} {\bibfnamefont {A.~S.}\ \bibnamefont
  {Barabash}},\ }\bibfield  {title} {\bibinfo {title} {{A Possibility for
  experimentally observing two neutrino double beta decay}},\ }\href@noop {}
  {\bibfield  {journal} {\bibinfo  {journal} {JETP Lett.}\ }\textbf {\bibinfo
  {volume} {51}},\ \bibinfo {pages} {207} (\bibinfo {year} {1990})}\BibitemShut
  {NoStop}%
\bibitem [{\citenamefont {Barabash}\ \emph {et~al.}(1995)\citenamefont
  {Barabash} \emph {et~al.}}]{Barabash:1995}%
  \BibitemOpen
  \bibfield  {author} {\bibinfo {author} {\bibfnamefont {A.~S.}\ \bibnamefont
  {Barabash}} \emph {et~al.},\ }\bibfield  {title} {\bibinfo {title} {{Two
  neutrino double beta decay of $^{100}$Mo to the first excited $0^+$ state in
  $^{100}$Ru}},\ }\href {https://doi.org/10.1016/0370-2693(94)01657-X}
  {\bibfield  {journal} {\bibinfo  {journal} {Phys. Lett. B}\ }\textbf
  {\bibinfo {volume} {345}},\ \bibinfo {pages} {408} (\bibinfo {year}
  {1995})}\BibitemShut {NoStop}%
\bibitem [{\citenamefont {Barabash}\ \emph {et~al.}(1999)\citenamefont
  {Barabash}, \citenamefont {Umatov}, \citenamefont {Gurriaran}, \citenamefont
  {Hubert},\ and\ \citenamefont {Hubert}}]{Barabash:1999}%
  \BibitemOpen
  \bibfield  {author} {\bibinfo {author} {\bibfnamefont {A.~S.}\ \bibnamefont
  {Barabash}}, \bibinfo {author} {\bibfnamefont {V.~I.}\ \bibnamefont
  {Umatov}}, \bibinfo {author} {\bibfnamefont {R.}~\bibnamefont {Gurriaran}},
  \bibinfo {author} {\bibfnamefont {F.}~\bibnamefont {Hubert}},\ and\ \bibinfo
  {author} {\bibfnamefont {P.}~\bibnamefont {Hubert}},\ }\bibfield  {title}
  {\bibinfo {title} {{2$\nu\beta\beta$ decay of $^{100}$Mo to the first 0$^+$
  excited state in $^{100}$Ru}},\ }\href@noop {} {\bibfield  {journal}
  {\bibinfo  {journal} {Phys. Atom. Nucl.}\ }\textbf {\bibinfo {volume} {62}},\
  \bibinfo {pages} {2039} (\bibinfo {year} {1999})}\BibitemShut {NoStop}%
\bibitem [{\citenamefont {Arnold}\ \emph {et~al.}(2007)\citenamefont {Arnold}
  \emph {et~al.}}]{NEMO_ES}%
  \BibitemOpen
  \bibfield  {author} {\bibinfo {author} {\bibfnamefont {R.}~\bibnamefont
  {Arnold}} \emph {et~al.} (\bibinfo {collaboration} {NEMO}),\ }\bibfield
  {title} {\bibinfo {title} {{Measurement of double beta decay of $^{100}$Mo to
  excited states in the NEMO 3 experiment}},\ }\href
  {https://doi.org/10.1016/j.nuclphysa.2006.09.021} {\bibfield  {journal}
  {\bibinfo  {journal} {Nucl. Phys. A}\ }\textbf {\bibinfo {volume} {781}},\
  \bibinfo {pages} {209} (\bibinfo {year} {2007})},\ \Eprint
  {https://arxiv.org/abs/hep-ex/0609058} {arXiv:hep-ex/0609058} \BibitemShut
  {NoStop}%
\bibitem [{\citenamefont {Kidd}\ \emph {et~al.}(2009)\citenamefont {Kidd},
  \citenamefont {Esterline}, \citenamefont {Tornow}, \citenamefont {Barabash},\
  and\ \citenamefont {Umatov}}]{Kidd}%
  \BibitemOpen
  \bibfield  {author} {\bibinfo {author} {\bibfnamefont {M.~F.}\ \bibnamefont
  {Kidd}}, \bibinfo {author} {\bibfnamefont {J.~H.}\ \bibnamefont {Esterline}},
  \bibinfo {author} {\bibfnamefont {W.}~\bibnamefont {Tornow}}, \bibinfo
  {author} {\bibfnamefont {A.~S.}\ \bibnamefont {Barabash}},\ and\ \bibinfo
  {author} {\bibfnamefont {V.~I.}\ \bibnamefont {Umatov}},\ }\bibfield  {title}
  {\bibinfo {title} {{New Results for Double-Beta Decay of $^{100}$Mo to
  Excited Final States of $^{100}$Ru Using the TUNL-ITEP Apparatus}},\ }\href
  {https://doi.org/10.1016/j.nuclphysa.2009.01.082} {\bibfield  {journal}
  {\bibinfo  {journal} {Nucl. Phys. A}\ }\textbf {\bibinfo {volume} {821}},\
  \bibinfo {pages} {251} (\bibinfo {year} {2009})},\ \Eprint
  {https://arxiv.org/abs/0902.4418} {arXiv:0902.4418 [nucl-ex]} \BibitemShut
  {NoStop}%
\bibitem [{\citenamefont {Belli}\ \emph {et~al.}(2010)\citenamefont {Belli}
  \emph {et~al.}}]{armonia}%
  \BibitemOpen
  \bibfield  {author} {\bibinfo {author} {\bibfnamefont {P.}~\bibnamefont
  {Belli}} \emph {et~al.},\ }\bibfield  {title} {\bibinfo {title} {{New
  observation of $2\beta2\nu$ decay of $^{100}$Mo to the $0_1^+$ level of
  $^{100}$Ru in the ARMONIA experiment}},\ }\href
  {https://doi.org/10.1016/j.nuclphysa.2010.06.010} {\bibfield  {journal}
  {\bibinfo  {journal} {Nuclear Physics A}\ }\textbf {\bibinfo {volume}
  {846}},\ \bibinfo {pages} {143} (\bibinfo {year} {2010})}\BibitemShut
  {NoStop}%
\bibitem [{\citenamefont {Arnold}\ \emph {et~al.}(2014)\citenamefont {Arnold}
  \emph {et~al.}}]{NEMO3-ES.2014}%
  \BibitemOpen
  \bibfield  {author} {\bibinfo {author} {\bibfnamefont {R.}~\bibnamefont
  {Arnold}} \emph {et~al.},\ }\bibfield  {title} {\bibinfo {title}
  {{Investigation of double beta decay of $^{100}$Mo to excited states of
  $^{100}$Ru}},\ }\href
  {https://doi.org/https://doi.org/10.1016/j.nuclphysa.2014.01.008} {\bibfield
  {journal} {\bibinfo  {journal} {Nuclear Physics A}\ }\textbf {\bibinfo
  {volume} {925}},\ \bibinfo {pages} {25} (\bibinfo {year} {2014})}\BibitemShut
  {NoStop}%
\bibitem [{\citenamefont {Belli}\ \emph {et~al.}(2020)\citenamefont {Belli}
  \emph {et~al.}}]{universe_2020}%
  \BibitemOpen
  \bibfield  {author} {\bibinfo {author} {\bibfnamefont {P.}~\bibnamefont
  {Belli}} \emph {et~al.},\ }\bibfield  {title} {\bibinfo {title} {{Double Beta
  Decay to Excited States of Daughter Nuclei}},\ }\href
  {https://www.mdpi.com/2218-1997/6/12/239} {\bibfield  {journal} {\bibinfo
  {journal} {Universe}\ }\textbf {\bibinfo {volume} {6}},\ \bibinfo {pages}
  {239} (\bibinfo {year} {2020})}\BibitemShut {NoStop}%
\bibitem [{\citenamefont {Armengaud}\ \emph
  {et~al.}(2017{\natexlab{a}})\citenamefont {Armengaud} \emph
  {et~al.}}]{Armengaud:2017b}%
  \BibitemOpen
  \bibfield  {author} {\bibinfo {author} {\bibfnamefont {E.}~\bibnamefont
  {Armengaud}} \emph {et~al.} (\bibinfo {collaboration} {EDELWEISS}),\
  }\bibfield  {title} {\bibinfo {title} {{Performance of the EDELWEISS-III
  experiment for direct dark matter searches}},\ }\href
  {https://doi.org/10.1088/1748-0221/12/08/P08010} {\bibfield  {journal}
  {\bibinfo  {journal} {JINST}\ }\textbf {\bibinfo {volume} {12}}\bibfield
  {number} {\bibinfo  {number} { (08)},\ \bibinfo {pages} {P08010 (2017)}},\
  }\Eprint {https://arxiv.org/abs/1706.01070} {arXiv:1706.01070
  [physics.ins-det]} \BibitemShut {NoStop}%
\bibitem [{\citenamefont {Armengaud}\ \emph
  {et~al.}(2017{\natexlab{b}})\citenamefont {Armengaud} \emph
  {et~al.}}]{Armengaud:2017}%
  \BibitemOpen
  \bibfield  {author} {\bibinfo {author} {\bibfnamefont {E.}~\bibnamefont
  {Armengaud}} \emph {et~al.} (\bibinfo {collaboration} {{LUMINEU}}),\
  }\bibfield  {title} {\bibinfo {title} {{Development of $^{100}$Mo-containing
  scintillating bolometers for a high-sensitivity neutrinoless double-beta
  decay search}},\ }\href@noop {} {\bibfield  {journal} {\bibinfo  {journal}
  {Eur. Phys. J. C}\ }\textbf {\bibinfo {volume} {77}},\ \bibinfo {pages} {785}
  (\bibinfo {year} {2017}{\natexlab{b}})}\BibitemShut {NoStop}%
\bibitem [{\citenamefont {Haller}\ \emph {et~al.}(1984)\citenamefont {Haller},
  \citenamefont {Palaio}, \citenamefont {Rodder}, \citenamefont {Hansen},\ and\
  \citenamefont {Kreysa}}]{Haller:1984}%
  \BibitemOpen
  \bibfield  {author} {\bibinfo {author} {\bibfnamefont {E.~E.}\ \bibnamefont
  {Haller}}, \bibinfo {author} {\bibfnamefont {N.~P.}\ \bibnamefont {Palaio}},
  \bibinfo {author} {\bibfnamefont {M.}~\bibnamefont {Rodder}}, \bibinfo
  {author} {\bibfnamefont {W.~L.}\ \bibnamefont {Hansen}},\ and\ \bibinfo
  {author} {\bibfnamefont {E.}~\bibnamefont {Kreysa}},\ }\bibfield  {title}
  {\bibinfo {title} {{NTD Germanium: A Novel Material for Low Temperature
  Bolometers}},\ }in\ \href@noop {} {\emph {\bibinfo {booktitle} {{Neutron
  Transmutation Doping of Semiconductor Materials}}}},\ \bibinfo {editor}
  {edited by\ \bibinfo {editor} {\bibfnamefont {R.~D.}\ \bibnamefont
  {Larrabee}}}\ (\bibinfo  {publisher} {Springer US},\ \bibinfo {address}
  {Boston, MA},\ \bibinfo {year} {1984})\ pp.\ \bibinfo {pages}
  {21--36}\BibitemShut {NoStop}%
\bibitem [{\citenamefont {Armengaud}\ \emph {et~al.}(2021)\citenamefont
  {Armengaud} \emph {et~al.}}]{Armengaud:2020c}%
  \BibitemOpen
  \bibfield  {author} {\bibinfo {author} {\bibfnamefont {E.}~\bibnamefont
  {Armengaud}} \emph {et~al.} (\bibinfo {collaboration} {CUPID}),\ }\bibfield
  {title} {\bibinfo {title} {{New Limit for Neutrinoless Double-Beta Decay of
  $^{100}$Mo from the CUPID-Mo Experiment}},\ }\href
  {https://doi.org/10.1103/PhysRevLett.126.181802} {\bibfield  {journal}
  {\bibinfo  {journal} {Phys. Rev. Lett.}\ }\textbf {\bibinfo {volume} {126}},\
  \bibinfo {pages} {181802} (\bibinfo {year} {2021})},\ \Eprint
  {https://arxiv.org/abs/2011.13243} {arXiv:2011.13243 [nucl-ex]} \BibitemShut
  {NoStop}%
\bibitem [{\citenamefont {Adams}\ \emph {et~al.}(2021)\citenamefont {Adams}
  \emph {et~al.}}]{Adams:2021}%
  \BibitemOpen
  \bibfield  {author} {\bibinfo {author} {\bibfnamefont {D.~Q.}\ \bibnamefont
  {Adams}} \emph {et~al.} (\bibinfo {collaboration} {CUORE}),\ }\bibfield
  {title} {\bibinfo {title} {{Search for double-beta decay of $\mathrm
  {^{130}Te}$ to the $0^+$ states of $\mathrm {^{130}Xe}$ with CUORE}},\ }\href
  {https://doi.org/10.1140/epjc/s10052-021-09317-z} {\bibfield  {journal}
  {\bibinfo  {journal} {Eur. Phys. J. C}\ }\textbf {\bibinfo {volume} {81}},\
  \bibinfo {pages} {567} (\bibinfo {year} {2021})},\ \Eprint
  {https://arxiv.org/abs/2101.10702} {arXiv:2101.10702 [nucl-ex]} \BibitemShut
  {NoStop}%
\bibitem [{\citenamefont {Alduino}\ \emph {et~al.}(2019)\citenamefont {Alduino}
  \emph {et~al.}}]{Alduino:2018_1}%
  \BibitemOpen
  \bibfield  {author} {\bibinfo {author} {\bibfnamefont {C.}~\bibnamefont
  {Alduino}} \emph {et~al.} (\bibinfo {collaboration} {CUORE}),\ }\bibfield
  {title} {\bibinfo {title} {{Double-beta decay of $^{130}\hbox {Te}$ to the
  first $0^+$ excited state of $^{130}\hbox {Xe}$ with CUORE-0}},\ }\href
  {https://doi.org/10.1140/epjc/s10052-019-7275-5} {\bibfield  {journal}
  {\bibinfo  {journal} {Eur. Phys. J. C}\ }\textbf {\bibinfo {volume} {79}},\
  \bibinfo {pages} {795} (\bibinfo {year} {2019})},\ \Eprint
  {https://arxiv.org/abs/1811.10363} {arXiv:1811.10363 [nucl-ex]} \BibitemShut
  {NoStop}%
\bibitem [{\citenamefont {Azzolini}\ \emph
  {et~al.}(2018{\natexlab{b}})\citenamefont {Azzolini} \emph
  {et~al.}}]{Azzolini:2018oph}%
  \BibitemOpen
  \bibfield  {author} {\bibinfo {author} {\bibfnamefont {O.}~\bibnamefont
  {Azzolini}} \emph {et~al.} (\bibinfo {collaboration} {CUPID-0}),\ }\bibfield
  {title} {\bibinfo {title} {{Search of the neutrino-less double beta decay of
  $^{82}$Se into the excited states of $^{82}$Kr with CUPID-0}},\ }\href@noop
  {} {\bibfield  {journal} {\bibinfo  {journal} {Eur. Phys. J. C}\ }\textbf
  {\bibinfo {volume} {78}},\ \bibinfo {pages} {888} (\bibinfo {year}
  {2018}{\natexlab{b}})}\BibitemShut {NoStop}%
\bibitem [{\citenamefont {Allison}\ \emph {et~al.}(2016)\citenamefont {Allison}
  \emph {et~al.}}]{Allison:2016}%
  \BibitemOpen
  \bibfield  {author} {\bibinfo {author} {\bibfnamefont {J.}~\bibnamefont
  {Allison}} \emph {et~al.},\ }\bibfield  {title} {\bibinfo {title} {{Recent
  developments in Geant4}},\ }\href
  {https://doi.org/10.1016/j.nima.2016.06.125} {\bibfield  {journal} {\bibinfo
  {journal} {Nucl. Instrum. Meth. A}\ }\textbf {\bibinfo {volume} {835}},\
  \bibinfo {pages} {186} (\bibinfo {year} {2016})}\BibitemShut {NoStop}%
\bibitem [{\citenamefont {Armengaud}\ \emph {et~al.}(2013)\citenamefont
  {Armengaud} \emph {et~al.}}]{Armengaud:2013}%
  \BibitemOpen
  \bibfield  {author} {\bibinfo {author} {\bibfnamefont {E.}~\bibnamefont
  {Armengaud}} \emph {et~al.} (\bibinfo {collaboration} {EDELWEISS}),\
  }\bibfield  {title} {\bibinfo {title} {{Background studies for the EDELWEISS
  dark matter experiment}},\ }\href
  {https://doi.org/10.1016/j.astropartphys.2013.05.004} {\bibfield  {journal}
  {\bibinfo  {journal} {Astropart. Phys.}\ }\textbf {\bibinfo {volume} {47}},\
  \bibinfo {pages} {1} (\bibinfo {year} {2013})},\ \Eprint
  {https://arxiv.org/abs/1305.3628} {arXiv:1305.3628 [physics.ins-det]}
  \BibitemShut {NoStop}%
\bibitem [{\citenamefont {Hamilton}(1940)}]{Hamilton:1940}%
  \BibitemOpen
  \bibfield  {author} {\bibinfo {author} {\bibfnamefont {D.~R.}\ \bibnamefont
  {Hamilton}},\ }\bibfield  {title} {\bibinfo {title} {{On Directional
  Correlation of Successive Quanta}},\ }\href
  {https://doi.org/10.1103/PhysRev.58.122} {\bibfield  {journal} {\bibinfo
  {journal} {Phys. Rev.}\ }\textbf {\bibinfo {volume} {58}},\ \bibinfo {pages}
  {122} (\bibinfo {year} {1940})}\BibitemShut {NoStop}%
\bibitem [{\citenamefont {Chauvie}\ \emph {et~al.}(2004)\citenamefont {Chauvie}
  \emph {et~al.}}]{livermore}%
  \BibitemOpen
  \bibfield  {author} {\bibinfo {author} {\bibfnamefont {S.}~\bibnamefont
  {Chauvie}} \emph {et~al.},\ }\bibfield  {title} {\bibinfo {title} {{Geant4
  low energy electromagnetic physics}},\ }in\ \href
  {https://doi.org/10.1109/NSSMIC.2004.1462612} {\emph {\bibinfo {booktitle}
  {{IEEE Symposium Conference Record Nuclear Science 2004.}}}},\ Vol.~\bibinfo
  {volume} {3}\ (\bibinfo {year} {2004})\ pp.\ \bibinfo {pages}
  {1881--1885}\BibitemShut {NoStop}%
\bibitem [{\citenamefont {Simkovic}\ \emph {et~al.}(2001)\citenamefont
  {Simkovic}, \citenamefont {Domin},\ and\ \citenamefont
  {Semenov}}]{Simkovic:2000}%
  \BibitemOpen
  \bibfield  {author} {\bibinfo {author} {\bibfnamefont {F.}~\bibnamefont
  {Simkovic}}, \bibinfo {author} {\bibfnamefont {P.}~\bibnamefont {Domin}},\
  and\ \bibinfo {author} {\bibfnamefont {S.~V.}\ \bibnamefont {Semenov}},\
  }\bibfield  {title} {\bibinfo {title} {{The Single state dominance hypothesis
  and the two neutrino double beta decay of $^{100}$Mo}},\ }\href
  {https://doi.org/10.1088/0954-3899/27/11/304} {\bibfield  {journal} {\bibinfo
   {journal} {J. Phys. G}\ }\textbf {\bibinfo {volume} {27}},\ \bibinfo {pages}
  {2233} (\bibinfo {year} {2001})},\ \Eprint
  {https://arxiv.org/abs/nucl-th/0006084} {arXiv:nucl-th/0006084} \BibitemShut
  {NoStop}%
\bibitem [{\citenamefont {Arnold}\ \emph {et~al.}(2019)\citenamefont {Arnold}
  \emph {et~al.}}]{Arnold:2019}%
  \BibitemOpen
  \bibfield  {author} {\bibinfo {author} {\bibfnamefont {R.}~\bibnamefont
  {Arnold}} \emph {et~al.} (\bibinfo {collaboration} {{NEMO-3}}),\ }\bibfield
  {title} {\bibinfo {title} {{Detailed studies of $^{100}$Mo two-neutrino
  double beta decay in NEMO-3}},\ }\href@noop {} {\bibfield  {journal}
  {\bibinfo  {journal} {Eur. Phys. J. C}\ }\textbf {\bibinfo {volume} {79}},\
  \bibinfo {pages} {440} (\bibinfo {year} {2019})}\BibitemShut {NoStop}%
\bibitem [{\citenamefont {Armengaud}\ \emph
  {et~al.}(2020{\natexlab{b}})\citenamefont {Armengaud} \emph
  {et~al.}}]{Armengaud:2020b}%
  \BibitemOpen
  \bibfield  {author} {\bibinfo {author} {\bibfnamefont {E.}~\bibnamefont
  {Armengaud}} \emph {et~al.} (\bibinfo {collaboration} {{CUPID-Mo}}),\
  }\bibfield  {title} {\bibinfo {title} {{Precise measurement of $2\nu \beta
  \beta $ decay of $^{100}$Mo with the CUPID-Mo detection technology}},\
  }\href@noop {} {\bibfield  {journal} {\bibinfo  {journal} {Eur. Phys. J. C}\
  }\textbf {\bibinfo {volume} {80}},\ \bibinfo {pages} {674} (\bibinfo {year}
  {2020}{\natexlab{b}})}\BibitemShut {NoStop}%
\bibitem [{\citenamefont {Alduino}\ \emph {et~al.}(2016)\citenamefont {Alduino}
  \emph {et~al.}}]{Alduino:2016}%
  \BibitemOpen
  \bibfield  {author} {\bibinfo {author} {\bibfnamefont {C.}~\bibnamefont
  {Alduino}} \emph {et~al.} (\bibinfo {collaboration} {{CUORE}}),\ }\bibfield
  {title} {\bibinfo {title} {{Analysis Techniques for the Evaluation of the
  Neutrinoless Double-Beta Decay Lifetime in $^{130}$Te with CUORE-0}},\
  }\href@noop {} {\bibfield  {journal} {\bibinfo  {journal} {Phys. Rev. C}\
  }\textbf {\bibinfo {volume} {93}},\ \bibinfo {pages} {045503} (\bibinfo
  {year} {2016})}\BibitemShut {NoStop}%
\bibitem [{\citenamefont {Azzolini}\ \emph
  {et~al.}(2018{\natexlab{c}})\citenamefont {Azzolini} \emph
  {et~al.}}]{Azzolini2018b}%
  \BibitemOpen
  \bibfield  {author} {\bibinfo {author} {\bibfnamefont {O.}~\bibnamefont
  {Azzolini}} \emph {et~al.} (\bibinfo {collaboration} {CUPID-0}),\ }\bibfield
  {title} {\bibinfo {title} {{Analysis of cryogenic calorimeters with light and
  heat read-out for double beta decay searches}},\ }\href@noop {} {\bibfield
  {journal} {\bibinfo  {journal} {Eur. Phys. J. C}\ }\textbf {\bibinfo {volume}
  {78}},\ \bibinfo {pages} {734} (\bibinfo {year}
  {2018}{\natexlab{c}})}\BibitemShut {NoStop}%
\bibitem [{\citenamefont {Gatti}\ and\ \citenamefont
  {Manfredi}(1986)}]{Gatti:1986}%
  \BibitemOpen
  \bibfield  {author} {\bibinfo {author} {\bibfnamefont {E.}~\bibnamefont
  {Gatti}}\ and\ \bibinfo {author} {\bibfnamefont {P.}~\bibnamefont
  {Manfredi}},\ }\bibfield  {title} {\bibinfo {title} {Processing the signals
  from solid-state detectors in elementary-particle physics},\ }\href@noop {}
  {\bibfield  {journal} {\bibinfo  {journal} {Riv. Nuovo Cim.}\ }\textbf
  {\bibinfo {volume} {9}},\ \bibinfo {pages} {1} (\bibinfo {year}
  {1986})}\BibitemShut {NoStop}%
\bibitem [{\citenamefont {Huang}\ \emph {et~al.}(2021)\citenamefont {Huang}
  \emph {et~al.}}]{Huang:2020}%
  \BibitemOpen
  \bibfield  {author} {\bibinfo {author} {\bibfnamefont {R.}~\bibnamefont
  {Huang}} \emph {et~al.} (\bibinfo {collaboration} {CUPID}),\ }\bibfield
  {title} {\bibinfo {title} {{Pulse Shape Discrimination in CUPID-Mo using
  Principal Component Analysis}},\ }\href
  {https://doi.org/10.1088/1748-0221/16/03/P03032} {\bibfield  {journal}
  {\bibinfo  {journal} {JINST}\ }\textbf {\bibinfo {volume} {16}}\bibfield
  {number} {\bibinfo  {number} { (03)},\ \bibinfo {pages} {P03032 (2021)}},\
  }\Eprint {https://arxiv.org/abs/2010.04033} {arXiv:2010.04033
  [physics.data-an]} \BibitemShut {NoStop}%
\bibitem [{\citenamefont {Caldwell}\ \emph {et~al.}(2009)\citenamefont
  {Caldwell}, \citenamefont {Koll{\'{a}}r},\ and\ \citenamefont
  {Kr{\"{o}}ninger}}]{Caldwell:2009}%
  \BibitemOpen
  \bibfield  {author} {\bibinfo {author} {\bibfnamefont {A.}~\bibnamefont
  {Caldwell}}, \bibinfo {author} {\bibfnamefont {D.}~\bibnamefont
  {Koll{\'{a}}r}},\ and\ \bibinfo {author} {\bibfnamefont {K.}~\bibnamefont
  {Kr{\"{o}}ninger}},\ }\bibfield  {title} {\bibinfo {title} {{BAT – The
  Bayesian analysis toolkit}},\ }\href@noop {} {\bibfield  {journal} {\bibinfo
  {journal} {Comput. Phys. Commun.}\ }\textbf {\bibinfo {volume} {180}},\
  \bibinfo {pages} {2197} (\bibinfo {year} {2009})}\BibitemShut {NoStop}%
\bibitem [{\citenamefont {Weidenspointner}\ \emph {et~al.}(2013)\citenamefont
  {Weidenspointner} \emph {et~al.}}]{Weidenspointner:2013}%
  \BibitemOpen
  \bibfield  {author} {\bibinfo {author} {\bibfnamefont {G.}~\bibnamefont
  {Weidenspointner}} \emph {et~al.},\ }\bibfield  {title} {\bibinfo {title}
  {{Validation of Compton scattering Monte Carlo simulation models}},\ }in\
  \href {https://doi.org/10.1109/NSSMIC.2013.6829500} {\emph {\bibinfo
  {booktitle} {{2013 IEEE Nuclear Science Symposium and Medical Imaging
  Conference and Workshop on Room-Temperature Semiconductor Detectors}}}}\
  (\bibinfo {year} {2013})\BibitemShut {NoStop}%
\bibitem [{\citenamefont {Cirrone}\ \emph {et~al.}(2010)\citenamefont
  {Cirrone}, \citenamefont {Cuttone}, \citenamefont {Di~Rosa}, \citenamefont
  {Pandola}, \citenamefont {Romano},\ and\ \citenamefont
  {Zhang}}]{Cirrone:2010}%
  \BibitemOpen
  \bibfield  {author} {\bibinfo {author} {\bibfnamefont {G.~A.~P.}\
  \bibnamefont {Cirrone}}, \bibinfo {author} {\bibfnamefont {G.}~\bibnamefont
  {Cuttone}}, \bibinfo {author} {\bibfnamefont {F.}~\bibnamefont {Di~Rosa}},
  \bibinfo {author} {\bibfnamefont {L.}~\bibnamefont {Pandola}}, \bibinfo
  {author} {\bibfnamefont {F.}~\bibnamefont {Romano}},\ and\ \bibinfo {author}
  {\bibfnamefont {Q.}~\bibnamefont {Zhang}},\ }\bibfield  {title} {\bibinfo
  {title} {{Validation of the Geant4 electromagnetic photon cross-sections for
  elements and compounds}},\ }\href
  {https://doi.org/10.1016/j.nima.2010.02.112} {\bibfield  {journal} {\bibinfo
  {journal} {Nucl. Instrum. Meth. A}\ }\textbf {\bibinfo {volume} {618}},\
  \bibinfo {pages} {315} (\bibinfo {year} {2010})}\BibitemShut {NoStop}%
\bibitem [{\citenamefont {Kotila}\ and\ \citenamefont
  {Iachello}(2012)}]{Kotila:2012}%
  \BibitemOpen
  \bibfield  {author} {\bibinfo {author} {\bibfnamefont {J.}~\bibnamefont
  {Kotila}}\ and\ \bibinfo {author} {\bibfnamefont {F.}~\bibnamefont
  {Iachello}},\ }\bibfield  {title} {\bibinfo {title} {{Phase-space factors for
  double-$\beta$ decay}},\ }\href@noop {} {\bibfield  {journal} {\bibinfo
  {journal} {Phys. Rev. C}\ }\textbf {\bibinfo {volume} {85}},\ \bibinfo
  {pages} {034316} (\bibinfo {year} {2012})}\BibitemShut {NoStop}%
\bibitem [{\citenamefont {Coraggio}\ \emph {et~al.}(2022)\citenamefont
  {Coraggio} \emph {et~al.}}]{Coraggio:2022vgy}%
  \BibitemOpen
  \bibfield  {author} {\bibinfo {author} {\bibfnamefont {L.}~\bibnamefont
  {Coraggio}} \emph {et~al.},\ }\bibfield  {title} {\bibinfo {title}
  {{Shell-model calculation of $^{100}$Mo double-$\beta$ decay}},\ }\href
  {https://doi.org/10.1103/PhysRevC.105.034312} {\bibfield  {journal} {\bibinfo
   {journal} {Phys. Rev. C}\ }\textbf {\bibinfo {volume} {105}},\ \bibinfo
  {pages} {034312} (\bibinfo {year} {2022})},\ \Eprint
  {https://arxiv.org/abs/2203.01013} {arXiv:2203.01013 [nucl-th]} \BibitemShut
  {NoStop}%
\bibitem [{\citenamefont {Barea}\ \emph
  {et~al.}(2015{\natexlab{b}})\citenamefont {Barea}, \citenamefont {Kotila},\
  and\ \citenamefont {Iachello}}]{BareaIBM2}%
  \BibitemOpen
  \bibfield  {author} {\bibinfo {author} {\bibfnamefont {J.}~\bibnamefont
  {Barea}}, \bibinfo {author} {\bibfnamefont {J.}~\bibnamefont {Kotila}},\ and\
  \bibinfo {author} {\bibfnamefont {F.}~\bibnamefont {Iachello}},\ }\bibfield
  {title} {\bibinfo {title}
  {{$0\ensuremath{\nu}\ensuremath{\beta}\ensuremath{\beta}$ and
  $2\ensuremath{\nu}\ensuremath{\beta}\ensuremath{\beta}$ nuclear matrix
  elements in the interacting boson model with isospin restoration}},\ }\href
  {https://doi.org/10.1103/PhysRevC.91.034304} {\bibfield  {journal} {\bibinfo
  {journal} {Phys. Rev. C}\ }\textbf {\bibinfo {volume} {91}},\ \bibinfo
  {pages} {034304} (\bibinfo {year} {2015}{\natexlab{b}})}\BibitemShut
  {NoStop}%
\bibitem [{\citenamefont {Pirinen}\ and\ \citenamefont
  {Suhonen}(2015)}]{Jouni2015}%
  \BibitemOpen
  \bibfield  {author} {\bibinfo {author} {\bibfnamefont {P.}~\bibnamefont
  {Pirinen}}\ and\ \bibinfo {author} {\bibfnamefont {J.}~\bibnamefont
  {Suhonen}},\ }\bibfield  {title} {\bibinfo {title} {{Systematic approach to
  $\ensuremath{\beta}$ and
  $2\ensuremath{\nu}\ensuremath{\beta}\ensuremath{\beta}$ decays of mass
  $A=100$--$136$ nuclei}},\ }\href {https://doi.org/10.1103/PhysRevC.91.054309}
  {\bibfield  {journal} {\bibinfo  {journal} {Phys. Rev. C}\ }\textbf {\bibinfo
  {volume} {91}},\ \bibinfo {pages} {054309} (\bibinfo {year}
  {2015})}\BibitemShut {NoStop}%
\bibitem [{\citenamefont {Bandac}\ \emph {et~al.}(2020)\citenamefont {Bandac}
  \emph {et~al.}}]{CROSS:2019xov}%
  \BibitemOpen
  \bibfield  {author} {\bibinfo {author} {\bibfnamefont {I.~C.}\ \bibnamefont
  {Bandac}} \emph {et~al.} (\bibinfo {collaboration} {CROSS}),\ }\bibfield
  {title} {\bibinfo {title} {{The $0\nu2\beta$-decay CROSS experiment:
  preliminary results and prospects}},\ }\href
  {https://doi.org/10.1007/JHEP01(2020)018} {\bibfield  {journal} {\bibinfo
  {journal} {JHEP}\ }\textbf {\bibinfo {volume} {01}},\ \bibinfo {pages} {{018
  (2020)}}},\ \Eprint {https://arxiv.org/abs/{1906.10233}} {arXiv:{1906.10233}
  [nucl-ex]} \BibitemShut {NoStop}%
\end{thebibliography}%
\appendix
\section{Optimization of categories}
\label{ap:opt_cat}
As explained in Section~\ref{Analysis}, some categories are divided up by their projected out energies. In particular, for the $2\nu\beta\beta$ decay analysis the difference in change of spin and Q-values, $Q_{2\beta, 0_{1}^{+}}=1904.1$ keV and $Q_{2\beta, 2_{1}^{+}}=2494.9$ keV, leads to an expected difference in the $E_{\beta\beta}$ spectral shape. This can be used to help reduce correlation between the decays to these two states.

We use a preliminary background model fit to optimize these choices. We emphasize that this fit is only used for the optimization of the choice of categories and not in the Bayesian analysis. We first optimize a choice of $\beta\beta$ energies to minimize the expected measurement error ($\sqrt{S+B}/S$) for $2\nu\beta\beta$ to $0_{1}^{+}$ \es decay. This optimization is performed on the 540 keV peak, however a consistent result is also obtained with 591 keV. 
A separate optimization is performed for the vertical line where the peak is in $E_2$  and a horizontal where the peak is in $E_1$ (see Fig. \ref{2Dist}). This leads to accepting only events with $E_2>220$ keV in the first case, and $E_1<1900$ keV in the second.

Next we optimize the choice of energies, $E_a$, such that the beta energy with a $\gamma$ in $E_1$ (horizontal line) is divided into two slices separated by $E_a$:
\begin{align}
   E_2 \in &[220 \ \mathrm{keV}- E_a], \ [E_a-500 \ \mathrm{keV}].
\end{align}
Similarly for the events with the $\gamma$ in $E_2$ (vertical line) we divide into slices separated by the energies $E_b,E_c$:
\begin{align}
   E_1 \in &[500 \ \mathrm{keV} - E_b], \ [E_b-E_{\text{c}}], \ [E_c-1900 \ \mathrm{keV}].
\end{align}
This optimization maximizes the limit setting sensitivity for $2\nu\beta\beta$ to $2_1^+$ e.s. We estimate this as $S/\sqrt{B}$ using the 540 keV peak, this is a proxy for the correlation between the two decays. These optimizations lead to $E_{a},E_b,E_c= 410,890,1190 $ keV respectively.

We use a similar optimization scheme for the 0\nbb decay analysis. In this case the relevant categories are horizontal lines with peaks in $E_1$ (see Fig. \ref{2Dist}). We therefore make cuts only on the $E_2$ variable. This is performed separately for the peaks at 1900 keV and $\sim$\,$2400$ keV, and results in the categories shown in the Table \ref{tab:0nu_sigs}. Here we optimize by maximizing the approximate sensitivity to $0_1^+$ decay for the 1904 keV peak, and the $2_1^+$ decay for the 2400 keV peaks.

\section{Signal model functions}~ 
\label{ap:functions}
To model the signal shape as described in Section \ref{Analysis} we use phenomenological functions which are often used for modeling the shape of signals in cryogenic calorimeters (for example see \cite{Alduino:2016,Armengaud:2020b}).
In particular we use a linear combination of:
\begin{align}
    &f_{\text{peak}}(E;\vec{p},\mu,\vec{R},\sigma)=\sum_{i=1}^3p_i \cdot \mathcal{N}(E,\mu,R_i\cdot \sigma(E)),\\
   &f_{\text{xray}}(E;\mu,E_X,\sigma) = \mathcal{N}(E,\mu- E_{X},\sigma(E)),\\
    &f_{\text{step}}(E;E_{min},\mu,s,\sigma)=N\cdot\text{Erfc}\Big(\frac{\mu-E}{\sqrt{2}\sigma}\Big),\\
    &f_{\text{step-neg}}= N\cdot\text{Erfc}\Big(\frac{E-\mu}{\sqrt{2}\sigma}\Big),\\
    &f_{\text{bkg}}(E;s;E_{min},E_{max})= N \cdot (1+s\cdot E),\\
    &f_{\text{bump}}=\mathcal{N}(E,\mu,\sigma).
\end{align}
Here $\mathcal{N}(E,\mu,\sigma)$ is a Gaussian with mean $\mu$ and standard deviation $\sigma$, $N$ is a normalization coefficient and Erfc is the complementary error function. The first function models a peak, either $\gamma$ or $\beta\beta$, the second models events where a Mo X-ray escapes the crystal. The third and fourth functions (step/step-neg.) are step functions to account for Compton scattering events. The step function accounts for the Compton scatter of a single $\gamma$ leading to a partial energy deposition and $E<E_{\text{photopeak}}$ and the step negative accounts for a combination of photo-absorption of one $\gamma$ and Compton scatter of the other and $E>E_{\text{photopeak}}$. We include a linear background and a single Gaussian to model some features in the data where a diagonal line crosses the projected out energy leading to a very small but broader peak (bump). For example see the category of $0\nu\beta\beta$ decay to $2_1^+$ \es with $E_1=1904$~keV and $E_{2} \in [500,650]$~keV, here a diagonal line crosses the box at around 591~keV. This line is caused by a 2494~keV $\beta\beta$ in one crystal and a 540~keV $\gamma$ shared between this crystal and another.

\section{Containment efficiency uncertainty}
\label{ap:cont}
\begin{figure*}[htbp]
    \centering
    \includegraphics[width=0.45\textwidth]{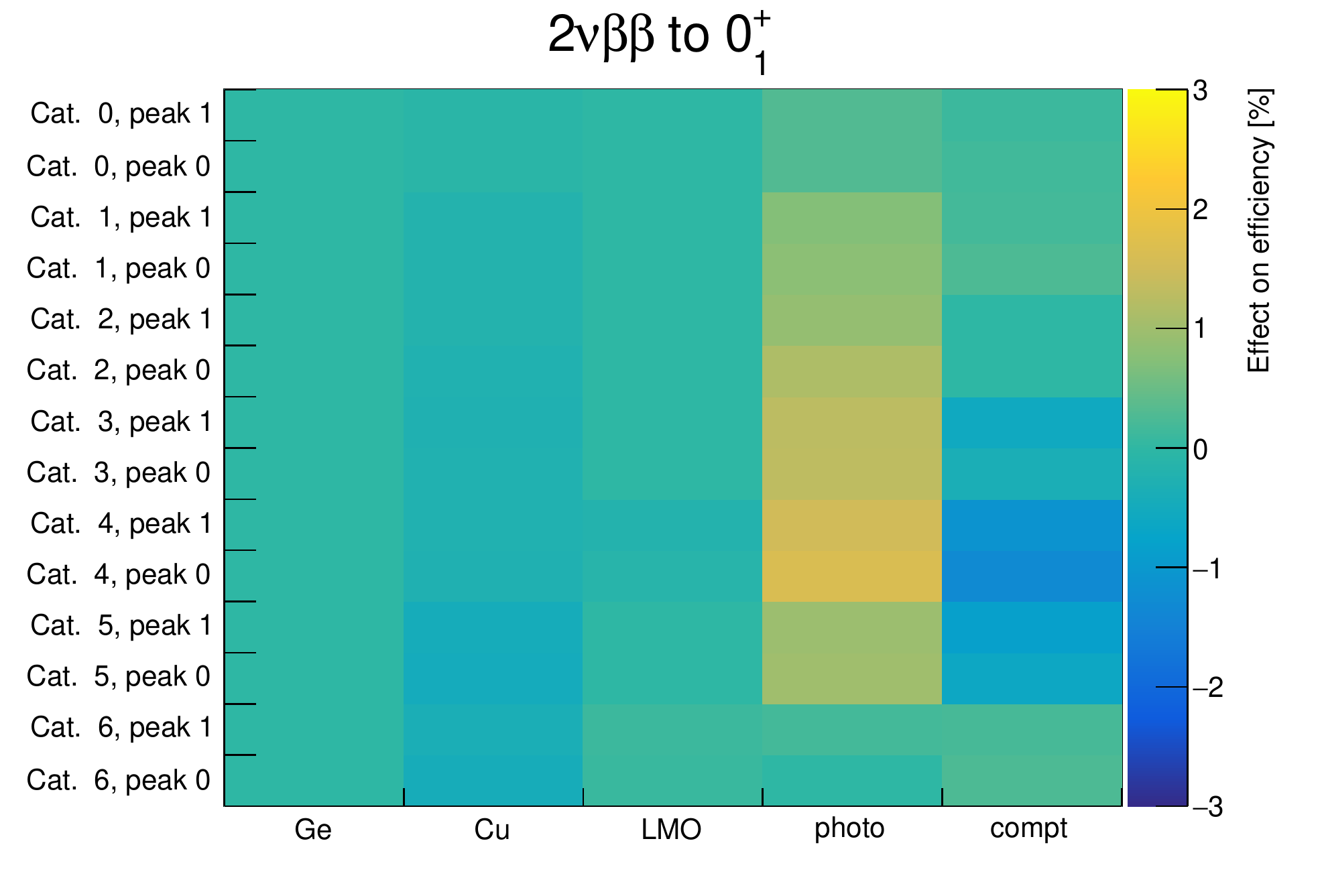}
        \includegraphics[width=0.45\textwidth]{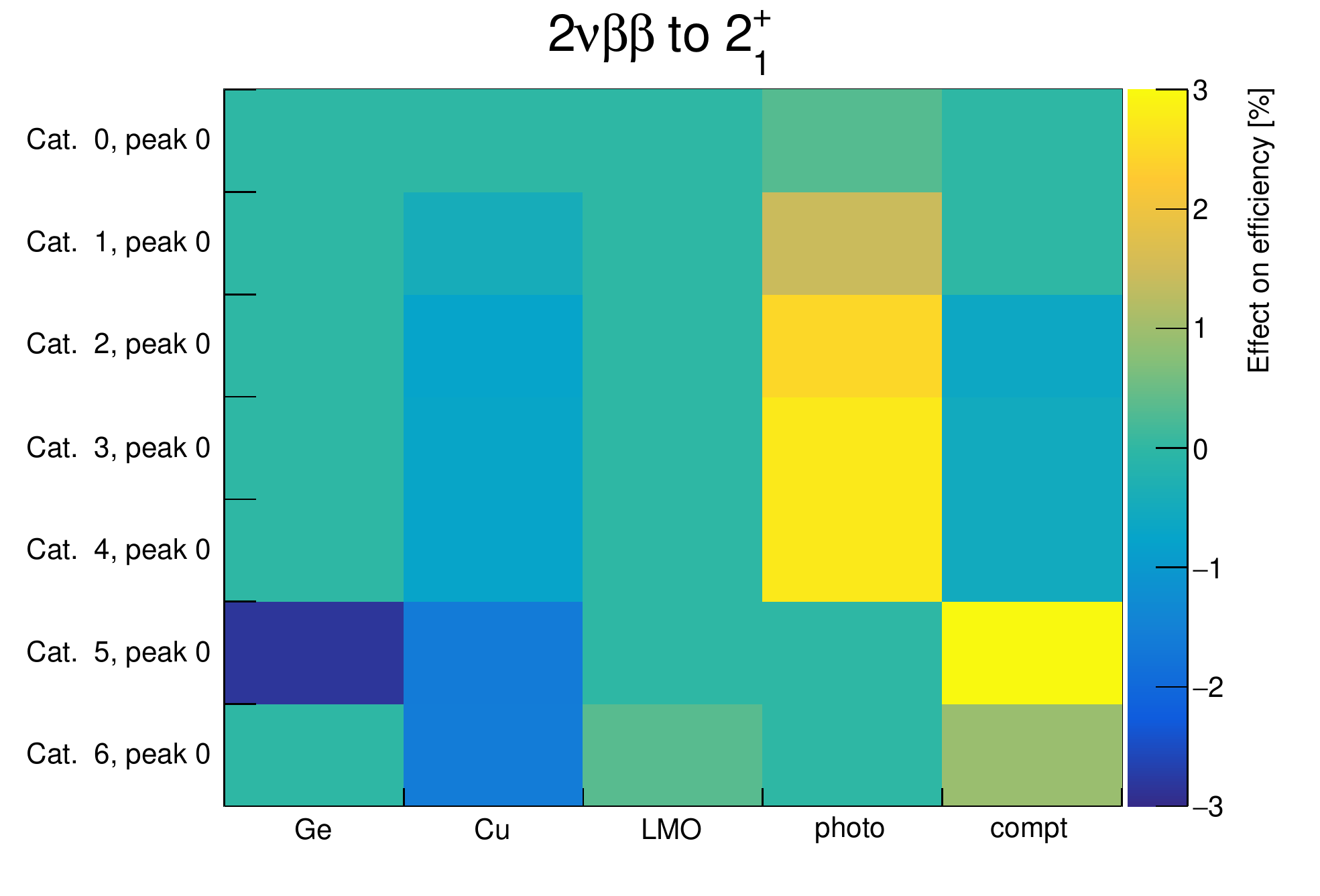}
            \includegraphics[width=0.45\textwidth]{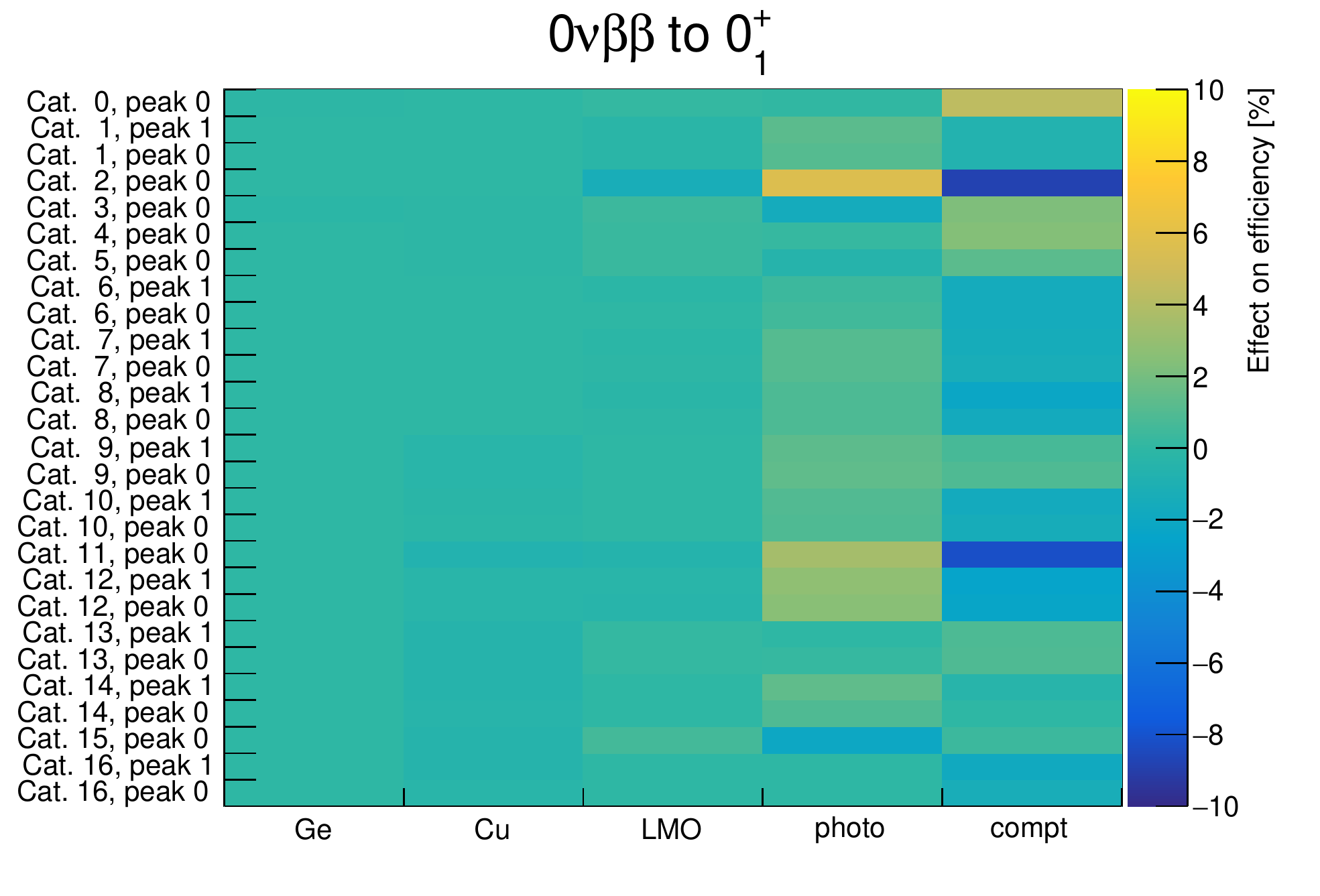}
        \includegraphics[width=0.45\textwidth]{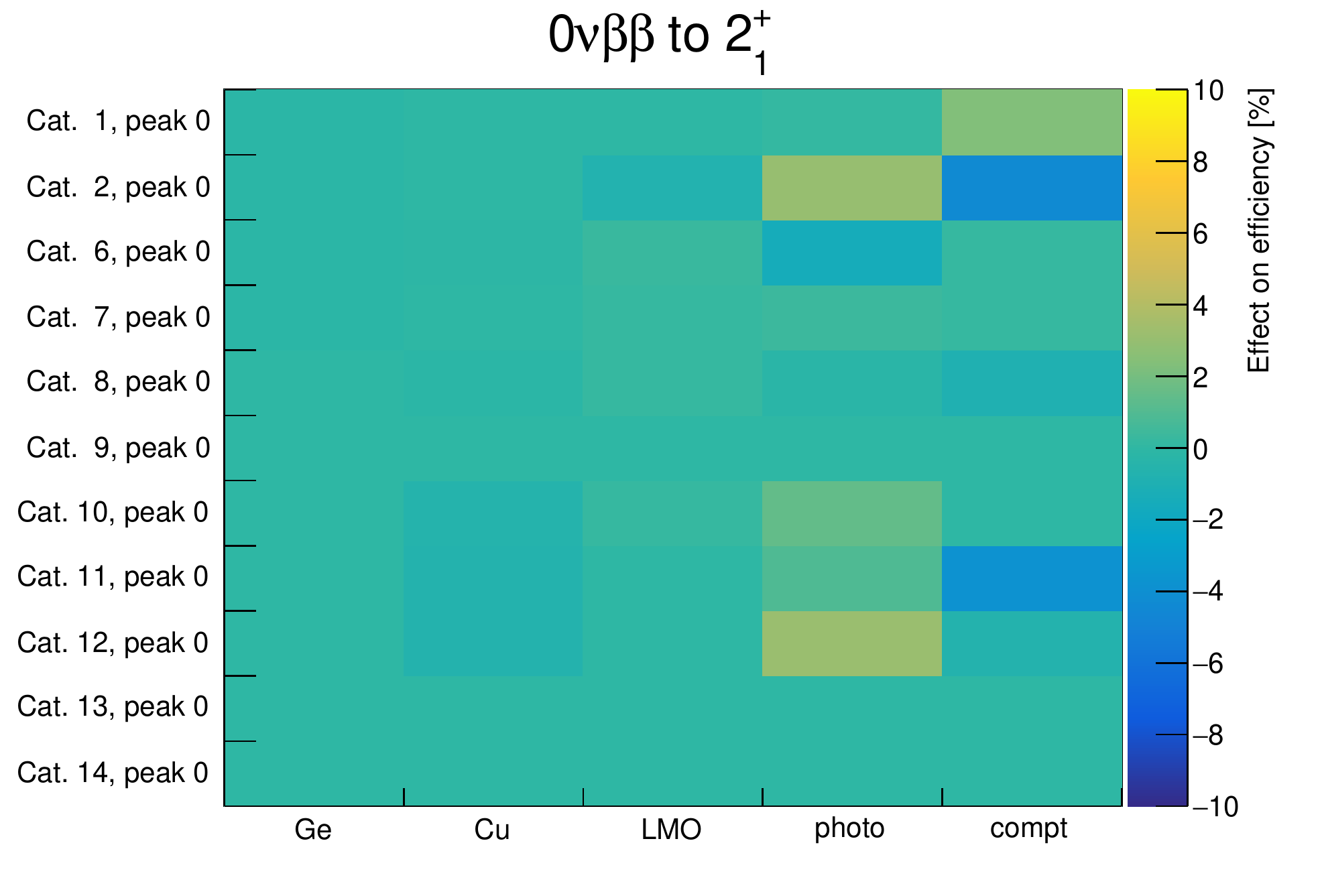}

    \caption{Plots showing the systematic uncertainty on containment efficiency for each peak, category, decay mode, and systematic effect. The z-axis scale is the percentage effect. Top left this is $2\nu\beta\beta$ decay to $0_1^+$ state and on the right for decay to $2_1^+$ state. Bottom left is $0\nu\beta\beta$ decay to $0_1^+$ state, while bottom right is $0\nu\beta\beta$ decay to $2_1^+$ state.}
    \label{fig:cont_2nu}
\end{figure*}

As mentioned in Section \ref{sec:cont}, several sets of MC simulations were used to estimate the uncertainty in the containment efficiency. For each peak, category and systematic test the percentage change in the containment efficiency was extracted. These values are shown in Fig. \ref{fig:cont_2nu}.

\section{Fits to each category}
\label{ap:fits}
\begin{figure*}
    \centering
    \includegraphics[width=0.47\textwidth]{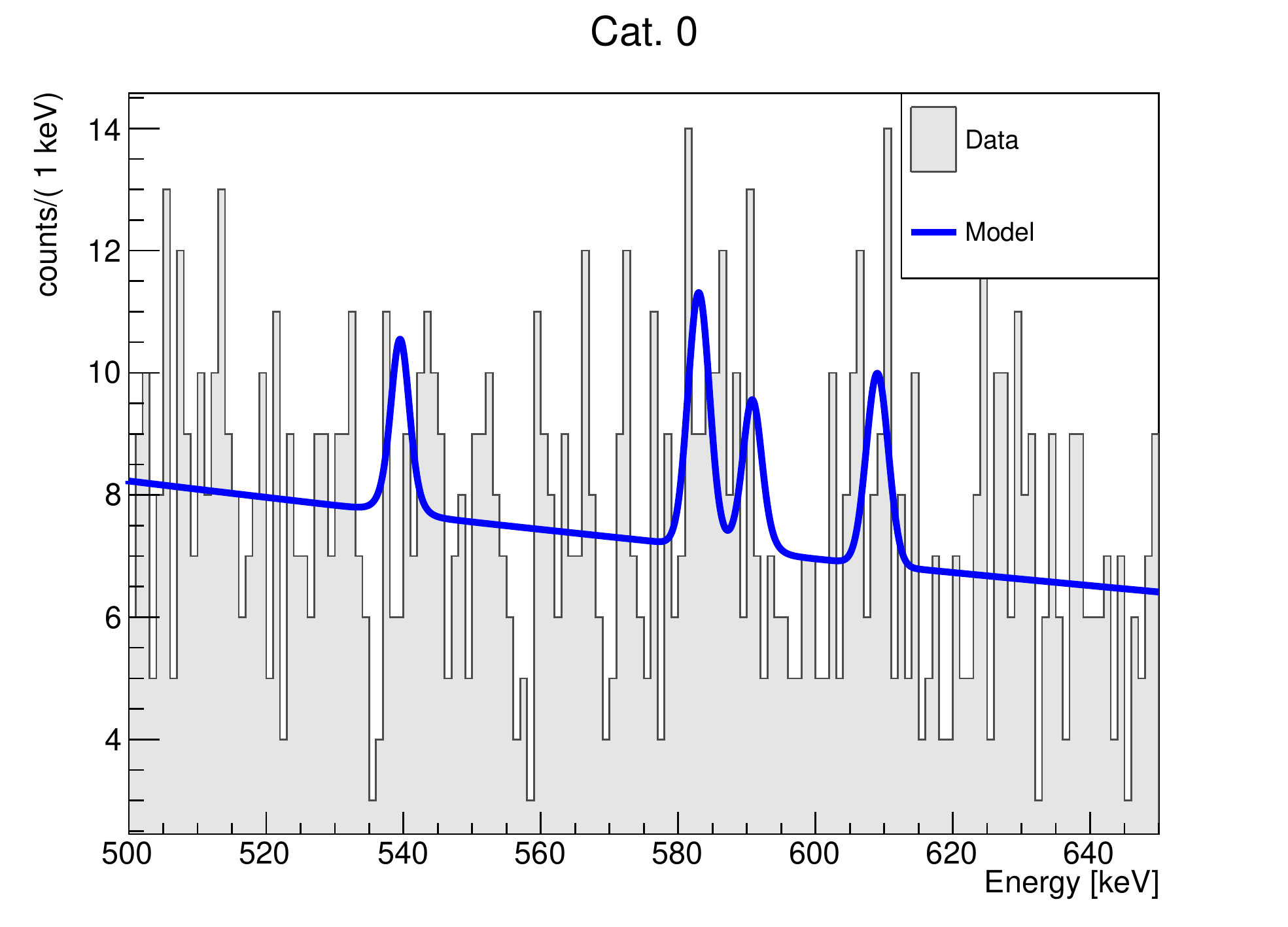}
        \includegraphics[width=0.47\textwidth]{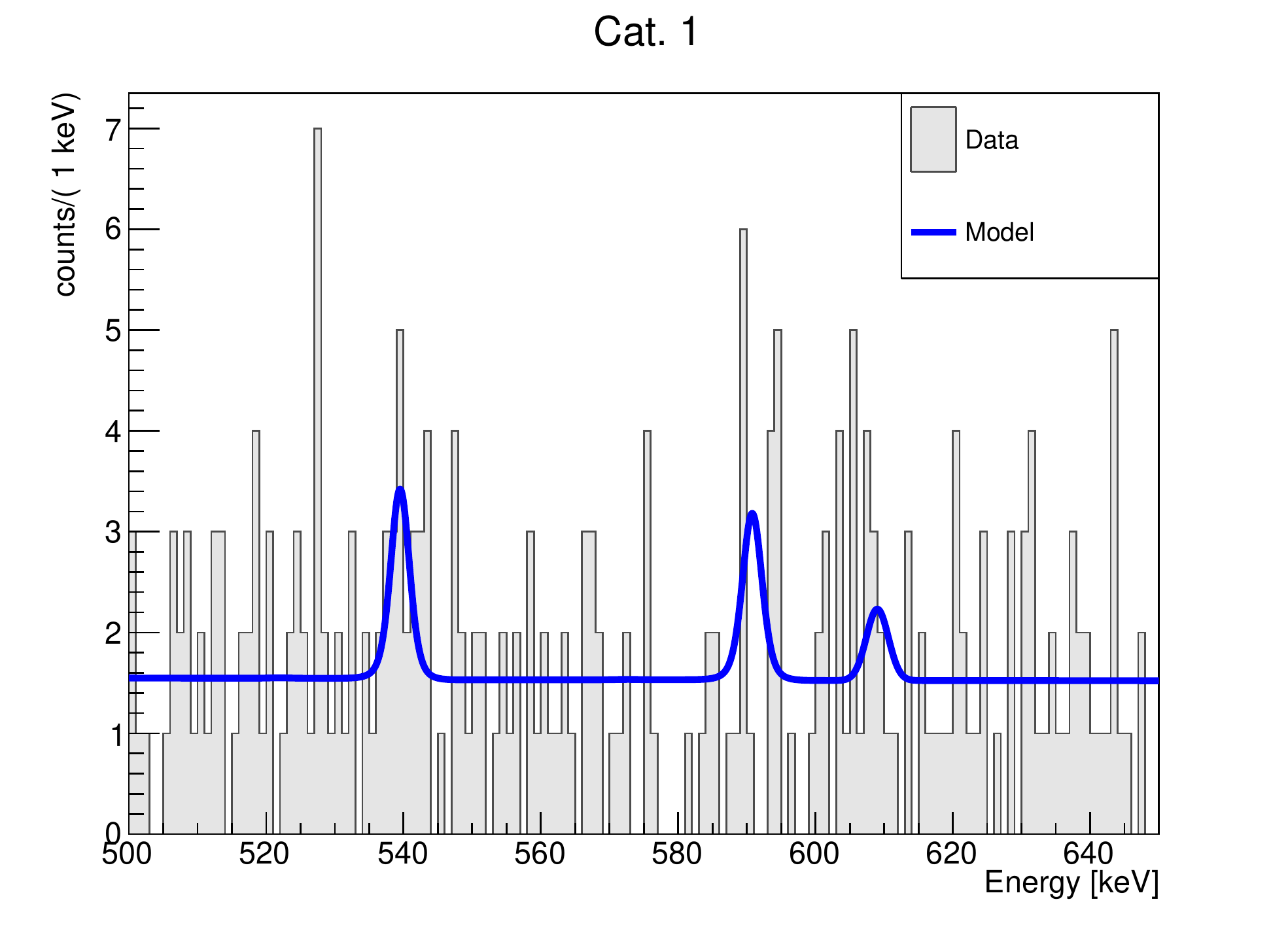}
    \includegraphics[width=0.47\textwidth]{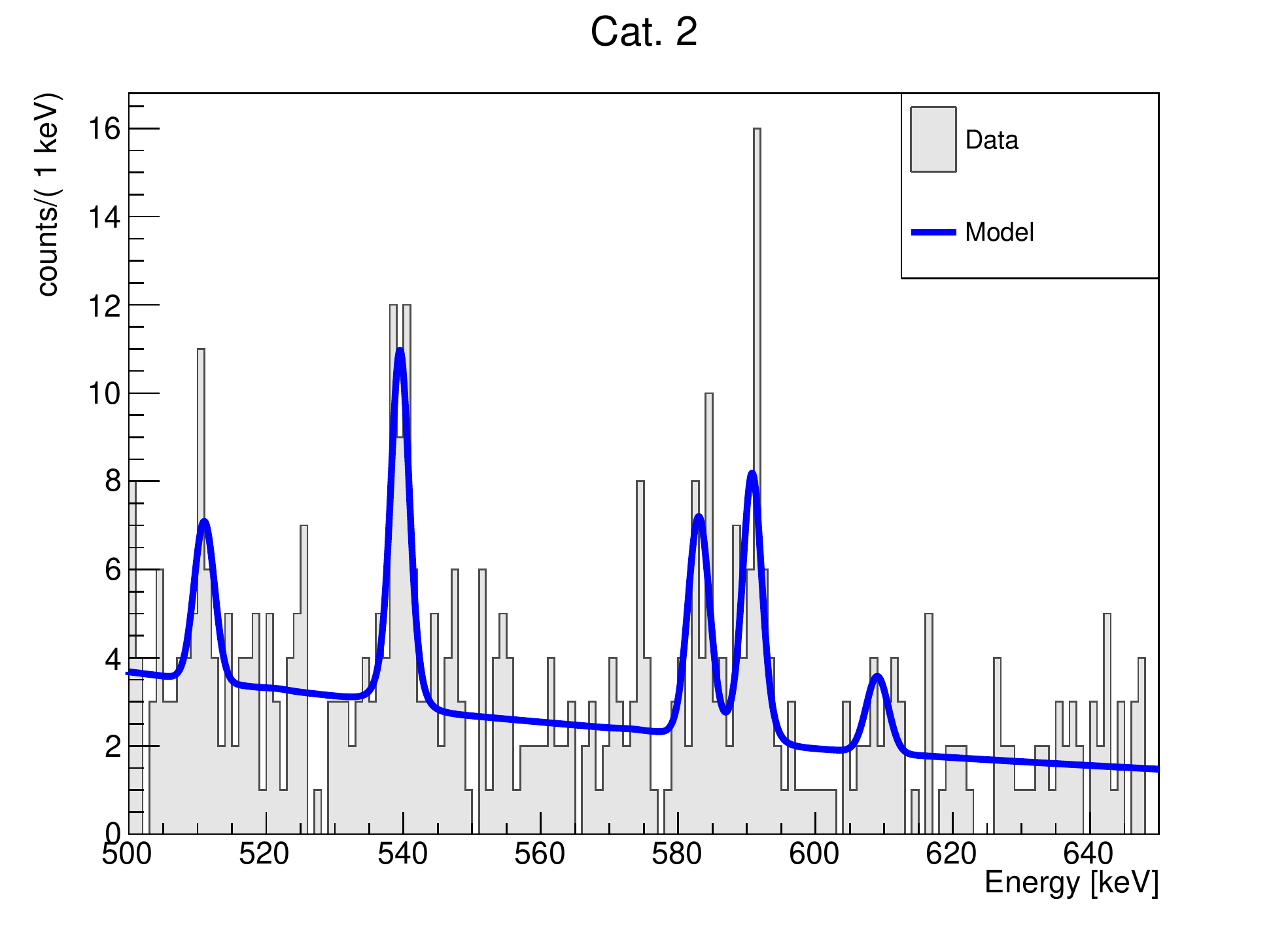}
    \includegraphics[width=0.47\textwidth]{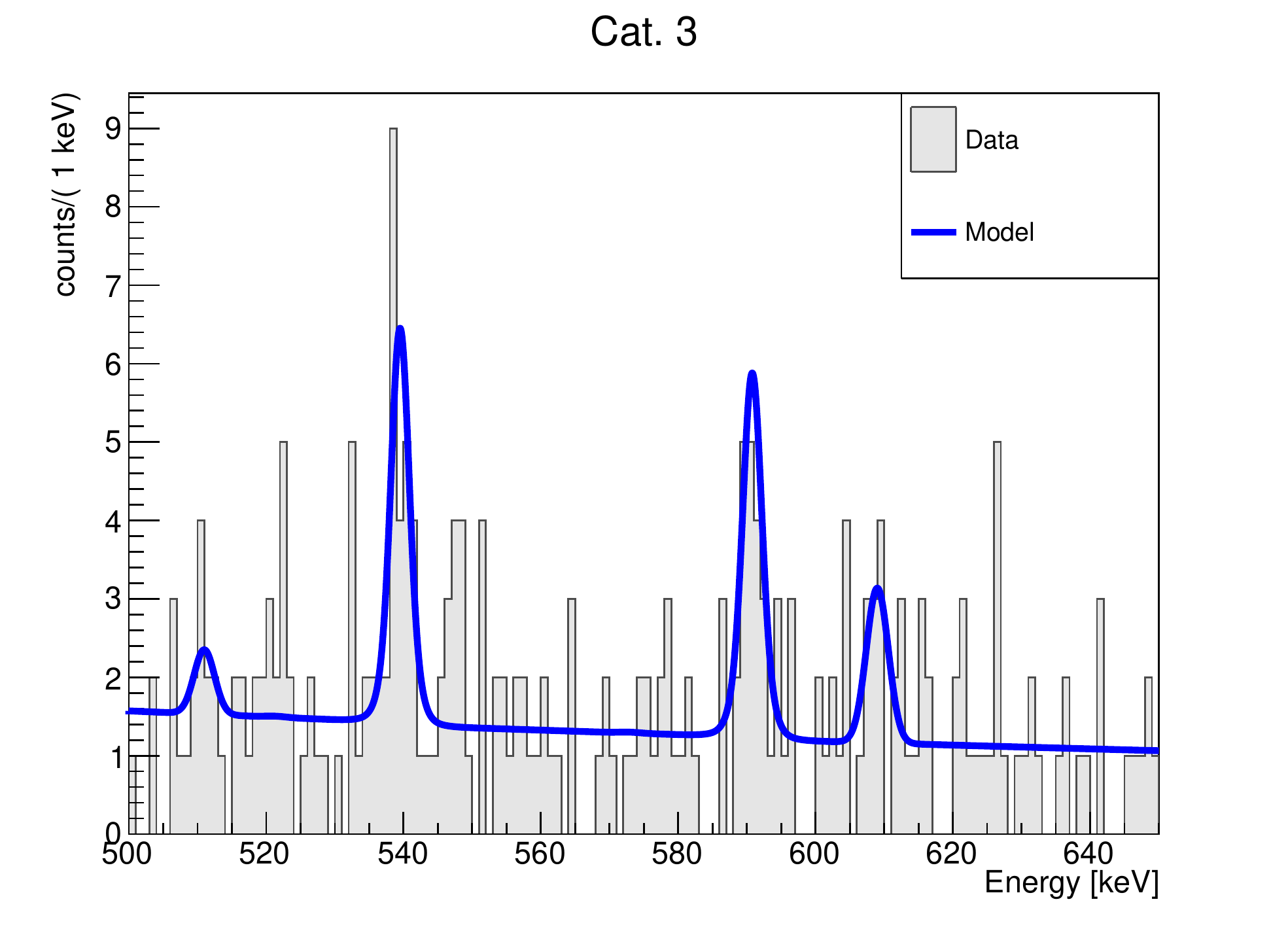}
    \includegraphics[width=0.47\textwidth]{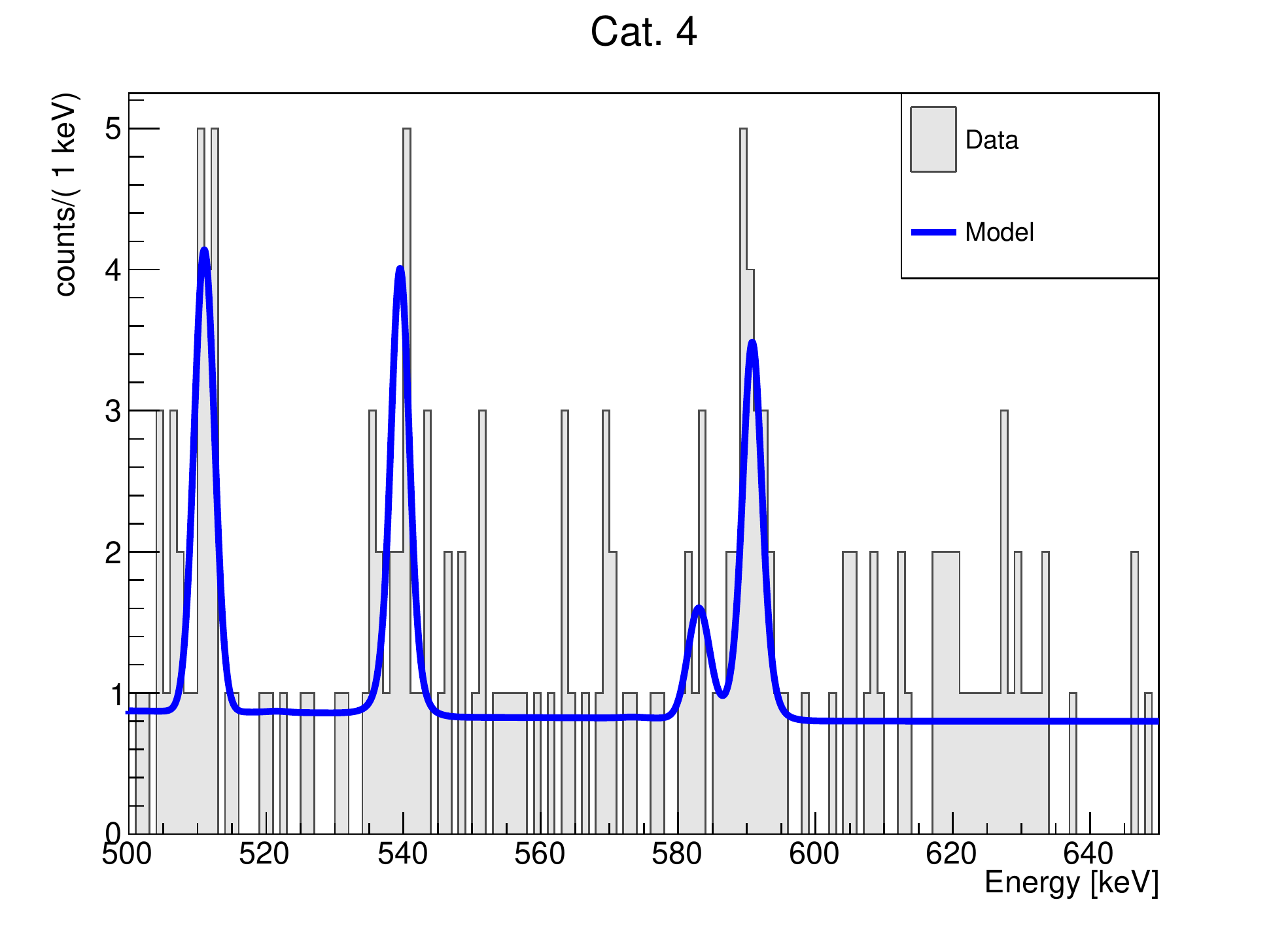}
    \includegraphics[width=0.47\textwidth]{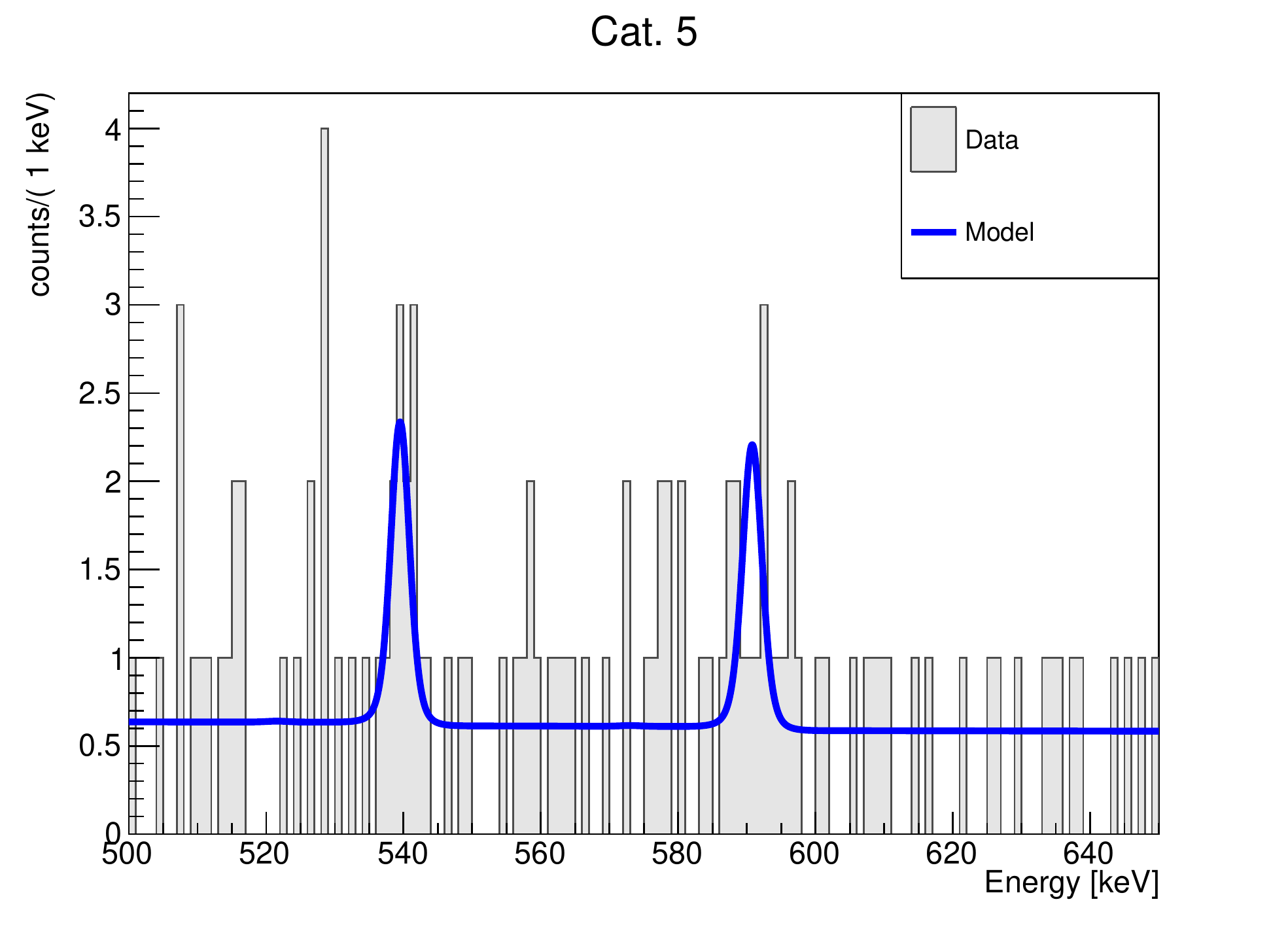}
    \includegraphics[width=0.47\textwidth]{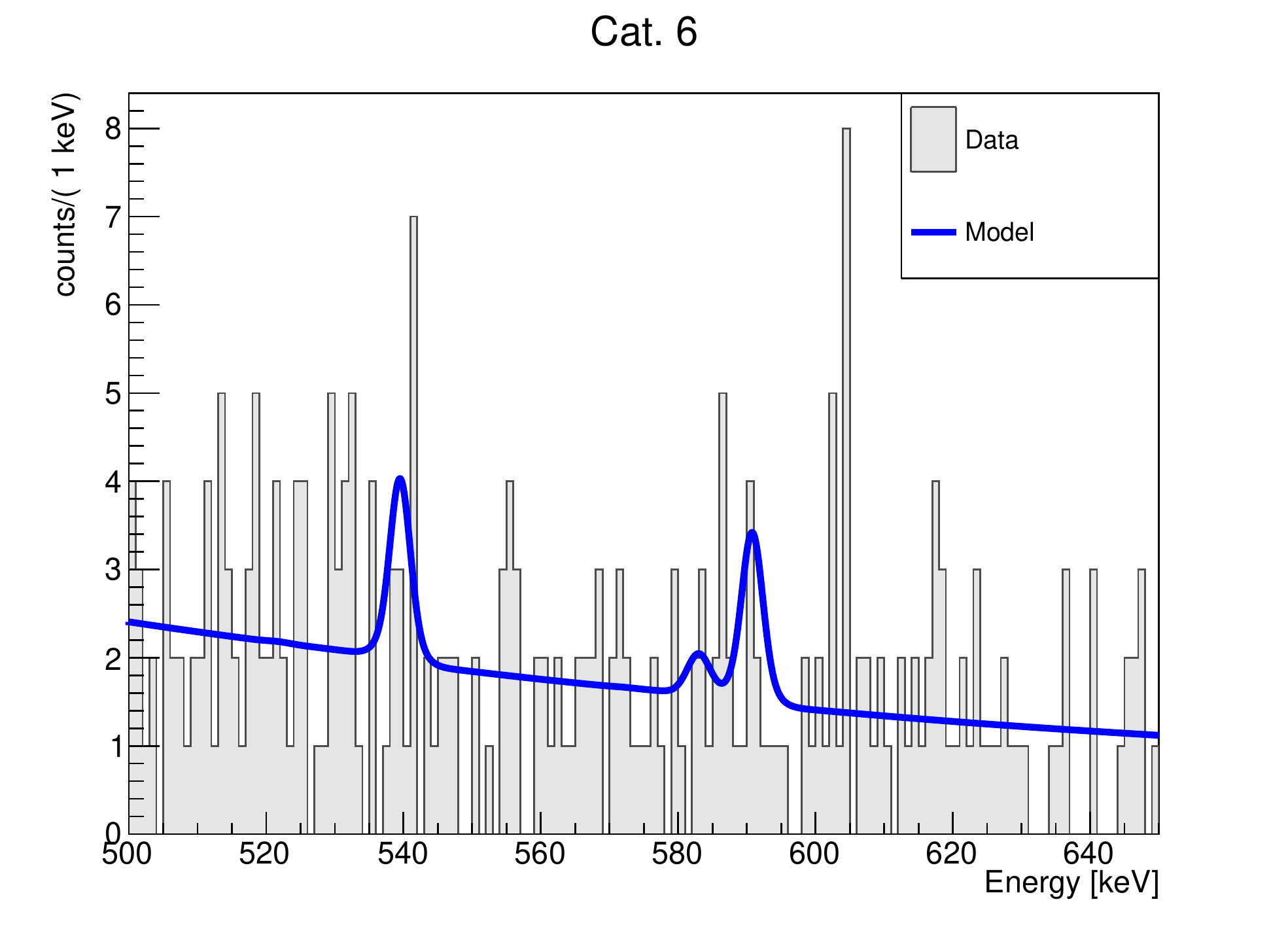}
      \caption{Fits of the seven $2\nu\beta\beta$ decay regions of interest, the experimental data is shown in grey while the best fit (global mode) is shown in blue. We see clearly a signal in each of the regions of interest.}
    \label{fig:2vbb_fits}
\end{figure*}

\begin{figure*}
\centering
\includegraphics[width=0.31\textwidth]{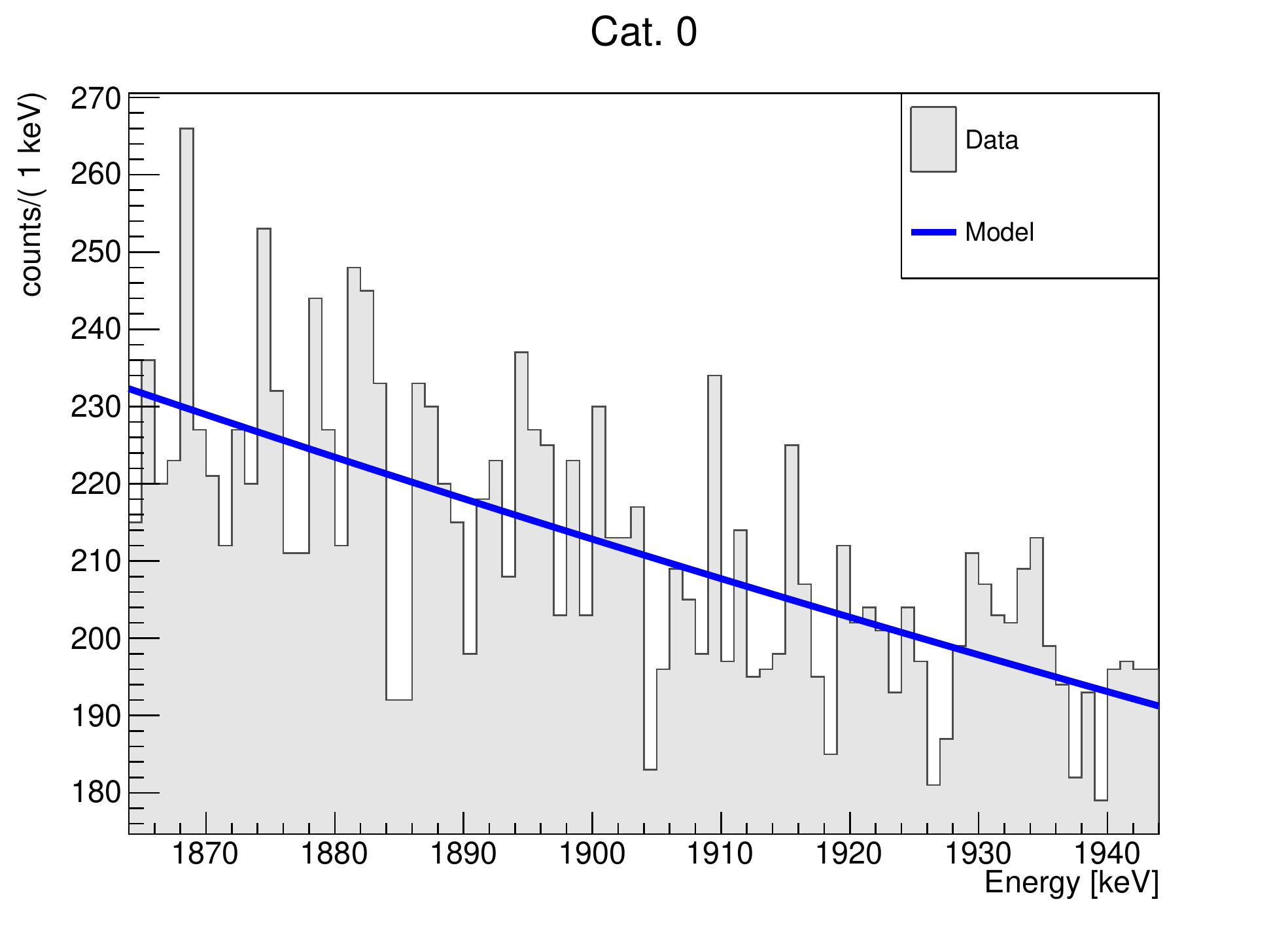}
\includegraphics[width=0.31\textwidth]{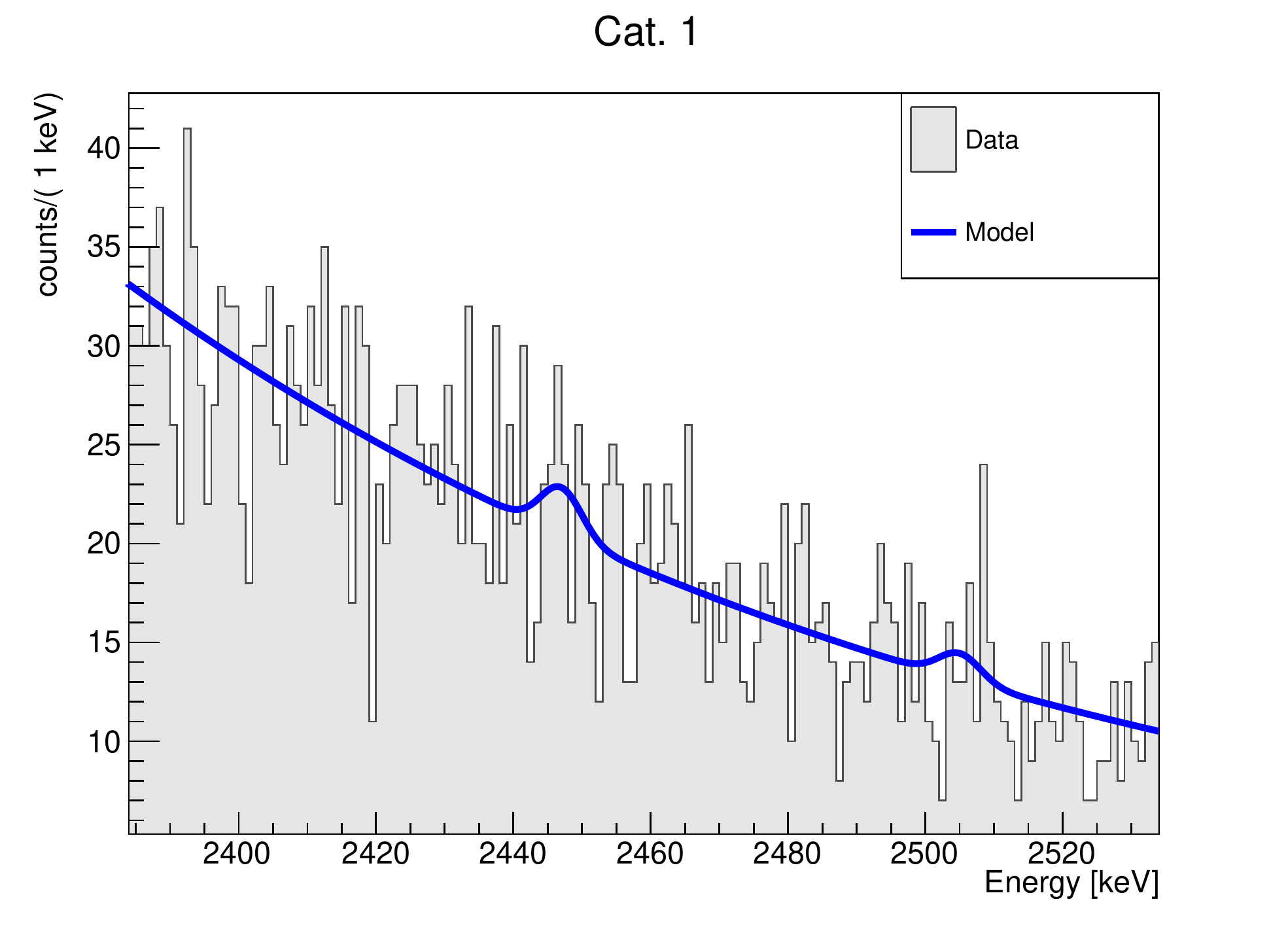}
\includegraphics[width=0.31\textwidth]{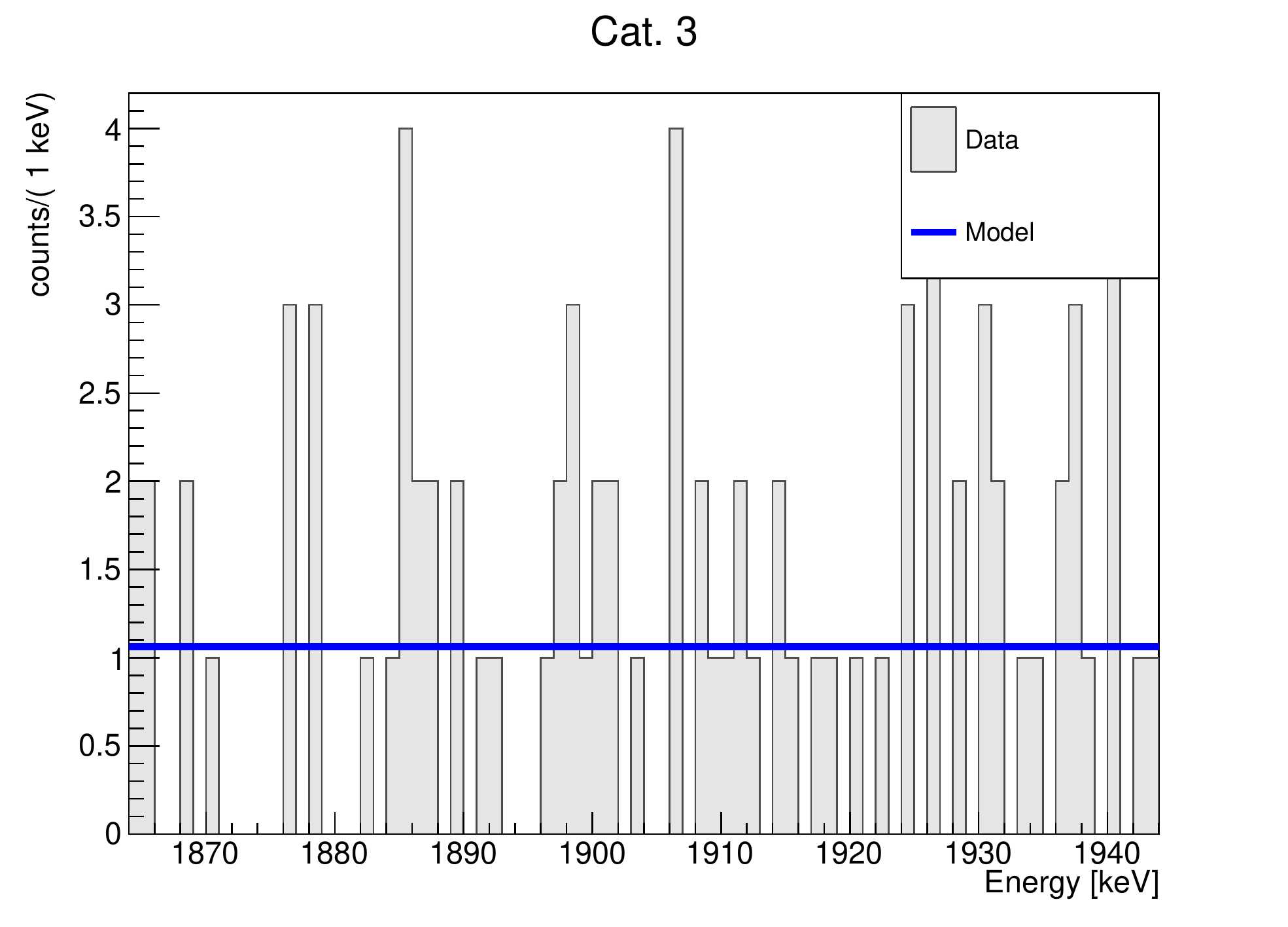}
\includegraphics[width=0.31\textwidth]{ImagesFinal/Final_0nu_0.pdf}
\includegraphics[width=0.31\textwidth]{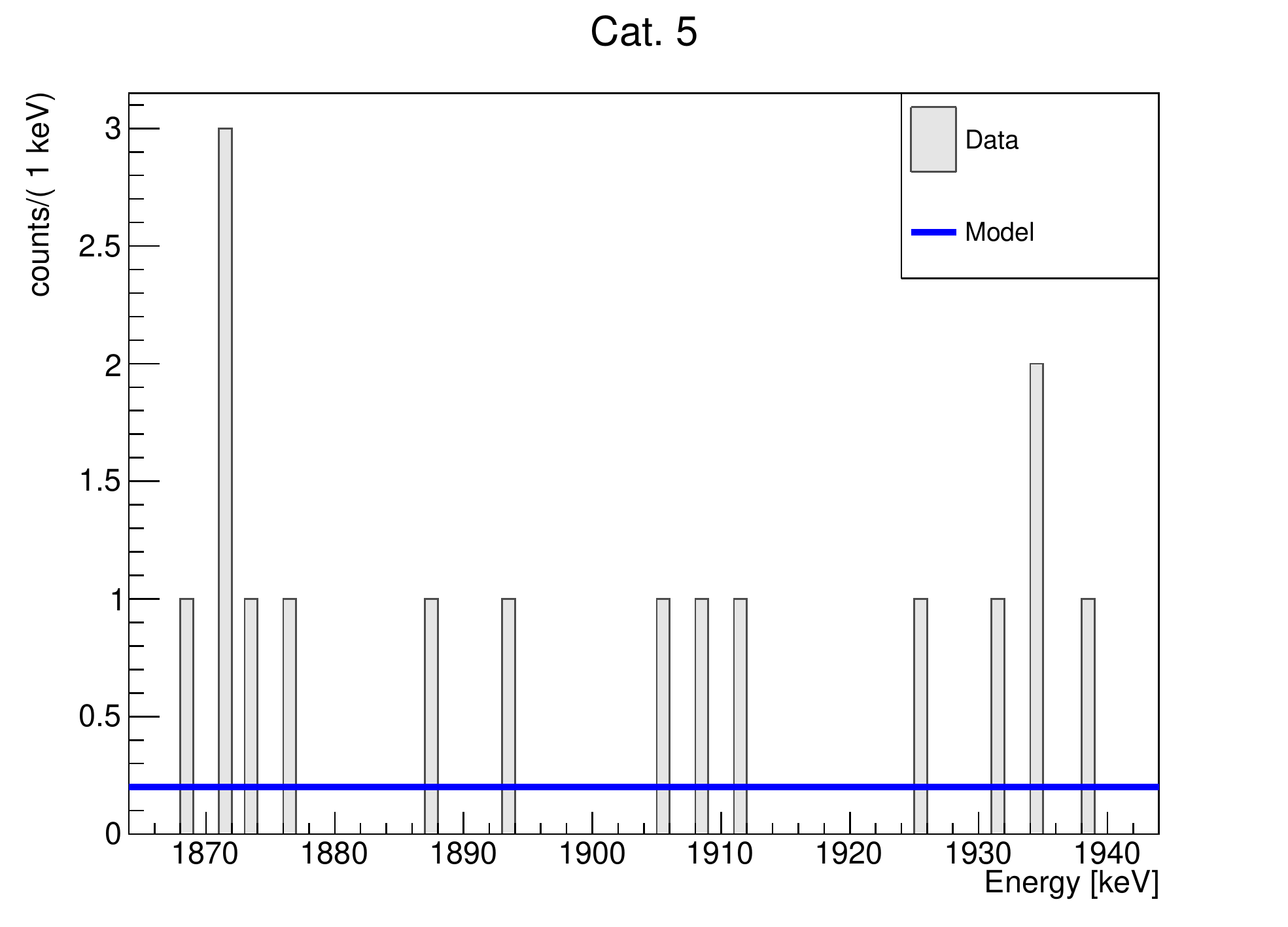}
\includegraphics[width=0.31\textwidth]{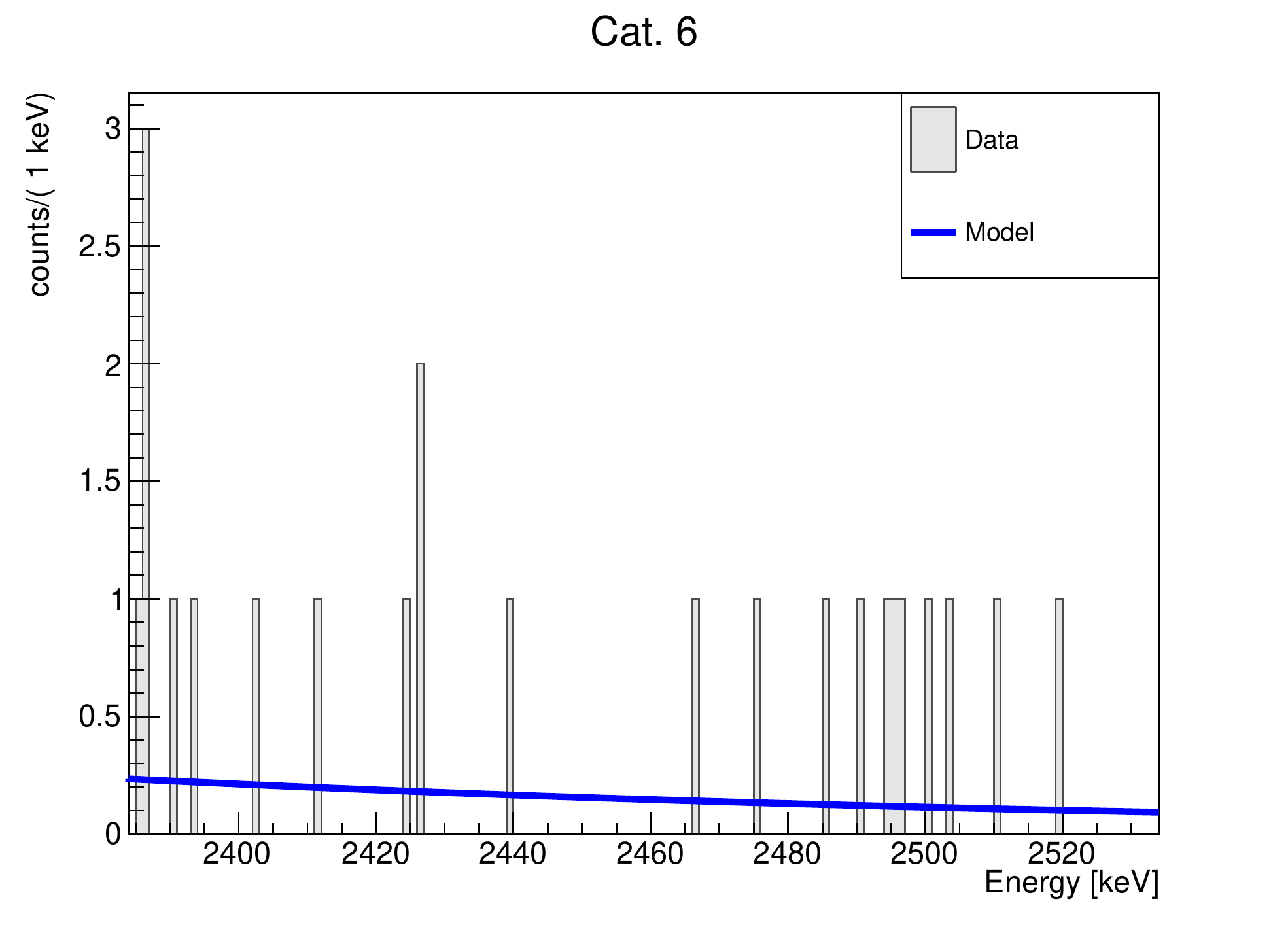}
\includegraphics[width=0.31\textwidth]{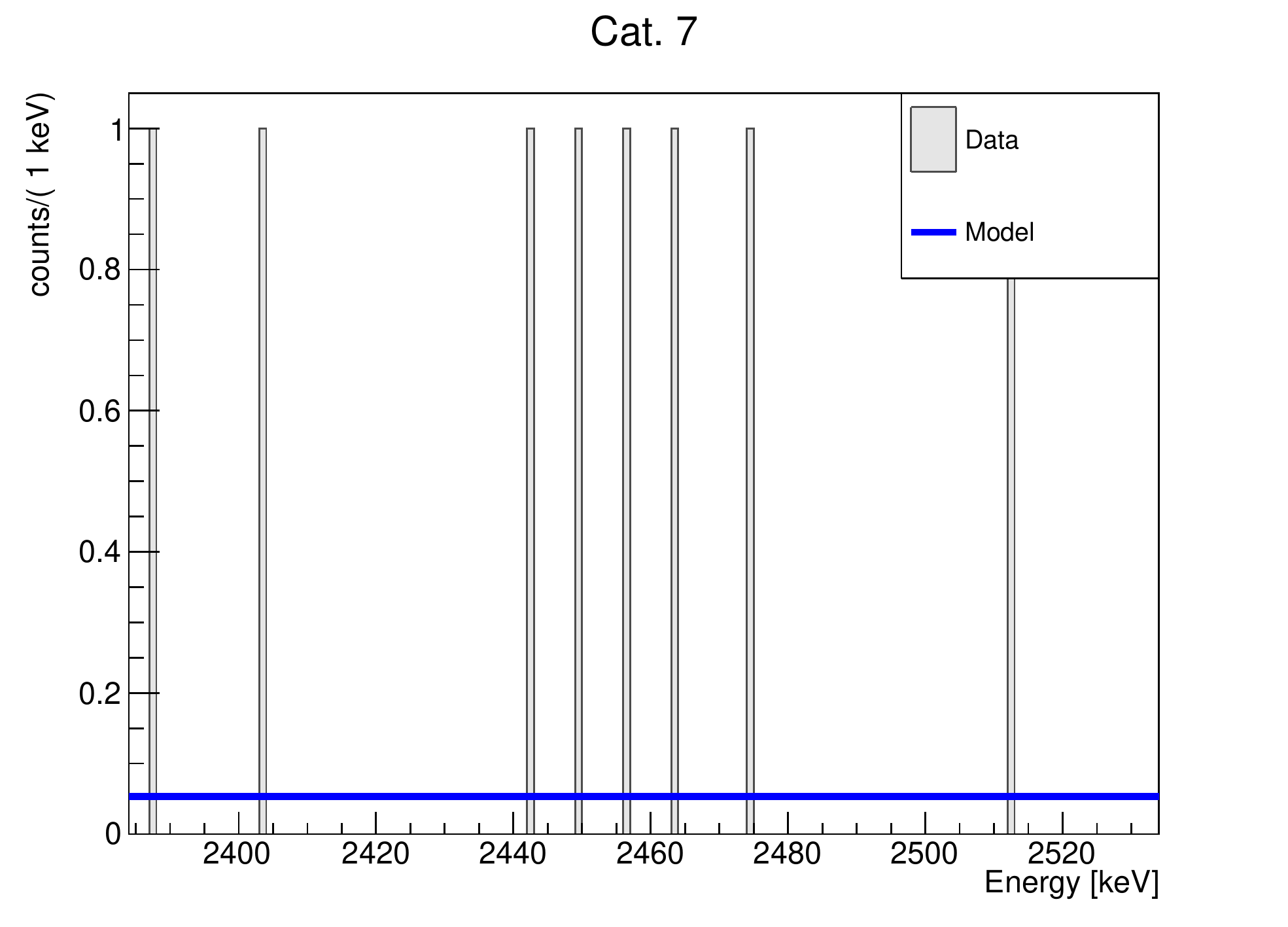}
\includegraphics[width=0.31\textwidth]{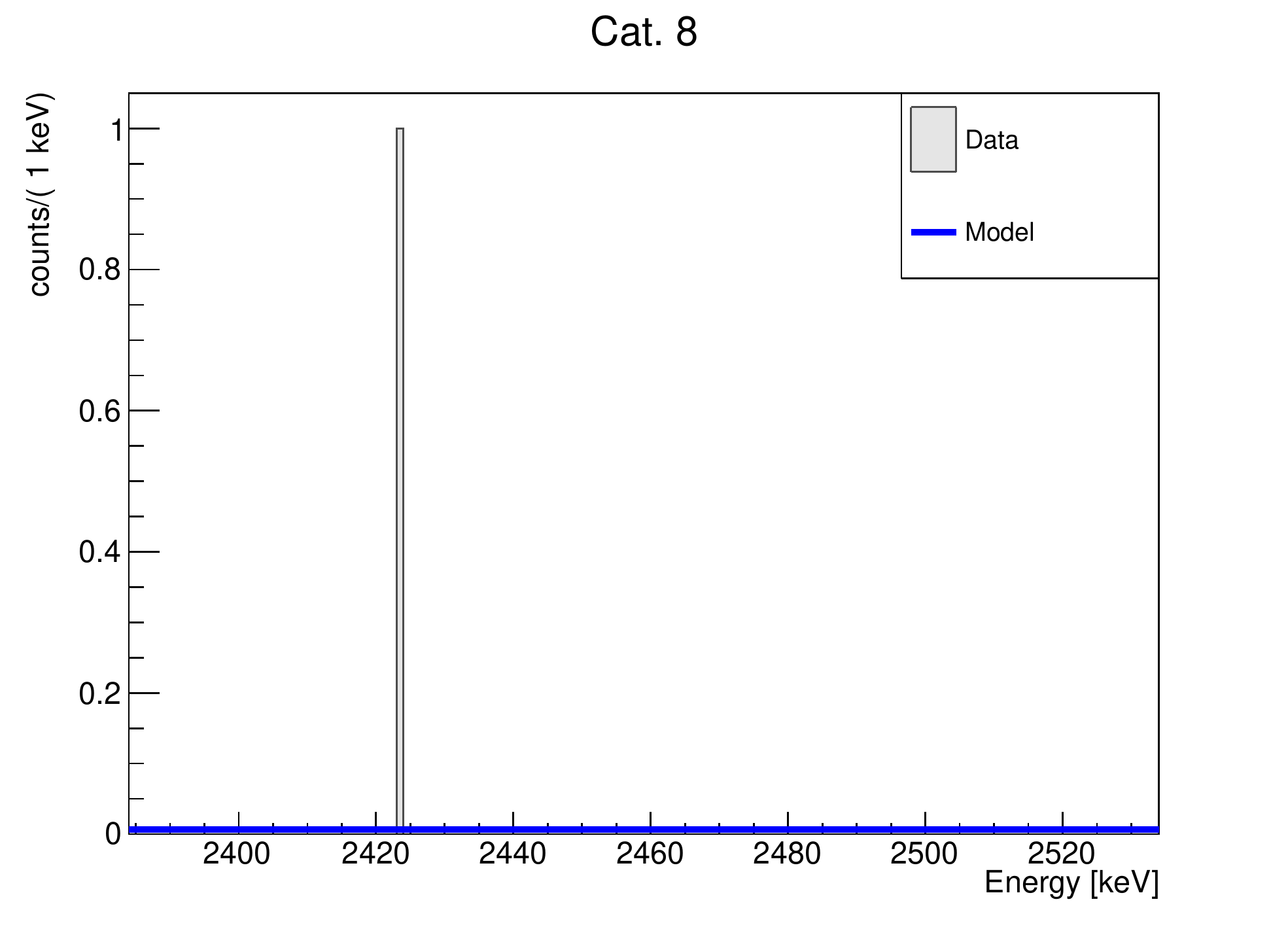}
\includegraphics[width=0.31\textwidth]{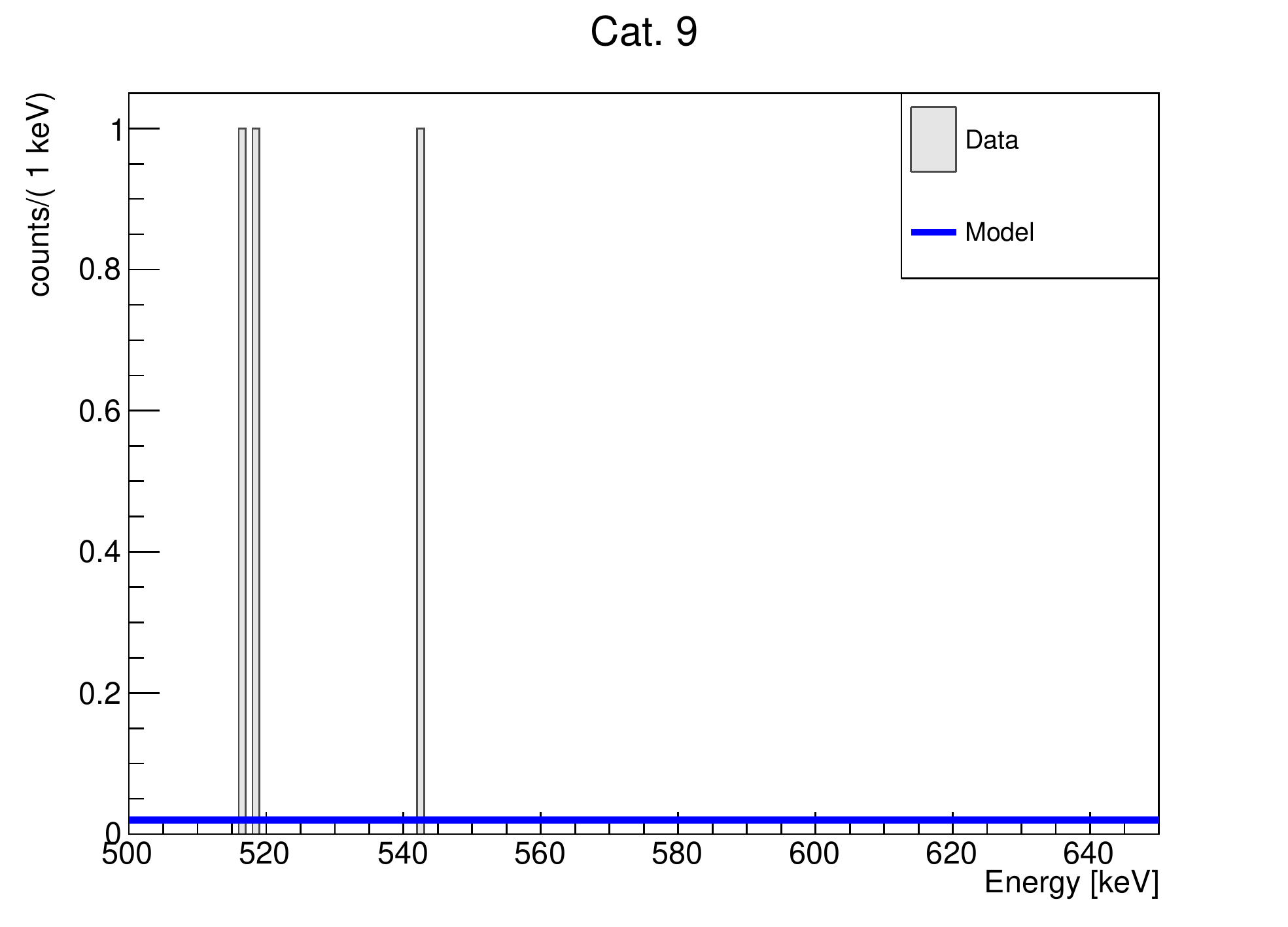}
\includegraphics[width=0.31\textwidth]{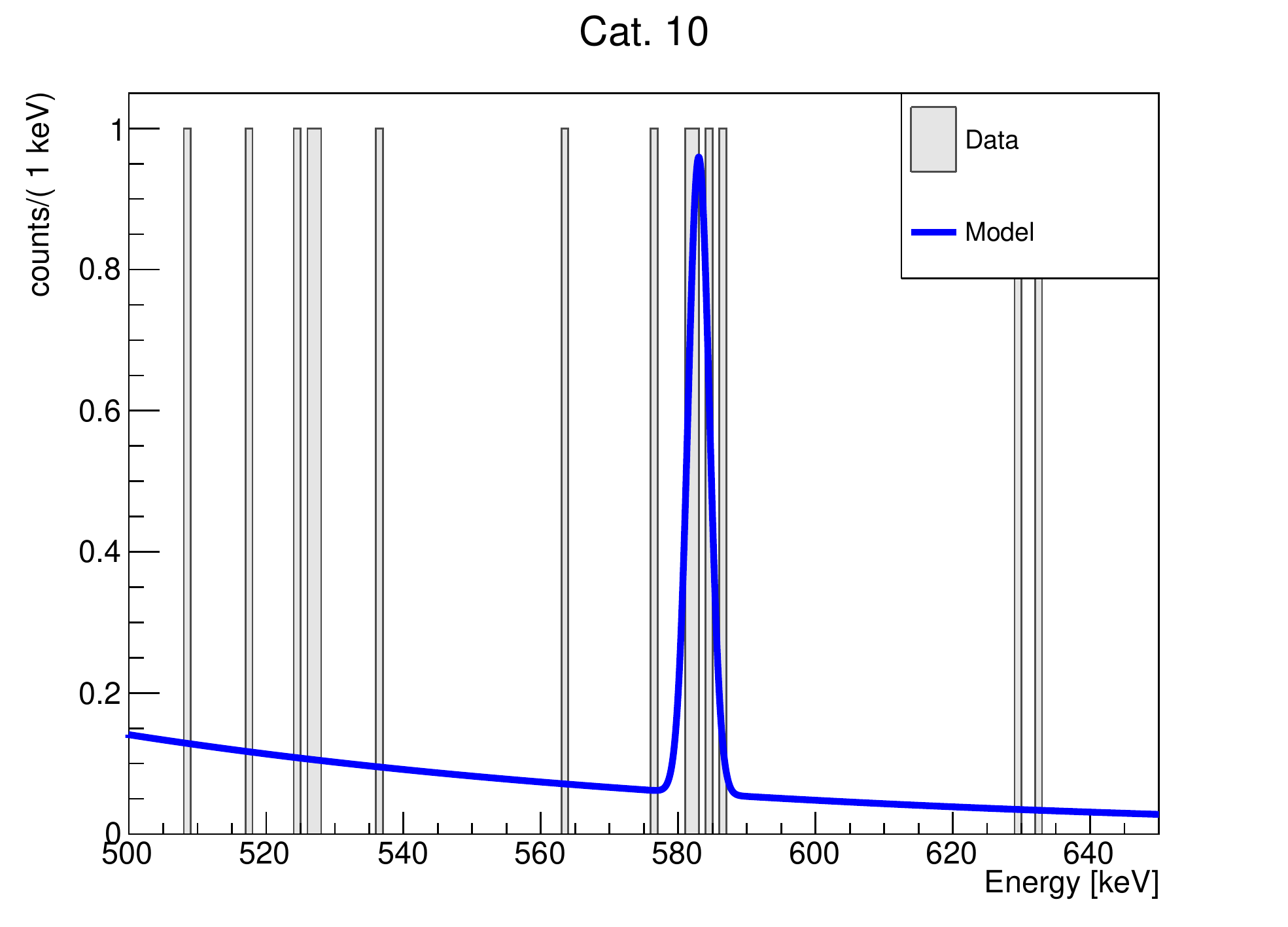}
\includegraphics[width=0.31\textwidth]{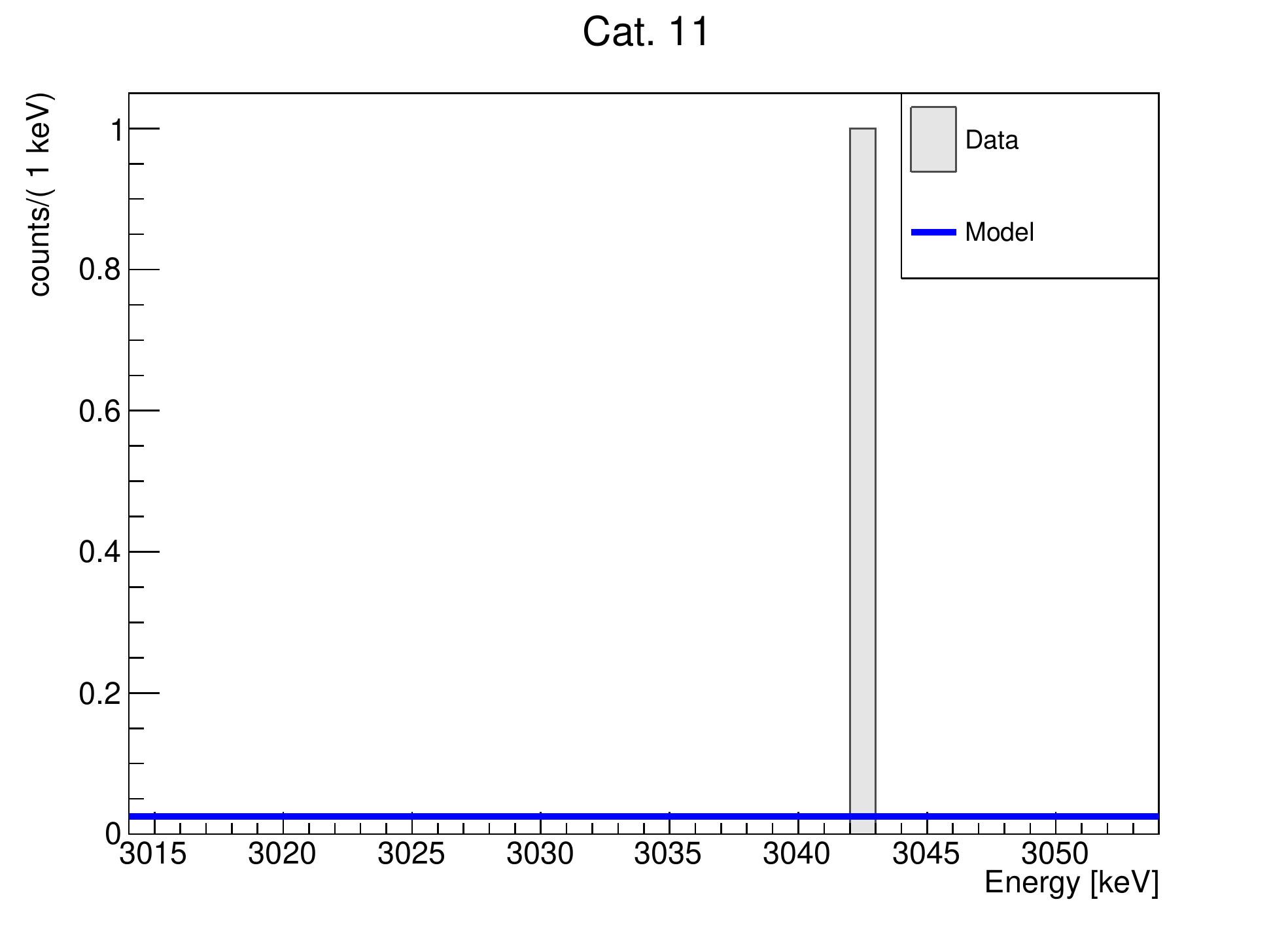}
\includegraphics[width=0.31\textwidth]{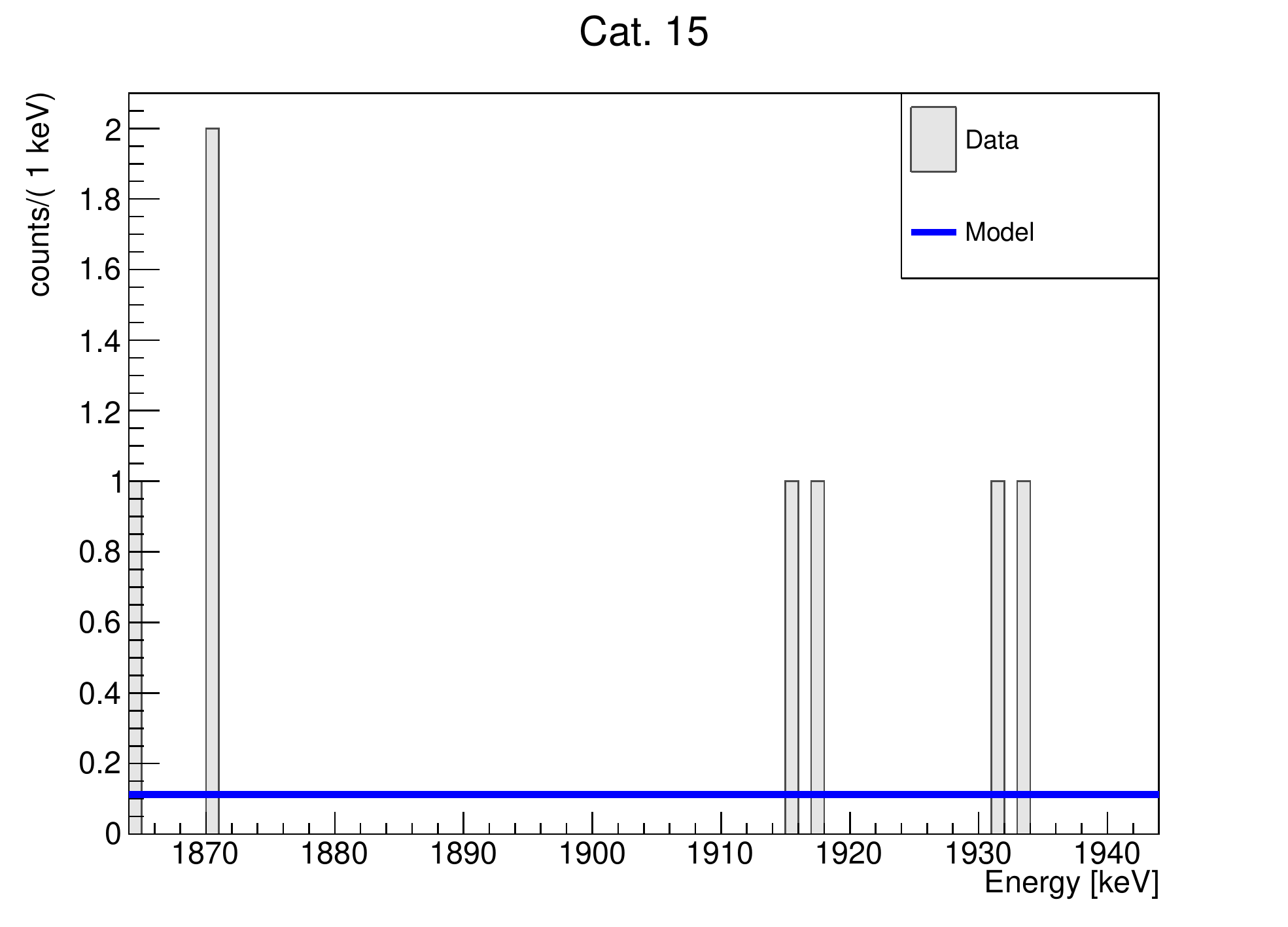}

    \caption{Fits of the $0\nu\beta\beta$ decay regions of interest, the experimental data is shown in black while the best fit (global mode) is shown in blue. We only show categories with non-zero counts in the fit region.}
    \label{fig:0vbb_fits}
\end{figure*}

We show the best fit reproduction of each category for both $2\nu\beta\beta$ and $0\nu\beta\beta$ decay analysis in Figs. \ref{fig:2vbb_fits} and \ref{fig:0vbb_fits}. Note that the spectra are binned for visualization purposes and we exclude plots with 0 counts in the experimental data. For the $2\nu\beta\beta$ decay fits we use the $H_1=(0_1^++B)$ model.

\end{document}